\documentclass[12pt,epsf]{article}
\pdfoutput=1

\usepackage{amsfonts}
\usepackage{amsmath}
\usepackage[colorlinks=true, urlcolor=dblue,linktocpage=true,linkcolor=dblue,citecolor=purple]{hyperref}
\usepackage{ulem}
\usepackage{tipa}
\usepackage{graphics,graphicx,xcolor}
\usepackage{setspace}
\usepackage{subcaption}
\usepackage{mathrsfs,amssymb}

\usepackage[letterpaper]{geometry}

\usepackage{amsthm}
\usepackage{tikz}
\usepackage{pgfplots}
\usepackage{xfrac}
\usepackage[toc,page]{appendix}
\usepackage{mathtools}
\usepackage{enumitem}
\usepackage{float}

\colorlet{dblue}{blue!70!black}

\newcommand{\mrm}{\mathrm}

\newcommand{\lb}{\left(}
\newcommand{\rb}{\right)}

\newcommand{\bL}{\mathbf{L}}

\newcommand\be{\begin{equation}}
\newcommand\ba{\begin{eqnarray}}
\newcommand\ee{\end{equation}}
\newcommand\ea{\end{eqnarray}}

\newcommand\happy{HaPPY tensor network~}
\newcommand\happies{HaPPY tensor networks~}
\newcommand\ads{$\mrm{AdS}_3$}

\newcommand\padic{$p$-adic }
\def\adscft#1{$\text{AdS}_{#1 + 1}/\text{CFT}_{#1}$}

\DeclarePairedDelimiter\abs{\lvert}{\rvert}

\DeclareMathOperator{\PL}{P^1}

\DeclareMathOperator{\SL}{SL}
\DeclareMathOperator{\SU}{SU}
\DeclareMathOperator{\SO}{SO}
\DeclareMathOperator{\PGL}{PGL}
\DeclareMathOperator{\PSL}{PSL}
\DeclareMathOperator{\U}{U}

\DeclareMathOperator{\Tr}{tr}

\DeclareMathOperator{\ord}{ord}
\DeclareMathOperator{\supp}{supp}

\def\R{\mathbb{R}}
\def\C{\mathbb{C}}
\def\Z{\mathbb{Z}}
\def\Q{\mathbb{Q}}

\def\FF{\mathbb{F}} 

\def\ket#1{\left|#1\right\rangle}
\def\Hilb{\mathscr{H}}
\def\inj{\hookrightarrow}

\def\Laplace{\bigtriangleup}
\def\goesto{\rightarrow}

\def\abr#1{\left\langle #1 \right\rangle}
\newtheorem{proposition}{Proposition}

\newcommand{\cH}{\mathcal{H}}
\newcommand{\cD}{\mathcal{D}}
\newcommand{\cF}{\mathcal{F}}
\newcommand{\cC}{\mathcal{C}}
\newcommand{\cO}{\mathcal{O}}
\newcommand{\fm}{\mathfrak{m}}

\def\bP{\mathrm{P}}
\def\bZ{\mathbb{Z}}
\def\bC{\mathbb{C}}
\def\bR{\mathbb{R}}
\def\bN{\mathbb{N}}
\def\bH{\mathbf{H}}
\def\bK{\mathbf{k}}

\newcommand{\bQ}{\mathbb{Q}}
\def\bF{\FF}

\numberwithin{equation}{section}
\usepackage{tikz-cd}
\DeclareMathOperator{\ev}{ev}

\newsavebox{\ramif}
\savebox{\ramif}{%
\begin{tikzpicture}[scale=0.48]
\tikzstyle{vertex}=[draw,scale=0.4,fill=black,circle]
\tikzstyle{ver2}=[draw,scale=0.6,circle]
\tikzstyle{ver3}=[draw,scale=0.4,circle]
\coordinate (C) at (0,0);
\foreach \x in {0,1,2} {
	\coordinate (a\x) at (\x*360/3 + 30:2);
	\coordinate (A\x) at (\x*360/3 + 30:1);
	\draw
		(a\x) -- (C);
	\draw
		(A\x) node[vertex] {};
	};
\foreach \x in {0,1,2} {
	\coordinate (aa\x) at (\x*360/3 - 30:2);
	\draw[dashed] (aa\x) -- (A\x);
	\foreach \y in {-1/2,1/2} {
		\coordinate (d\x\y) at (\x*360/3 - 30 +\y*360/12:3);
		\draw[dashed] (d\x\y) -- (aa\x);
		\foreach \z in {-1/2,1/2} {
			\coordinate (e\x\y\z) at (\x*360/3 - 30 +\y*360/12 + \z*360/24:4);
			\draw[dashed] (e\x\y\z) -- (d\x\y);
			};
		};
	};
\foreach \y in {0,1,...,5} 
		{
		\coordinate (b\y) at (\y*360/6:4);
		\pgfmathparse{floor(\y/2)};
		\draw
			(b\y) -- (a\pgfmathresult) node[pos=0.45,vertex](B\y) {};
		};
\foreach \y in {0,1,...,5} 
		{
		\pgfmathparse{mod(\y+1,2) ? "+" : "-"};
		\coordinate (c\y) at (\y*360/6 \pgfmathresult 360/18:4);
		\pgfmathparse{floor(\y/2)};
		\draw[dashed] (c\y) -- (B\y) {};
		};
\draw[dotted] (0,0) circle[radius=1];
\draw[dotted] (0,0) circle[radius=2];
\draw[dotted] (0,0) circle[radius=3];
\draw[dotted] (0,0) circle[radius=4];
\end{tikzpicture}	}
\newsavebox{\unramif}
\savebox{\unramif}{%
\begin{tikzpicture}[scale=0.48]
\tikzstyle{vertex}=[draw,scale=0.4,fill=black,circle]
\tikzstyle{ver2}=[draw,scale=0.6,circle]
\tikzstyle{ver3}=[draw,scale=0.4,circle]
\coordinate (C) at (0,0);
\foreach \x in {0,2,4} {
	\coordinate (a\x) at (\x*360/5:2);
	\draw
		(a\x) -- (C);
	\foreach \y in {0,1,2,3} {
		\coordinate (b\x\y) at (\x*360/5 + \y*360/20 - 1.5*360/20:4);
		\pgfmathparse{mod(\y,2) ? 
			:  "dashed"};
		\edef\type{\pgfmathresult};
		\draw[\type] (b\x\y) -- (a\x);
		};
	};
\foreach \x in {1,3} {
	\coordinate (a\x) at (\x*360/5:2);
	\draw[dashed] (a\x) -- (C);
	\foreach \y in {0,1,2,3} {
		\coordinate (b\x\y) at (\x*360/5 + \y*360/20 - 1.5*360/20:4);
		\draw[dashed] (b\x\y) -- (a\x);
		};
	};
\draw[dotted] (0,0) circle[radius=1];
\draw[dotted] (0,0) circle[radius=2];
\draw[dotted] (0,0) circle[radius=3];
\draw[dotted] (0,0) circle[radius=4];
\end{tikzpicture}	}
\newsavebox{\basefld}
\savebox{\basefld}{%
\begin{tikzpicture}[scale=0.48]
\tikzstyle{vertex}=[draw,scale=0.4,fill=black,circle]
\tikzstyle{ver2}=[draw,scale=0.6,circle]
\tikzstyle{ver3}=[draw,scale=0.4,circle]
\coordinate (C) at (0,0);
\foreach \x in {0,1,2} {
	\coordinate (a\x) at (\x*360/3 + 30:2);
	\draw (a\x) -- (C);
	};
\foreach \x in {0,1,...,5} {
	\coordinate (b\x) at (\x*360/6:4);
	\pgfmathparse{floor(\x/2)}
	\draw (b\x) -- (a\pgfmathresult);
	};
\draw[dotted] (0,0) circle[radius=1];
\draw[dotted] (0,0) circle[radius=2];
\draw[dotted] (0,0) circle[radius=3];
\draw[dotted] (0,0) circle[radius=4];
\end{tikzpicture}	}

\usepackage{setspace}

\begin{document}

\begin{spacing}{1.3}
\begin{titlepage}

\begin{center}
{\Large \bf Tensor networks, $p$-adic fields, and algebraic curves: \\ arithmetic and the $\text{AdS}_3/\text{CFT}_2$ correspondence}

\vspace*{6mm}

Matthew Heydeman,$^*$ Matilde Marcolli,$^\dagger$ Ingmar A. Saberi,$^{*\ddagger}$ and Bogdan Stoica$^{*\S}$

\vspace{5mm}
{
\setstretch{0.95}
\textit{
$^*$ Walter Burke Institute for Theoretical Physics,\\ California Institute of Technology, 452-48, Pasadena, CA 91125, USA}\\ 

\vspace{2mm}

\textit{
$^\dagger$ Department of Mathematics,\\ California Institute of Technology, 253-37, Pasadena, CA 91125, USA}\\ 

\vspace{2mm}

\textit{
$^\ddagger$ 
Present address: Mathematisches Institut, Ruprecht-Karls-Universit\"at Heidelberg,\\ Im Neuenheimer Feld 205, 69120 Heidelberg, Germany
}\\ 

\vspace{2mm}

\textit{
$^\S$ 
Present address:
Martin A. Fisher School of Physics,\\ Brandeis University, Waltham, MA 02453, USA\\
and \\
Department of Physics, Brown University, Providence, RI 02912, USA
}\\ 
}

\vspace{5mm}

{\tt  mheydema@caltech.edu, matilde@caltech.edu, saberi@mathi.uni-heidelberg.de, bstoica@brandeis.edu}

\vspace*{3mm}
\end{center}
\begin{abstract}
{
One of the many remarkable properties of conformal field theory in two dimensions is its connection to algebraic geometry. Since every compact Riemann surface is a projective algebraic curve, many constructions of interest in physics (which \textsl{a priori} depend on the analytic structure of the spacetime) can be formulated in purely algebraic language. This opens the door to interesting generalizations, obtained by taking another choice of field: for instance, the $p$-adics. We generalize the AdS/CFT correspondence according to this principle; the result is a formulation of holography in which the bulk geometry is discrete---the Bruhat--Tits tree for $\mathrm{PGL}(2,\mathbb{Q}_p)$---but the group of bulk isometries nonetheless agrees with that of boundary conformal transformations and is not broken by discretization. We suggest that this forms the natural geometric setting for tensor networks that have been proposed as models of bulk reconstruction via quantum error correcting codes; in certain cases, geodesics in the Bruhat--Tits tree reproduce those constructed using quantum error correction. Other aspects of holography also hold: Standard holographic results for massive free scalar fields in a fixed background carry over to the tree, whose vertical direction can be interpreted as a renormalization-group scale for modes in the boundary CFT.
Higher-genus bulk geometries (the BTZ black hole and its generalizations) can be understood straightforwardly in our setting, and the Ryu-Takayanagi formula for the entanglement entropy appears naturally.
}
\end{abstract}
\vspace{5mm}

\begin{flushleft}{\qquad CALT-TH 2016--013}\end{flushleft}

\end{titlepage}
\end{spacing}

\vskip 1cm

\setcounter{tocdepth}{2}

\tableofcontents

\vfill\eject

\begin{spacing}{1.3}

\section{Introduction}


Much attention has been paid of late to ideas that allow certain features of conformal field theory, such as long-range correlations, to be reproduced in lattice systems or other finitary models. As an example, the multiscale entanglement renormalization ansatz (or MERA), formulated by Vidal in~\cite{Vidal}, provides an algorithm to compute many-qubit quantum states whose entanglement properties are similar to those of the vacuum state in a conformal field theory. In Vidal's method, the states of progressively more distant qubits are entangled using successive layers of a self-similar network of finite tensors.

These proposals can typically be thought of as constructing analogues of the CFT vacuum state using a quantum circuit with an additional ``spatial direction,'' consisting of the successive computational layers of the circuit, and corresponding roughly to the distance scale up to which long-range entanglement has been introduced. As such, they are strongly suggestive of the AdS/CFT correspondence~\cite{GKP,Maldacena,Witten}, in which a $d$-dimensional conformal field theory is related to a gravitational theory in $d+1$-dimensional negatively curved spacetime,  and the extra direction can be interpreted as a renormalization scale (or equivalently a length scale) from the perspective of the boundary theory.
Furthermore, the construction of the layers (in which the number of tensors scales exponentially with the number of layers) bears comparison with the geometry of hyperbolic space. 
It was thus natural to search for a connection with holography. In~\cite{Swingle}, Swingle proposed that MERA might be a natural discretization of AdS/CFT, in which the holographic direction (or renormalization scale) corresponds to the successive layers of the tensor network, and individual tensors are associated to ``primitive cells'' of the bulk geometry. 
However, successive work~\cite{Bao} identified constraints that prevent such an AdS/MERA correspondence from fully reproducing all features of the bulk physics.

Motivated by this similarity, further new connections between quantum information theory and holography were made in \cite{ADH}, which pointed out that bulk reconstruction and bulk locality in the AdS/CFT correspondence bear strong similarities to the properties of quantum error-correcting codes. 
This intuition was used in~\cite{HaPPY} to construct a family of ``holographic'' quantum codes, associated to hyperbolic tilings. In these codes, bulk qubits are thought of as the logical inputs, the boundary qubits at the periphery of the tiling constitute the encoded state, and the error-correcting properties of the code mimic features of holography such as the Ryu-Takayanagi formula~\cite{RT}.

In this paper, we propose that discrete holographic models should be understood as approximating bulk geometry in a fundamentally different way. We are guided by considering a new and orthogonal direction in which the $\text{AdS}_3/\text{CFT}_2$ correspondence can be generalized, and construct a family of lattice field theories along these lines.
Unlike tensor network models, our models are fully dynamical theories, with path integral descriptions. Discrete bulk geometries (based on the $p$-adic numbers) appear naturally. Despite this, essential and basic features of AdS/CFT, such as bulk isometries and boundary conformal symmetry (which are destroyed by a naive discretization), have analogues and can be fully understood in the discrete setting.

The bulk geometries relevant to the $\text{AdS}_3/\text{CFT}_2$ correspondence are well understood. 
The most well-known black hole solution is that of Ba\~nados, Teitelboim, and Zanelli~\cite{BTZ}; this solution was generalized to a family of higher-genus Euclidean black holes by Krasnov~\cite{Kras}.
These solutions can be understood in general using the technique of \textsl{Schottky uniformization}, which presents a higher-genus black hole as the quotient of empty $\text{AdS}_3$ by a particular discrete subgroup of its isometries. 
 
In \cite{ManMar}, a holographic correspondence was established for these three-dimensional geometries.
This correspondence expresses the conformal two point correlation
function on the conformal boundary at infinity (a Riemann surface $X_\Gamma$ of genus $g$)
in terms of geodesic lengths in the bulk space (a hyperbolic handlebody $\cH_\Gamma$ of
genus $g$). The formula relating the boundary theory to gravity in the bulk is based
on Manin's result~\cite{Man91} on the Arakelov Green's function.

However, we consider $\text{AdS}_3/\text{CFT}_2$ 
not merely because it is a simple setting for holography. 
For us, the crucial property of conformal field theory in two dimensions is its strong ties to algebraic geometry. These occur because every compact Riemann surface is a projective algebraic curve, so that many of the analytic concepts that arise in physics can (in two-dimensional contexts) be reformulated in purely algebraic terms. 
Once a concept can be formulated algebraically, it has many natural generalizations, obtained by changing the field of numbers one is considering. For instance, given a Riemann surface as the zero locus of a polynomial equation with rational coefficients, one can ask for the set of solutions over~$\C$, over~$\R$, over more exotic fields like the $p$-adics, or even over the integers. 

The aforementioned holographic formula---and the whole geometric setting of the correspondence, consisting of the Euclidean hyperbolic space $\text{AdS}_3$, its conformal
boundary $\bP^1(\bC)$, and quotients by actions of Schottky groups 
$\Gamma \subset \PSL(2,\bC)$---has a natural analogue in which the field is the $p$-adic numbers~$\Q_p$.
The bulk space
becomes the Bruhat--Tits tree of $\Q_p$, which is a manifestly discrete infinite graph of uniform valence. 
Its conformal boundary at infinity is $\PL(\Q_p)$, which can be thought of as the spacetime for an unusual class of CFTs. Black hole solutions are understood to be quotients of this geometry by $p$-adic Schottky groups $\Gamma \subset \PGL(2,\Q_p)$; these are known as \textsl{Mumford curves} in the mathematics literature. The results of Drinfeld and Manin \cite{DriMan} on periods
of $p$-adic Schottky groups provide the corresponding holographic formula in this
non-archimedean setting.
We will give what we hope are intuitive introductions to these possibly unfamiliar concepts in the bulk of the paper.

Conformal field theory on $p$-adic spacetime has previously been developed, for the most part, in the context of the $p$-adic string theory (see, for instance, \cite{BrFr} and references therein), but has also been considered abstractly~\cite{Melzer}. However, our perspective on the subject will be somewhat different: rather than using the $p$-adics as a worldsheet to construct real-space string amplitudes, 
our
goal in this paper is to further develop the original holographic correspondence of \cite{ManMar} 
for the higher-genus black holes, informed by recent developments in the understanding of the AdS/CFT correspondence.
We will emphasize the large extent to which algebraic structure allows familiar ideas, concepts, and arguments from ordinary $\text{AdS}_3/\text{CFT}_2$ can be carried over---in many cases line by line---to the $p$-adic setting. In addition to the holographic formulas of Manin and~Marcolli, the standard semiclassical holographic analysis of scalar fields propagating without backreaction in anti-de~Sitter space applies almost without alteration to the Bruhat--Tits tree. We discuss this in detail in \S\ref{scalars}. 

In some cases, intuitions about how holography works in the archimedean case are supported even more sharply over the $p$-adics. For example, one normally thinks of the holographic direction as corresponding to a renormalization-group scale. Over the $p$-adics, as shown in~\S\ref{reconstruction}, boundary modes contribute to the reconstruction of bulk functions only up to a height determined by their wavelength, and reconstruct precisely to zero above this height in the tree. This result is foreshadowed in the literature by the observation that renormalization-group methods become exact in the context of hierarchical models (see, for instance,~\cite{WilsonKogut,LernerMissarov}).

One of the most important new ideas in the AdS/CFT correspondence is the study of entanglement entropy in boundary states and its connection (via the Ryu-Takayanagi formula) to the geometry of the bulk. While there is no definitive calculation at this point, we argue that, at least for the $p$-adic free boson CFT, an analogue of the familiar logarithmic scaling of the ground-state entanglement entropy is likely to hold. Given such a formula, the Ryu-Takayanagi formula follows immediately from simple considerations of the geometry of the tree.

Tensor network models are often of interest because they reproduce our expectations about ground-state entanglement entropy, and in some cases (like the holographic quantum code of Pastawski \textsl{et~al.}~\cite{HaPPY}) also satisfy formulas similar to Ryu-Takayanagi that relate the entanglement entropy to the size of paths or surfaces in the interior of the network. Given that our models exhibit a discrete bulk spacetime, a Ryu-Takayanagi formula, and a meaningful (and unbroken) group of bulk isometries/boundary conformal mappings, we suggest that the $p$-adic geometry is the natural one to consider in attempting to link tensor network models to spacetimes. We offer some ideas in this direction in~\S\ref{happy}. 

Finally, on an even more speculative note, it is natural to wonder if the study of $p$-adic models of holography can be used to learn about the real case. So-called ``adelic formulas'' relate quantities defined over the various places (finite and infinite) of~$\Q$; it was suggested in \cite{Man89a} that fundamental physics should
be adelic in nature, with product formulae that relate the archimedean side of
physics to a product of the contributions of all the $p$-adic counterparts. We briefly speculate about adelic formulas for the entanglement entropy in~\S\ref{EE}; one might hope that such formulas could be used to prove inequalities for entanglement entropy like those considered in~\cite{cone}, using ultrametric properties of the $p$-adics. We hope to further develop the adelic perspective, and return to these questions, in future work.

\section{Review of necessary ideas}

\subsection{Basics of $p$-adic numbers}
\label{sec:basics}

We begin with a lightning review of elementary properties of the $p$-adic fields. Our treatment here is far from complete; 
for a more comprehensive exposition, the reader is referred to~\cite{Koblitz}, or to another of the many books that treat $p$-adic techniques.

When one constructs the continuum of the real numbers from the rationals, one completes with respect to a metric: the distance between two points $x,y\in\Q$ is
\begin{equation}
d(x,y) = \abs{x-y}_\infty,
\end{equation}
where $\abs{\cdot}_\infty$ is the usual absolute value.
There are Cauchy sequences of rational numbers for which successive terms become arbitrarily close together, but the sequences do not approach any limiting rational numbers. The real numbers ``fill in the gaps,'' such that every Cauchy sequence of rational numbers converges to a real limit by construction. This property is known as metric completeness.

The $p$-adic fields $\Q_p$ are completions of~$\Q$ with respect to its other norms; there is one such norm for every prime $p$. These $p$-adic norms are defined by
\begin{align}
\abs{x}_p &= p^{-\ord_p(x)} ;\\
\ord_p(x) &= n \text{ when } x = p^n(a/b)\text{ with }a,b\perp p.
\end{align}
Every rational number $x$ has a unique prime factorization, and the (possibly negative) integer $n$ labels the power of $p$ which divides $x$. Two norms denoted $a$, $b$ are considered equivalent if $|x|_a = |x|_b^\gamma$ for some positive real constant $\gamma$; by a theorem of Ostrowski, every possible norm on~$\Q$ is equivalent either to one of the $p$-adic norms, the usual ($\infty$-adic) norm, or to the trivial norm for which $|x|_0=1$ $\forall x \neq 0$. Thus, the nontrivial norms (or possible completions) are labeled by the primes together with~$\infty$. It is common to refer to the different possible completions as the different ``places'' of~$\Q$. 

A number is $p$-adically small when it is divisible by a large power of~$p$; one can think of the elements of~$\Q_p$ as consisting of decimal numbers written in base~$p$, which can extend infinitely far \textsl{left} (just as real numbers can be thought of as ordinary decimals extending infinitely far \textsl{right}). $\Q_p$ is uncountable and locally compact with respect to the topology defined by its metric; as usual, a basis for this topology is the set of open balls,
\begin{equation}
B_\epsilon(x) = \{ y \in \Q_p: \abs{x-y}_p < \epsilon \} .
\end{equation}

The ring of integers $\Z_p$ of~$\Q_p$ is also the unit ball about the origin:
\begin{equation}
\Z_p = \{ x \in \Q_p : \abs{x}_p \leq 1 \}. 
\end{equation}
It can be described as the inverse limit of the system of base-$p$ decimals with no fractional part and finite (but increasingly many) digits:
\begin{equation}
\Z_p = \varprojlim \left( \cdots \goesto \Z/p^{n+1}\Z \goesto \Z/p^n\Z \goesto \Z/p^{n-1}\Z \goesto \cdots \right) . 
\end{equation}
$\Z_p$ is a discrete valuation ring; its unique maximal ideal is $\fm = p\Z_p$, and the quotient of $\Z_p$ by~$\fm$ is the finite field~$\FF_p$. In general, for any finite extension of~$\Q_p$, the quotient of its ring of integers by its maximal ideal is a finite field~$\FF_{p^n}$; we give more detail about this case in~\S\ref{matilde-tree}.

\subsection{The Bruhat--Tits tree and its symmetries}
\label{matt-tree}

In this section, we will describe the Bruhat--Tits Tree $T_p$ and its symmetries. It should be thought of as a hyperbolic (though discrete) bulk space with conformal boundary~$\PL(\mathbb{Q}_p)$. Since these trees are a crucial part of the paper and may be unfamiliar to the reader, our treatment is informal, and aims to build intuition. Out of necessity, our discussion is also brief; for a more complete treatment, the reader may consult notes by Casselman~\cite{Casselman} for constructions and properties related to the tree, or~\cite{Zabrodin} for analysis on the tree and connections to the $p$-adic string. 

We begin with a description of the boundary and its symmetries, which are completely analogous to the global conformal transformations of $\PL(\mathbb{C})$. We then turn our attention to the bulk space $T_p$, focusing on its construction as a coset space and the action of $\PGL(2, \mathbb{Q}_p)$ on the vertices. Despite the fractal topology of the $p$-adic numbers, we will find (perhaps surprisingly) that many formulas from the real or complex cases are related to their $p$-adic counterparts by the rule $\abs{\cdot}_\infty \rightarrow \abs{\cdot}_p$.

\begin{figure}[t]
\includegraphics[width=13cm]{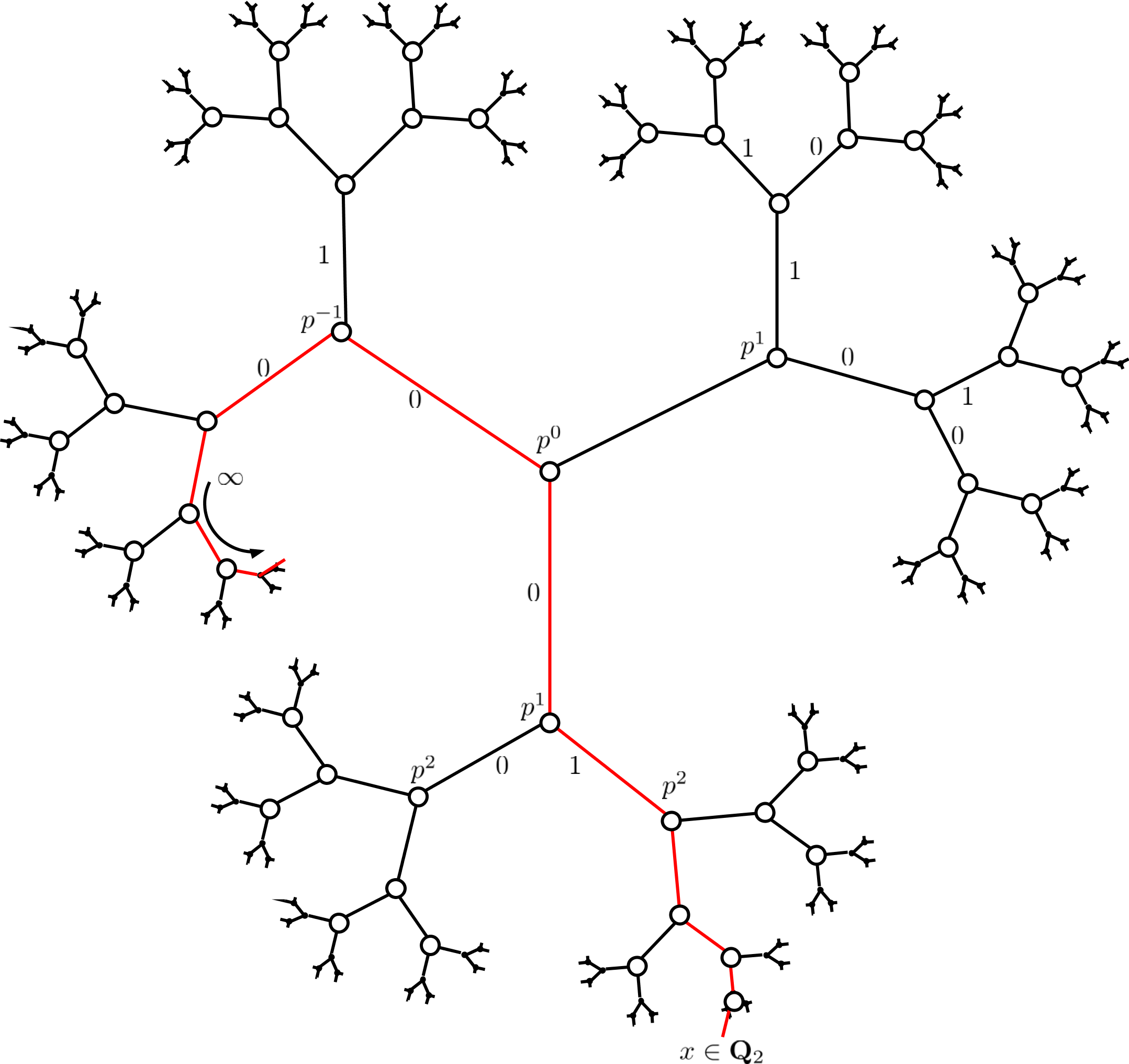} 
\centering
\caption{The standard representation of the Bruhat--Tits tree. The point at infinity and the center are arbitrary as the tree is homogeneous. Geodesics such as the highlighted one are infinite paths through the tree from $\infty$ to the boundary which uniquely specify elements of $\mathbb{Q}_p$. This path as a series specifies the digits of the decimal expansion of $x \in \mathbb{Q}_2$ in this example. At the $n$th vertex, we choose either $0$ or $1$ corresponding to the value of $x_n$ in the $p^n$th term of $x$. Negative powers of $p$ correspond to larger $p$-adic norms as we move towards the point $\infty$.}
\label{standardtree}
\end{figure}

\subsubsection{Conformal group of $\PL(\mathbb{Q}_p)$}

The global conformal group on the boundary is $\SL(2,\mathbb{Q}_p)$, which consists of matrices of the form
\begin{align}
A = \begin{pmatrix}
a & b \\
c & d
\end{pmatrix}, \text{ with }{a,b,c,d} \in \mathbb{Q}_p, \ ad-bc = 1.
\end{align}
This acts on points $x \in \PL(\mathbb{Q}_p)$ by fractional linear transformations,
\begin{align}
x \rightarrow \frac{ax + b}{cx+d}.
\end{align}
It can be checked that matrix multiplication corresponds to composition of such maps, so that the group action is well-defined. This is analogous to the $\SL(2,\mathbb{C})$ action on the Riemann sphere $\PL(\mathbb{C})$. (We will sometimes also refer to $\PGL(2,\Q_p)$; the two differ only in minor details.)

The existence of a local conformal algebra for~$\mathbb{Q}_p$, in analogy with the Virasoro symmetry in two-dimensional conformal field theory or general holomorphic mappings of~$\PL(\C)$, is a subtle question. It is difficult to find definitions of a $p$-adic derivative or an infinitesimal transformation that are satisfactory for this purpose. In particular, since the ``well-behaved'' complex-valued functions on~$\Q_p$ are in some sense locally constant, there are no interesting derivations that act on the space of fields~\cite{Melzer}. In this paper, we will concern ourselves only with global symmetries, which can still be used to constrain the properties of $p$-adic conformal field theories. We speculate about the possibility of enhanced conformal symmetry in~\S\ref{sec:Drinfeld}.

The determinant condition implies that there are three free $p$-adic numbers which specify an element of $\SL(2,\mathbb{Q}_p)$. A convenient way to decompose a general $\SL(2, \mathbb{Q}_p)$ transformation is to view it as the product of a special conformal transformation, a rotation, a dilatation, and a translation:
\begin{align}
 \begin{pmatrix}
1 & 0 \\
cp^{-m}a^{-1} & 1
\end{pmatrix} 
 \begin{pmatrix}
a & 0 \\
0 & a^{-1}
\end{pmatrix} 
 \begin{pmatrix}
p^m & 0 \\
0 & p^{-m}
\end{pmatrix} 
 \begin{pmatrix}
1 & bp^{-m}a^{-1} \\
0 & 1
\end{pmatrix} =
\begin{pmatrix}
p^ma & b \\
c & p^{-m}a^{-1}(1+bc)
\end{pmatrix},
\end{align}
where $a, b, c \in \mathbb{Q}_p$ and $|a|_p =1$. One can verify that the product is an arbitrary element of $\SL(2, \mathbb{Q}_p)$, where the determinant condition has been used to eliminate the $d$ parameter. This represents a translation by $bp^{-m}a^{-1}$, a dilatation by $p^{2m}$, a rotation by $a^2$, and a special conformal translation by $cp^{-m}a^{-1}$. We have separated the diagonal subgroup into multiplication by elements of the unit circle, $a \in \mathbb{U}_p \subset \mathbb{Z}_p$, which do not change the $p$-adic norm (and thus are ``rotations'' in a $p$-adic sense), and multiplication by powers of $p$ which scale the $p$-adic norm (and so correspond to dilatations). Representations of the multiplicative group of unit $p$-adics provide an analogue of the spin quantum number; we discuss this further in~\S\ref{sec:spin}. It is worth stressing that these transformations are finite, and so we are characterizing the symmetry group rather than the algebra.

As is often the case in real conformal field theories, we can focus on the dilation subgroup. A diagonal matrix in $\SL(2, \mathbb{Q}_p)$ and its action on the coordinate is
\begin{align}
\begin{pmatrix}
\alpha & 0 \\
0 & \alpha^{-1}
\end{pmatrix} , \hspace{3mm} x \rightarrow x' = \alpha^2 x.
\end{align}
This has the effect of changing the $p$-adic norm by
\begin{align}
|x'|_p = |\alpha|_p^2 |x|_p .
\end{align}
So if $|\alpha|_p \neq 1$, this will scale the size of coordinate. This parallels the complex case in which a dilatation changes the complex norm by $|z'| = |\alpha|^2|z|$. It will turn out to be the case that 2-point functions of spinless operators of dimension $\Delta$ in $p$-adic conformal field theory will depend only on the $p$-adic norm of their separation. Schematically,
\begin{equation}
\langle \phi(x) \phi(y) \rangle \approx \frac{1}{|x-y|_p^{2\Delta}}.
\end{equation}
Dilations will thus affect correlation functions of the $p$-adic conformal field theory exactly as in the complex case. 

\subsubsection{$\PGL(2, \mathbb{Q}_p)$ action on the tree $T_p$}

We have seen that fractional linear transformations of the boundary coordinate work as in the real case. The action of the symmetry on the bulk space $T_p$ is slightly more complicated to describe. Were we working in the archimedean theory, we would identify  $\PSL(2,\mathbb{R})$ as the isometry group of the hyperbolic upper half space  $\mathbb{H} =  \SL(2, \mathbb{R}) / \SO(2)$. Here~$\SO(2)$ is a maximal compact subgroup. Similarly, in the context of \adscft{2}, we can think of the hyperbolic upper-half 3-space as a quotient space of the isometry group by its maximal compact subgroup: $\bH^3 = \SL(2,\C)/\SU(2)$.

Following this intuition, we define the \textsl{Bruhat--Tits tree} to be the quotient of the $p$-adic conformal group by its maximal compact subgroup:
\begin{equation}
 T_p =  \PGL(2, \mathbb{Q}_p) / \PGL(2, \mathbb{Z}_p).
\end{equation}
In contrast with the archimedean examples, $T_p$ is a discrete space: it is a homogeneous infinite tree, with vertices of valence $p+1$, whose boundary can be identified with the $p$-adic projective line.
We expect isometries to correspond to rigid transformations of the vertices. Formally, the tree represents the incidence relations of equivalence classes of lattices in $\mathbb{Q}_p \times \mathbb{Q}_p$.  As outlined in the appendix of~\cite{BrFr}, the group $\PGL(2, \mathbb{Q}_p)$ acts by matrix multiplication on the lattice basis vectors and takes one between equivalence classes. These transformations are translations and rotations of the points in the tree; they preserve distances, which are measured in the tree by just counting the number of edges along a given path. Since any two vertices in a connected tree are joined by exactly one path, this is well-defined; all paths are geodesics.

A standard way of representing $T_p$ is depicted in Fig.~\ref{standardtree} for the case $p=2$. This is a regular tree with $p+1$ legs at each vertex; the exponential growth in the number of vertices with distance from a base point reflects the ``hyperbolic'' nature of distance in the tree. Since paths are unique, there is a one-to-one correspondence between infinite paths in the tree starting at $\infty$ and elements of~${\Q}_p$. (This can be viewed like a $p$-adic version of stereographic projection.) 

The choice of the apparent center and geodesic corresponding to infinity are arbitrary. Just as in the archimedean case, we must fix three boundary points to identify a $p$-adic coordinate on the projective line, corresponding to $0$, $1$, and~$\infty$. Once these arbitrary choices are made, the geodesics joining them form a Y in the bulk, whose center is taken to be the centerpoint of the tree. 
We can then understand the geodesic connecting~$\infty$ to~$x$ as labeling the unique $p$-adic decimal expansion for $x = p^{\gamma}(x_0+ x_1p + x_2 p^2 + \dots)$, where each of the $x_n$ take values in $0,1,\dots p-1$ corresponding to the $p$ possible choices to make at each vertex. Each vertex of the tree is naturally marked with a copy of the finite field~$\FF_p$, identified with one ``digit'' of a $p$-adic number.


Viewing the tree as the space of $p$-adic decimal expansions may in some ways be more useful than the definition in terms of equivalence classes of lattices. Geometrically moving closer or further from the boundary corresponds to higher or lower precision of $p$-adic decimal expansions. Even with no reference to quantum mechanics or gravity, we see some hint of holography and renormalization in the tree- a spatial direction in the bulk parameterizes a scale or precision of boundary quantities. This is explored more fully in~\S\ref{sec:modes}.

It is worth strongly emphasizing that the notion of dimension is quite confusing in the context of the tree. Many familiar intuitions go awry. For example, one might expect the unit circle $\{x \in \Q_p: |x|_p = 1\}$ to be a codimension-one object; open subsets of the unit circle in the subspace topology would then play the role of the intervals on which entanglement entropy is defined in two-dimensional CFT. However, following these steps for the tree quickly reveals that there is no difference between such a boundary ``interval'' and any other boundary open set! Indeed, the unit circle is an open set of positive measure.

One might reasonably therefore ask why we choose to emphasize the connection of our model with holomorphic AdS$_3$/CFT$_2$, rather than e.g.\ with AdS$_2$/CFT$_1$. One answer is that, from our algebraic perspective, these two instances are not very different: after all, they differ only by a choice of number field (respectively $\C$ or~$\R$). So either comparison to the archimedean case is warranted. Another answer we might give is that a key difference between the two cases has to do with the structure of the multiplicative group of units of the field: in~$\R$, this consists only of scale transformations (together with a $\Z_2$ reflection), whereas in~$\C$ it is the product of the scale transformations with a $\U(1)$ factor, the complex numbers of unit modulus, that gives rise to spin. In this sense, $\Q_p$ is more analogous to~$\C$: the unit circle (as a multiplicative group) is a nontrivial infinite group, whose representations are likely to play a role in the extension of our considerations here to fields of higher spin. (We remark on this possibility further in~\S\ref{sec:spin}). Yet a third answer would be that the free boson gives a conformally invariant theory only in two Euclidean dimensions, and it does in our setting as well.

We now illustrate some examples of $\PGL(2, \mathbb{Q}_p)$ transformations on the tree. First note that the choice of the center node is arbitrary. We can take this point to be the equivalence class of unit lattices modulo scalar multiplication. One can show that this equivalence class (or the node it corresponds to) is invariant under the $\PGL(2, \mathbb{Z}_p)$ subgroup, so these transformations leave the center fixed and rotate the branches of the tree about this point.

\begin{figure}[t]
\includegraphics[width=11cm]{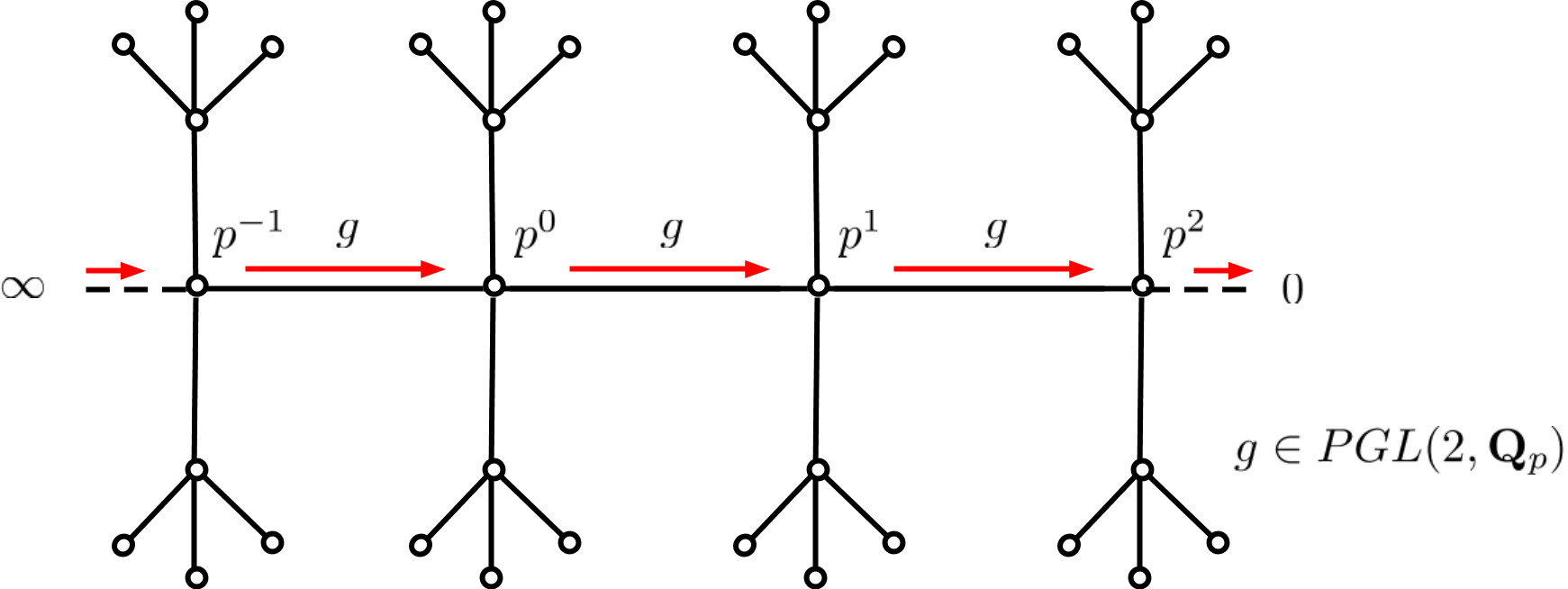} 
\centering
\caption{An alternative representation of the Bruhat--Tits Tree (for $p=3$) in which we have unfolded the tree along the $0$ geodesic. The action of elements of $\PGL(2, \mathbb{Q}_p)$ acts by translating the entire tree along different possible geodesics. In this example we translate along the $0$ geodesic, which can be thought of as multiplication of each term in a $p$-adic decimal expansion by $p$. This map has two fixed points at $0$ and $\infty$. In this ``unfolded'' form, a point in $\PL(\mathbb{Q}_p)$ is specified by a geodesic that runs from $\infty$ and follows the $0$ geodesic until some level in the tree where it leaves the $0$ geodesic towards the boundary. The $p$-adic norm is simply $p$ to the inverse power of the point where it leaves the $0$ geodesic (so leaving ``sooner'' leads to a larger norm, and later to a smaller norm).}
\label{lineartreetranslation}
\end{figure}

More interesting is a generator such as 
\begin{equation}
g = \begin{pmatrix}
p & 0 \\
0 & 1
\end{pmatrix}
\in \PGL(2, \mathbb{Q}_p).
\end{equation}
This transformation (and others in $\PGL(2, \mathbb{Q}_p)$) act by translating the entire tree along a given geodesic (one can see this either from the lattice incidence relations, or from translating or shifting the $p$-adic decimal series expansion). This is illustrated in Fig.~\ref{lineartreetranslation}. We can think of these transformations as the lattice analogs of translations and dilatations of the real hyperbolic plane. 

\subsubsection{Integration measures on $p$-adic spaces}
\label{sec:measures}

Just as is the case for~$\C$, there are two natural measures on~$\Q_p$ (or more properly, on the projective line over~$\Q_p$); they can be understood intuitively by thinking of~$\Q_p$ as the boundary of~$T_p$. The first is the Haar measure~$d\mu$, which exists for all locally compact topological groups. With respect to either measure, the size of the set of $p$-adic integers is taken to be 1:
\begin{equation}
\mu(\Z_p) = 1.
\end{equation}
The Haar measure is then fixed by multiplicativity and translation invariance; any open ball has measure equal to the $p$-adic norm of its radius. It is helpful to think of $\Q_p$ as being ``flat'' when considered with this measure. 

The other measure, the Patterson-Sullivan measure, is the $p$-adic analogue of the Fubini-Study metric on~$\PL(\C)$. It is most easily defined with reference to the tree, in which we fix a basepoint~$C$ (to be thought of as the unique meeting point of the geodesics joining $0$, $1$, and~$\infty$ when a coordinate is chosen on the boundary). Recall that the open balls in~$\Q_p$ correspond to the endpoints of branches of the tree below a vertex~$v$. In the Patterson-Sullivan measure, 
\begin{equation}
d\mu_0(B_v) = p^{-d(C,v)}. \label{PattersonSullivan}
\end{equation}
The two measures are related by 
\begin{alignat}{2}
d\mu_0(x) &= d\mu(x), \qquad &&|x|_p\leq 1; \nonumber \\ 
d\mu_0(x) &= \frac{d\mu(x)}{|x|_p^2}, \qquad &&|x|_p>1. \label{measures}
\end{alignat}
(Later on, we will at times use the familiar notation~$dx$ to refer to the Haar measure.) 
The most intuitive way to picture the Patterson-Sullivan measure is to imagine the tree pointing ``radially outward'' from its centerpoint, as in Fig.~\ref{standardtree}. This should be contrasted with a picture such as Fig.~\ref{ryufigure}, which is drawn in a natural way from the standpoint of the Haar measure. It is then easy to understand the transformation rule~\eqref{measures}; it says that when all geodesics point downward from infinity and the boundary is ``flat'' at the lower end of the picture, points far from zero (outside~$\Z_p$) can only be reached by geodesics that travel upward from~$C$ before turning back down towards the boundary. 

\subsubsection{Finite extensions of $p$-adic fields}
\label{sec:extensions}

Some basic facts regarding the geometry of the Bruhat--Tits
tree $T_p$ of $\bQ_p$ have been recalled throughout~\S\ref{matt-tree}.
More generally, though, the geometry we consider here applies to any
finite extension $\bK$ of the $p$-adic field $\bQ_p$ without any essential changes. 
We recall a couple of standard facts about finite extensions of local fields; the reader is referred e.g.\ to~\cite{Koblitz} for details. Let $n = [\bK:\Q_p]$ denote the degree of the extension. Firstly, there exists a unique norm on~$\bK$ as a vector space over~$\Q_p$, extending the standard $p$-adic norm. This is \textsl{not} identical with the usual ``norm map'' of a field extension! Rather,
\begin{equation}
|\alpha| = |\mathbb{N}_{\bK/\Q_p}(\alpha)|_p^{1/n}.
\end{equation}
(Remember that the field norm on~$\C$ is the \textsl{square} of the absolute value.)
By analogy with the ordinary $p$-adic field, we can define an extension of the ``order'' to all of~$\bK$:
\begin{equation}
\ord_p(\alpha) = - \log |\alpha| \in \frac{1}{n}\Z.
\end{equation}
The image of~$\bK$ under the map~$\ord_p$ is an additive subgroup $(1/e_\bK)\Z \subset (1/n)\Z$, for some integer $e_\bK \mid n$; this number is called the \textsl{ramification index} of~$\bK$. 

The ring of integers of~$\bK$ is a discrete valuation ring, with a unique maximal ideal that is easy to describe using this norm:
\begin{equation}
\cO_\bK = \{ \alpha \in \bK : |\alpha| \leq 1\}; \quad
\fm = \{ \alpha \in \bK : |\alpha| < 1 \}.
\end{equation}
(For~$\Q_p$, $\cO = \Z_p$ and $\fm = p\Z_p$.) Furthermore, the residue field $\cO_\bK$ is a finite extension field of $\FF_p = \Z_p / p\Z_p$, of degree $f$ between $1$ and~$n$. 
In fact, one can prove that $n = e_\bK f$, and further that there always exists
 an intermediate subfield $\bL$ of~$\bK$, fitting into the diagram
\begin{equation}
\begin{tikzcd}[row sep=small,column sep=small]
\bK \arrow[dash,"n"']{dd} \arrow[dash,"e_\bK"]{dr}&  \\
& \bL \arrow[dash,"f"]{dl} \\
\Q_p & \\
\end{tikzcd},
\end{equation}
such that $\bL$ is unramified over~$\Q_p$ and $\bK$ is totally ramified over~$\bL$. 

\begin{figure}
\begin{center}
\begin{tikzpicture}[scale=0.8]
	\draw (-1.75,3.5) -- node[midway,below left] {$e = 1$; $f=2$} (-1.25,2.5);
	\draw (1.75,3.5) -- node[midway,below right] {$e = 2$; $f=1$} (1.25,2.5);
	\node at (-6,6) {\llap{unram.}}; 
	\node at (-3,6) {\usebox{\unramif}};
	\node at (6,6) {\rlap{ram.}}; 
	\node at (3,6) {\usebox{\ramif}};
	\node at (-3,0) {$\mathbb{Q}_2$};
	\node at (0,0) {\usebox{\basefld}};
\end{tikzpicture} 
\end{center}
\caption{Obtaining trees for ramified and unramified quadratic extensions from the Bruhat--Tits tree of~$\Q_p$.}
\label{Extensions}
\end{figure}
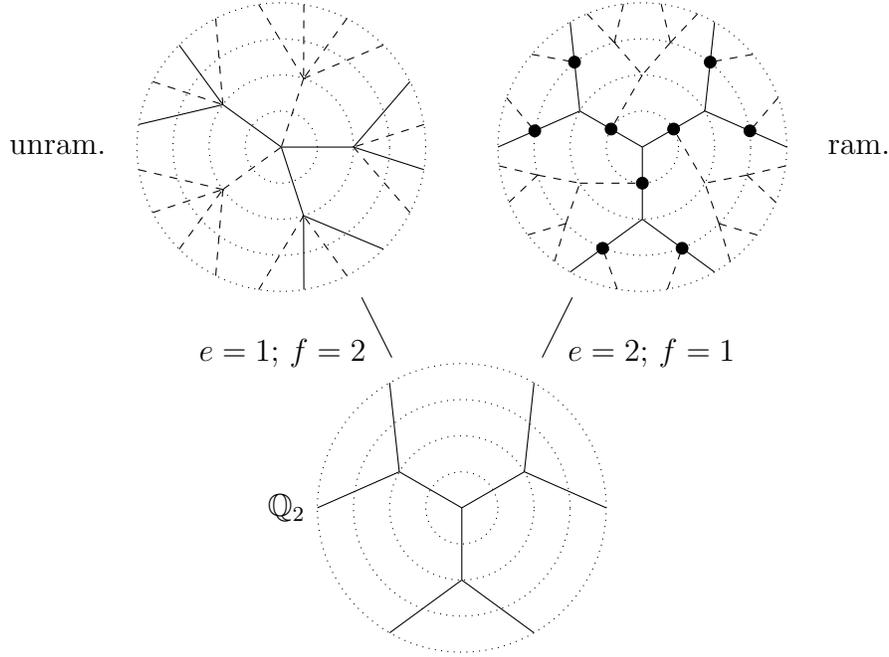

By identifying 
$(\cO_{\bQ_p}/\fm^r)\otimes \cO_\bK = \cO_{\bK}/\fm^{r e_{\bK}}$, for any positive $r$, 
we see that the Bruhat--Tits tree $T_\bK$ for
a finite extension $\bK$ of $\bQ_p$ is obtained from the Bruhat--Tits tree of $\bQ_p$ by
adding $e_{\bK}-1$ new vertices in each edge of $T_{\bQ_p}$ (expressing the difference in the set of values of the $\ord_p$ map on~$\bK$) and increasing the valence
of all vertices to $p^f+1$ (so that the neighborhood of each vertex can still be identified with the projective line over the finite residue field $\cO_\bK/\fm$). We illustrate these processes in Figure~\ref{Extensions}; for more details, the reader is referred to~\cite{Man76}.
The set of
vertices $V(T_\bK)$ of the Bruhat--Tits tree $T_\bK$ of $\bK$ is the set of equivalence
classes of free rank two $\cO_\bK$-modules, under the equivalence $M_1\sim M_2$
if $M_1=\lambda M_2$, for some $\lambda \in \bK^*$. For a pair of such modules
with $M_2\subset M_1$, one can define a distance function $d(M_1,M_2)=|l-k|$,
where $M_1/M_2=\cO_\bK/\fm^l \oplus \cO_\bK/\fm^k$. This distance is independent
of representatives in the equivalence relation. There is an edge in $E(T_\bK)$
connecting two vertices in $V(T_\bK)$ whenever the corresponding classes
of modules have distance one. The resulting tree $T_\bK$ is an infinite homogeneous
tree with vertices of valence $q+1$, where $q=\# \cO_\bK/\fm = p^f$ is the cardinality
of the residue field. The boundary at infinity of the Bruhat--Tits tree is identified with $\bP^1(\bK)$.
One can think of the Bruhat--Tits tree as a network,
with a copy of the finite field $\bF_q$ (or better of the projective line $\bP^1(\bF_q)$) 
associated to each vertex; this will be the guiding viewpoint in our approach 
to non-archimedean tensor networks. 

\subsection{Schottky uniformization of Riemann surfaces}


In this section, we review Schottky uniformization, which allows one to think of a higher-genus Riemann surface as a quotient of the projective line by a particular discrete subgroup of its M\"obius transformations.

A Schottky group of rank $g\geq 1$ is a discrete subgroup of 
$\PSL(2,\bC)$ which is purely loxodromic and isomorphic to a free group on $g$ generators.
The group $\PSL(2,\bC)$ acts on $\bP^1(\bC)$ by fractional linear transformations,
$$ \gamma=\begin{pmatrix} a & b \\ c & d \end{pmatrix}: z \mapsto \frac{az+b}{cz+d}. $$

The loxodromic condition means that each nontrivial element $\gamma \in \Gamma\setminus \{ 1 \}$ 
has two distinct fixed points $z_\gamma^\pm$ (one attractive and one repelling)
in $\bP^1(\bC)$. The closure in $\bP^1(\bC)$ of the set of all fixed points of elements 
in $\Gamma$ is the limit set $\Lambda_\Gamma$ of $\Gamma$, the set of 
all limit points of the action of $\Gamma$ on $\bP^1(\bC)$. In the case $g=1$ the limit
set consists of two points, which we can choose to identify with $\{ 0, \infty \}$,
while for $g>1$ the set $\Lambda_\Gamma$ is a Cantor set of Hausdorff
dimension $0\leq \dim_H(\Lambda_\Gamma) < 2$. The Hausdorff dimension
is also the exponent of convergence of the Poincar\'e series
of the Schottky group: $\sum_{\gamma \in \Gamma} |\gamma^\prime|^s$ converges
for $s> \dim_H(\Lambda_\Gamma)$~\cite{Bow}.

It is well known that any compact smooth Riemann surface $X$ admits a Schottky uniformization,
namely $X=\Omega_\Gamma /\Gamma$, where $\Gamma \subset \PSL(2,\bC)$ is a Schottky
group of rank equal to the genus $g=g(X)$ of the Riemann surface, and 
$\Omega_\Gamma = \bP^1(\bC) \setminus \Lambda_\Gamma$ is the 
{\it domain of discontinuity} of the action of $\Gamma$ on $\bP^1(\bC)$.
There is a well known relation between Schottky and Fuchsian uniformizations of
compact Riemann surfaces of genus $g\geq 2$; see~\cite{TakZog}.

A {\it marking} of a rank $g$ Schottky group $\Gamma \subset \PSL(2,\bC)$ is a
choice of a set of generators $\{ \gamma_1, \ldots, \gamma_g \}$ of $\Gamma$ and
a set of $2g$ open connected regions $D_i$ in $\bP^1(\bC)$, with $C_i =\partial D_i$ the
boundary Jordan curves homemorphic to $S^1$,
with the following properties:
\begin{enumerate}
\item the closures of the $D_i$ are pairwise disjoint
\item $\gamma_i(C_i)\subset C_{g+i}$
\item $\gamma_i (D_i) \subset \bP^1(\bC)\setminus D_{g+i}$.
\end{enumerate}
The marking is {\it classical} if all the $C_i$ are circles. (All Schottky groups
admit a marking, but not all admit a classical marking.)
A fundamental domain $F_\Gamma$ for the action of the Schottky group $\Gamma$ on the
domain of discontinuity $\Omega_\Gamma \subset \bP^1(\bC)$ can be constructed
by taking
$$ F_\Gamma = \bP^1(\bC) \setminus \cup_{i=1}^g (D_i \cup \bar D_{g+i}). $$
This satisfies $\cup_{\gamma\in \Gamma} \gamma(F_\Gamma)=\Omega_\Gamma$.
In the case of genus $g=1$, with $\Gamma =q^\bZ$, for some $q\in \bC$ with $|q|>1$,
the region $F_\Gamma$ constructed in this way is an annulus $A_q$, with $D_1$ 
the unit disk in $\bC$ and $D_2$ the disk around $\infty$ given by complement in 
$\bP^1(\bC)$ of the disk centered at zero of radius $|q|$, so that $q^\bZ A_q=\bC^*=\bP^1(\bC)
\setminus \{0,\infty \}=\Omega_{q^\bZ}$. The resulting quotient $E_q=\bC^*/q^\bZ$ is the Tate 
uniformization of elliptic curves.

\subsection{Hyperbolic handlebodies and higher genus black holes}


The action of $\PSL(2,\bC)$  by fractional linear transformations on $\bP^1(\bC)$
extends to an action by isometries on the real $3$-dimensional hyperbolic space
$\bH^3$, with $\bP^1(\bC)$ its conformal boundary at infinity. In coordinates $(z,y)\in \bC\times 
\bR^*_+$ in $\bH^3$, the action of $\PSL(2,\bC)$ by isometries of the hyperbolic metric is given by
$$ \gamma=\begin{pmatrix} a & b \\ c & d \end{pmatrix}: (z,y) \mapsto \left( \frac{(az+b)\overline{(cz+d)} + a\bar c y^2}{|cz+d|^2+|c|^2 y^2}, \frac{y}{|cz+d|^2+|c|^2 y^2}\right). $$


Given a rank $g$ Schottky group $\Gamma \subset \PSL(2,\bC)$, we can consider its action on
the conformally compactified hyperbolic $3$-space $\overline{\bH^3}=\bH^3 \cup \bP^1(\bC)$.
The only limit points of the action are on the limit set $\Lambda_\Gamma$ that is contained
in the conformal boundary $\bP^1(\bC)$, hence a domain of discontinuity for this action is
given by
$$ \bH^3 \cup \Omega_\Gamma \subset \overline{\bH^3}=\bH^3 \cup \bP^1(\bC). $$
The quotient of $\bH^3$ by this action is a $3$-dimensional hyperbolic handlebody of genus $g$
$$ \cH_\Gamma = \bH^3 / \Gamma, $$
with conformal boundary at infinity given by the Riemann surface
$X_\Gamma = \Omega_\Gamma/ \Gamma$,
$$ \overline{\cH}_\Gamma = \cH_\Gamma \cup X_\Gamma =  ( \bH^3 \cup \Omega_\Gamma )/ \Gamma. $$

\begin{figure}[t]
\includegraphics[width=11cm]{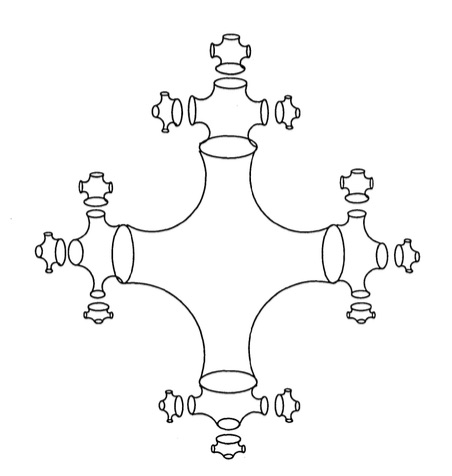} 
\centering
\caption{Fundamental domains for the action of $\Gamma$ on $\bH^3$. \label{FundDomH3Fig}}
\end{figure}

Given a marking of a rank $g$ Schottky group $\Gamma$ (for simplicity we will assume the
marking is classical), let $D_i$ be the discs in $\bP^1(\bC)$ of the marking, and let $\cD_i$
denote the geodesic domes in $\bH^3$ with boundary $C_i=\partial D_i$, namely the $\cD_i$
are the open regions of $\overline{\bH^3}$ with boundary $S_i \cup D_i$, where the $S_i$ are
totally geodesic surfaces in $\bH^3$ with boundary $C_i$ that project to $D_i$ on the conformal 
boundary. Then a fundamental domain for the action of $\Gamma$ on $\bH^3 \cup \Omega_\Gamma$
is given by
$$ \cF_\Gamma =F_\Gamma \cup (\bH^3 \setminus \cup_{i=1}^g (\cD_i\cup \bar\cD_{g+i}). $$
The boundary curves $C_i$ for $i=1,\ldots, g$ provide a collections of $A$-cycles, that
give half of the generators of the homology of the Riemann surface $X_\Gamma$: the
generators that become trivial in the homology of the handlebody $\bar\cH_\Gamma$.
The union of fundamental domains $\gamma(\cF_\Gamma)$ for $\gamma \in \Gamma$ can
be visualized as in Fig.~\ref{FundDomH3Fig}.

In the case of genus $g=1$ with $\Gamma=q^\bZ$, acting on $\bH^3$ by 
$$ \begin{pmatrix} q^{1/2} & 0 \\ 0 & q^{-1/2} \end{pmatrix} (z,y)= (qz, |q|y), $$
with limit set $\{0,\infty\}$ the fundamental domain $\cF_\Gamma$ consists of
the space in the upper half space $\bH^3$ contained in between the two spherical
domes of radius $1$ and $|q|>1$. The generator $q$ of the group acts on the
geodesic with endpoints $0$ and $\infty$ as a translation by $\log |q|$. The
quotient $\bH^3/q^\bZ$ is a hyperbolic solid torus, with the Tate uniformized elliptic
curve $E_q=\bC^*/q^\bZ$ as its conformal boundary at infinity, and with a unique
closed geodesic of length $\log q$. It is well known (see \cite{BKSW}, \cite{MalStr},
and~\S2.3 of \cite{ManMar}) that the genus one handlebodies $\cH_{q^\bZ}$ 
are the Euclidean BTZ black holes \cite{BTZ}, where the cases with $q\in \bC\setminus \bR$ 
correspond to spinning black holes. The geodesic length $\log |q|$ is the area
of the event horizon, hence proportional to the black hole entropy. 

\begin{figure}[t]
\includegraphics[width=11cm]{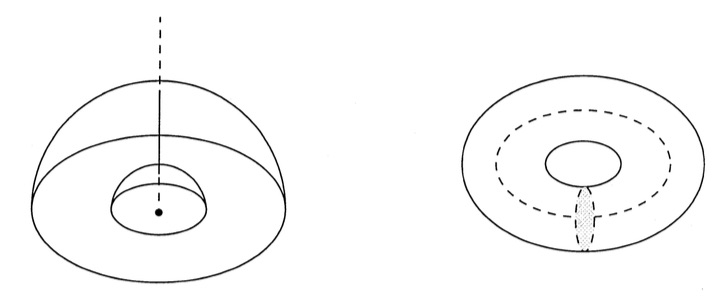} 
\centering
\caption{Fundamental domain and quotient for the Euclidean BTZ black hole. Compare with the $p$-adic BTZ geometry, shown in Fig.~\ref{BTZpadicFig}.
\label{EuclideanBTZFig}}
\end{figure}

The case of higher genus hyperbolic handlebodies correspond to generalizations
of the BTZ black hole to the higher genus asymptotically $AdS_3$ black holes 
considered in \cite{Kras} and \cite{ManMar}.

In these more general higher genus black hole, because of the very different nature
of the limit set (a fractal Cantor set instead of two points) the structure of the black
hole event horizon is significantly more complicated. In the Euclidean BTZ black hole,
the only infinite geodesic that remains confined into a compact region inside the 
hyperbolic solid torus $\cH_{q^\bZ}$ for both $t\to \pm \infty$ is the unique closed
geodesic (the image in the quotient of the geodesic in $\bH^3$ given by the vertical
line with endpoints $0$ and $\infty$. On the other hand, in the higher genus cases,
the geodesics in the hyperbolic handlebody $\cH_\Gamma=\bH^3/\Gamma$ can
be classified as:
\begin{enumerate}
\item {\it Closed geodesics}: these are the images in the quotient $\cH_\Gamma$ of 
geodesics in $\bH^3$ with endpoints $\{ z_\gamma^+, z_\gamma^- \}$, the attractive
and repelling fixed points of some element $\gamma\in \Gamma$.
\item {\it Bounded geodesics:} these images in the quotient $\cH_\Gamma$ of 
geodesics in $\bH^3$ with endpoints on $\Lambda_\Gamma$. If the endpoints
are not a pair of fixed points of the same element of $\Gamma$ the geodesic in
the quotient is not closed, but it remains forever confined within a compact region
inside $\cH_\Gamma$, the {\it convex core} $\cC_\Gamma$.
\item {\it Unbounded geodesics:} these are images in the quotient $\cH_\Gamma$ of 
geodesics in $\bH^3$ with at least one of the two endpoints in $\Omega_\Gamma$.
These are geodesics in $\cH_\Gamma$ that wander off (in at least one time direction $t\to \infty$ or $t\to -\infty$)
towards the conformal boundary $X_\Gamma$ at infinity and eventually
leave every compact region in $\cH_\Gamma$.
\end{enumerate}
The convex core $\cC_\Gamma \subset \cH_\Gamma$ is the quotient 
by $\Gamma$ of the geodesic hull in $\bH^3$ of the limit set $\Lambda_\Gamma$. 
It is a compact region of finite hyperbolic
volume in $\cH_\Gamma$, and it is a deformation retract of $\cH_\Gamma$. A natural replacement for
the event horizon of the BTZ black hole in these higher genus cases can be identified in terms
of the convex core $\cC_\Gamma$, where we think of $\cC_\Gamma$ as the region from which geodesic
trajectories cannot escape and must remain forever confined. The complement $\cH_\Gamma \setminus \cC_\Gamma$
is homeomorphic to $\partial \cC_\Gamma \times \bR_+$ (see \cite{Canary} for a more general treatment of
convex cores of Kleinian groups and ends of hyperbolic $3$-manifolds).
The boundary $\partial \cC_\Gamma$ is the event horizon of the higher genus black hole, with the black hole entropy 
proportional to the area of $\partial \cC_\Gamma$. 

In \cite{Man89} and \cite{Man91}, Manin proposed to interpret the tangle of bounded geodesics inside the
hyperbolic handlebody $\cH_\Gamma$ as a model for the missing ``closed fiber at infinity''
in Arakelov geometry. This interpretation was based on the calculation of the Arakelov Green
function \cite{Man91}, and the analogy with the theory of Mumford curves \cite{Mum} and the
computations of \cite{DriMan} for $p$-adic Schottky groups. The results of \cite{Man91} 
and their holographic interpretation in \cite{ManMar}, as well as the parallel theory of
Mumford curves and periods of $p$-adic Schottky groups, will form the basis for our development
of a $p$-adic and adelic form of the AdS/CFT correspondence. The interpretation of the
tangle of bounded geodesics in $\cH_\Gamma$ as ``closed fiber at infinity'' of Arakelov
geometry was further enriched with a cohomological interpretation in \cite{ConsMar1}
(see also \cite{ConsMar2}, \cite{ConsMar3} for the $p$-adic counterpart). 

\subsection{Bruhat--Tits trees, $p$-adic Schottky groups, and Mumford curves}
\label{sec:Mumford}

The theory of Schottky uniformization of Riemann surfaces as conformal boundaries of 
hyperbolic handlebodies has a non-archimedean parallel in the theory of Mumford
curves, uniformized by $p$-adic Schottky groups, seen as the boundary at infinity of
a quotient of a Bruhat--Tits tree. In the context of the $p$-adic string theory, such geometries were studied by Chekhov, Mironov, and Zabrodin~\cite{ChekhovZabrodin} in order to compute multiloop scattering amplitudes.


The reader should beware that there is an unavoidable clash of notation: $q$ is the standard notation for the modular parameter of an elliptic curve, but is also used to denote a prime power $q=p^r$ in the context of finite fields or extensions of the $p$-adics. While both uses will be made in this paper, particularly in this section and in~\S\ref{sec:Drinfeld}, we prefer not to deviate from standard usage; it should be apparent from context which is intended, and hopefully no confusion should arise.

There is an action of $\PGL(2,\bK)$ on the set of vertices $V(T_\bK)$ that preserves 
the distance, hence it acts as isometries of the tree $T_\bK$. A $p$-adic Schottky
group is a purely loxodromic finitely generated torsion free subgroup of $\PGL(2,\bK)$.
The Schottky group $\Gamma$ is isomorphic to a free group on $g$-generators, with
$g$ the rank of $\Gamma$.

In this $p$-adic setting the loxodromic condition means that every nontrivial
element $\gamma$ in $\Gamma$ has two fixed points $z^\pm_\gamma$ on the
boundary $\bP^1(\bK)$. Equivalently, an element $\gamma$ is loxodromic
if the two eigenvalues have different $p$-adic valuation. The closure of the
set of fixed points $z_\gamma^\pm$, or equivalently the set of accumulation
points of the action of $\Gamma$ on $T_\bK \cup \bP^1(\bK)$ is the limit
set $\Lambda_\Gamma$ of the Schottky group $\Gamma$. The complement
$\bP^1(\bK)\setminus \Lambda_\Gamma =\Omega_\Gamma(\bK)$ is
the domain of discontinuity of the action of $\Gamma$ on the boundary. 

There is a unique geodesic $\ell_\gamma$ in $T_\bK$ with endpoints $\{ z_\gamma^-, z_\gamma^+\}$,
the {\it axis} of a loxodromic element $\gamma$. The subgroup $\gamma^\bZ$ acts on $T_\bK$ by
translations along $\ell_\gamma$. There is a smallest subtree $T_\Gamma\subset T_\bK$
that contains all the axes $\ell_\gamma$ of all the nontrivial elements $\gamma \in \Gamma$.
The boundary at infinity of the subtree $T_\Gamma$ is the limit set $\Lambda_\Gamma$.
$T_\Gamma$ is the non-archimedean analog of the geodesic hull of the limit
set of a Schottky group in~$\bH^3$. 

The quotient $X_\Gamma (\bK)=\Omega_\Gamma(\bK)/\Gamma$ is a Mumford curve with
its $p$-adic Schottky uniformization, \cite{Mum}. The quotient $T_\bK/\Gamma$
consists of a finite graph $T_\Gamma /\Gamma$ with infinite trees appended at
the vertices of $T_\Gamma /\Gamma$, so that the boundary at infinity of the graph
$T_\bK/\Gamma$ is the Mumford curve $X_\Gamma(\bK)$. The finite graph
$G_\bK=T_\Gamma/\Gamma$ is the dual graph of the special fiber $X_q$ (a curve over $\bF_q$
which consists of a collection of $\bP^1(\bF_q)$ at each vertex of $G_\bK$,
connected along the edges). A family of finite graphs $G_{\bK,n}$, for $n\in \bN$, is obtained 
by considering neighborhoods $T_{\Gamma,n}$ of $T_\Gamma$ inside $T_\bK$
consisting of $T_\Gamma$ together with all vertices in $T_\bK$ that are at a distance at 
most $n$ from some vertex in $T_\Gamma$ and the edges between them (these trees
are preserved by the action of $\Gamma$), and taking the quotients 
$G_{\bK,n}=T_{\Gamma,n}/\Gamma$. The endpoints (valence one vertices) 
in $G_{\bK,n}$ correspond to reduction mod $\fm^n$ and the set of points $X(\cO_\bK/\fm^n)$, 
see Section 1.3 of \cite{Man89}.
One sees in this way, geometrically, how the $\bK$-points in the Mumford curve $X_\Gamma(\bK)$
are obtained as limits, going along the infinite ends of the graph $T_\bK/\Gamma$, which
correspond to successively considering reduction mod $\fm^n$.
Conversely, one can view the process of going into the tree from its boundary $X_\Gamma(\bK)$
towards the graph $G_{\bK}$ in the middle of $T_\bK/\Gamma$ as applying reductions mod $\fm^n$.
We will see later in the paper how this process should be thought of physically as a form
of renormalization. The finite graph $G_\bK=T_\Gamma/\Gamma$ is the non-archimedean
analog of the convex core $\cC_\Gamma$ of the hyperbolic handlebody $\cH_\Gamma$, while 
the infinite graph $T_\bK/\Gamma$ is the non-archimedean analog of $\cH_\Gamma$ itself,
with the Mumford curve $X_\Gamma(\bK)$ replacing the Riemann surface $X_\Gamma=X_\Gamma(\bC)$
as the conformal boundary at infinity of $T_\bK/\Gamma$. 

Geodesics in the bulk space $T_\bK/\Gamma$ correspond to images in the quotient of
infinite paths without backtracking in the tree $T_\bK$, with endpoints at infinity on $\bP^1(\bK)$.
Again, one can subdivide these in several cases. When the endpoints are the attractive and
repelling fixed points $z_\gamma^\pm$ of some element $\gamma \in \Gamma$, the
path in $T_\bK/\Gamma$ is a closed loop in the finite graph $G_{\bK}$. If the endpoints
are both in $\Lambda_\Gamma$ but not the fixed points of the same group element,
then the geodesic is a finite path in $G_{\bK}$ that is not a closed loop (but which winds
around several closed loops in $G_{\bK}$ without a fixed periodicity). If at least one of
the endpoints is in $\Omega_\Gamma(\bK)$, then the path in $T_\bK/\Gamma$ eventually
(for either $t\to +\infty$ or $t\to -\infty$) leaves the finite graph $G_{\bK}$ and wonders off
along one of the attached infinite trees towards the boundary $X_\Gamma(\bK)$ at infinity. 
We still refer to these cases as closed, bounded, and unbounded geodesics, as in the
archimedean case.  We refer the reader to \cite{GerPut}, \cite{Man76}, \cite{Mum}, \cite{Put}
for a more detailed account of the geometry of Mumford curves. 

\section{Tensor networks}
\label{happy}

Motivated by the idea that the Bruhat--Tits tree $T_p$ is a discrete (while still geometric) analogue of anti-de~Sitter space, we will use this section to consider some relations between tensor networks that have been considered in the literature and the tree. One might imagine that many such relations can be drawn, and we have made no effort to be exhaustive; indeed, 
there are several distinct proposals connecting tensor networks to discrete analogues of holography in the literature. Our purpose in this section is to propose one such model based on p-adic geometry, in which the bulk is naturally discrete from the outset.


In this section the basic Hilbert spaces in the bulk and the boundary will be those of finite dimensional qudits and the primary object of study will be the entanglement structure. In ~\S\ref{scalars}, the finite dimensional Hilbert spaces are replaced with those of a field theory valued in $\bR$ or $\bC$. We will find many aspects of holography hold in this field theoretic model and provide evidence for an exact correspondence. This connection puts the tensor network models of holography on a more equal footing with dynamical models, since both are defined from the same discrete spacetime.\footnote{We should also note that, in certain aspects, tensor networks from the tree are very different from their Archimedean counterparts. For example, the Ryu-Takayanagi area, the tensor network, and the action all live in the same (non-integer) `dimension'. This is likely an idiosyncrasy of the tree, that could potentially be alleviated by going to Drinfeld's \padic upper half-plane (see Sec.~\ref{sec:Drinfeld}).}

Throughout this paper, we will remain agnostic as to whether the tree and its boundary should be thought of as a stand-alone quantum system, or as a kind of ``section'' inside some more complicated object, such as a Bruhat-Tits building of higher rank. We hope to return to this latter point of view in future work.

We will focus our attention on the networks used by Pastawski, Yoshida, Harlow, and Preskill~\cite{HaPPY}, (or ``HaPPY''), in their construction of holographic quantum error-correcting codes. Such codes are easy to describe and admit many variations; in the simplest case, they are associated to a regular hyperbolic tiling of the plane. We will refer to such tilings by their \textsl{Schl\"afli symbol;} the notation $\{s,n\}$ refers to a tiling in which $n$ regular polygons, each with $s$ sides, meet at every vertex. A simple calculation shows that the tiling is hyperbolic whenever
\begin{equation}
n > \frac{2s}{s-2}.
\end{equation}
For instance, if pentagons are used, $n=4$ is the smallest possible choice ($n=3$ would give the dodecahedron). If the tiles are heptagons or larger, any $n\geq3$ gives a hyperbolic tiling. 

Due to constraints of space, we will not fully review the HaPPY construction here; for details, the reader is referred to the original paper. The key point is that each tile carries a \textsl{perfect tensor}, which has an even number of indices, each of which refers to a qudit Hilbert space of fixed size. Such tensors are characterized by the property that any partition of the indices into two equal sets yields a maximally entangled state; we review perfect tensors in more detail, and construct a family of them associated to finite fields, in~\S\ref{sec:codes}. Due to the appearance of finite fields in the construction of the tree, we feel this is the most natural family of tensors to consider.

The gist of our argument is that the natural ``geometric'' setting of a \happy (for certain uniform tilings) is the Bruhat--Tits tree corresponding to a prime power $q=p^n$. This means considering an extension $\bK$ of $\bQ_p$, of finite degree
$[\bK: \bQ_p]<\infty$, with $n = [\bK : \bQ_p]/e_{\bK}$ where $e_{\bK}$ is the ramification index of the extension $\bK$ 
of $\bQ_p$. Passing to the extension $\bK$ corresponds to modifying the Bruhat--Tits tree of $\bQ_p$ by adding new
edges at each vertex, so that the valence of all vertices is $p^n+1$, and inserting $e_{\bK}-1$ new vertices along each
edge. The latter property accounts for how the geodesic lengths in the tree change when passing to a field extension.
It is customary to normalize the distance in the Bruhat--Tits tree of $\bK$ accordingly, by dividing by a factor $e_\bK$. 
We are motivated in this argument not only by the algebraic similarity between the constructions of~$T_p$ and~$\text{AdS}_3$, but also by the fact that field-theoretic models of holography can be defined on the tree which exhibit it as the natural discrete setting for the AdS/CFT correspondence. 
In particular, the Bruhat--Tits tree with all edges of equal length can be thought of as a discrete analog of (empty) Euclidean \ads, and we conjecture that it is dual to the vacuum of the CFT living on $\mathbb{Q}_p$. It is therefore logical to guess that \happies naturally encode information about the entanglement structure of the conformal field theories living at the finite places, and not about the CFT living at the Archimedean place.

The precise connection we identify is that, at least for certain choices, the tiling used in HaPPY's code (when thought of as a graph) has a spanning tree that is a Bruhat--Tits tree. In some sense, therefore, the tree represents the union of as many geodesics as can be marked on the tiling without creating closed paths in the bulk. 


In HaPPY's original paper, a ``greedy algorithm'' (related to reconstruction of the quantum state input at the bulk or ``logical'' qubits of the code) is used to define a region of the bulk, the perimeter of which is then called a ``geodesic.'' We will show in~\S\ref{sec:wedge} that tree geodesics can be understood to correspond to these ``greedy'' geodesics for the $\{5,q+2\}$ tilings.

An analogue of the Ryu-Takayanagi formula holds for these codes, essentially because the length of the geodesic counts the number of bonds (contracted tensor indices) that cut across it, and---due to the properties of perfect tensors---each contributes a constant amount (the logarithm of the qudit dimension) to the entanglement entropy. 
For us, it will be crucial to note that the length of a (unique boundary anchored) tree geodesic is related to the $p$-adic size of the boundary region it defines. We will elaborate on this in~\S\ref{EE}; for now, we will simply remark on a few features of the formula that we will need in this section.

At the Archimedean place, entanglement entropy measures the entanglement between the degrees of freedom living on a spatial domain $A$ of a QFT and those living on the complement $A^c$. In \ads, the domain $A$ is usually taken to be an interval, or a collection of intervals. The finite place analogue of an interval is just an open ball (as defined in~\S\ref{sec:basics}); the notion of ``codimension'' is counterintuitive in the $p$-adic setting! One can see that there is no topological difference between (for instance) an open subset of the unit circle, which would be an interval in the normal case, and a generic open subset.

If the Ryu-Takayanagi formula holds, the entanglement entropy between an open ball $A\subset \Q_p$ and its complement $A^c$ is given by the length of a geodesic $\gamma\lb x_A,y_A\rb$ in the Bruhat--Tits tree connecting boundary points $x_A$ and $y_A$:
\be
S_\mrm{EE}(A) = \# \cdot \mrm{length}\lb \gamma\lb x_A,y_A\rb\rb.
\label{eq:Ryu1}
\ee
It should be emphasized that the $p$-adics are not an ordered field, and so there are no two unique points at the ``edges'' of~$A$. Our definition of the required pair of boundary points is simply a choice of $x_A, y_A \in A$ such that
\begin{equation}
|x_A - y_A|_p = \epsilon, \quad A = B_\epsilon(x_A) = B_\epsilon(y_A).
\label{eq:bdypts}
\end{equation} 
Conversely, to any pair $x_A\neq y_A \in \Q_p$, we can associate a unique open set $A$ such that~\eqref{eq:bdypts} is satisfied. The length of the geodesic in~\eqref{eq:Ryu1} depends only on the set~$A$ and not on the choice of~$x_A$ and~$y_A$; this is easy to see, since $A$ is the boundary of a cone in the tree below a particular vertex $v$, and the requirements on the pair $(x_A,y_A)$ are equivalent to the condition that $\gamma(x_A,y_A)$ pass through~$v$. The entanglement entropy (considered as a function of a pair of points) thus depends on the points only though the quantity $|x_A - y_A|_p$, just like a two-point function of operator insertions in the CFT. We will return to this point later.

Just as in the case of \ads, entanglement entropy is a logarithmically divergent quantity. The divergence arises because if $x_A$ and $y_A$ are in~$\PL(\Q_p)$, the number of legs on the geodesic is infinite. To regularize this divergence, we introduce a cutoff $\epsilon_p$ such that the length of the geodesic is
\be
\mrm{length}\lb \gamma\lb x_A,y_A\rb\rb = 2 \log_p \left| {\frac{x_A-y_A}{\epsilon_p}} \right|_p,
\ee
with $|\cdot|_p$ the \padic norm. (For details about this, see~\S\ref{sec:RT}.) This gives the entanglement entropy between $A$ and its complement in $\mathbb{Q}_p$ as\footnote{For the length of the geodesic between $x_A$ and $y_A$ to warrant the interpretation of entanglement entropy, it must be the case that the tensor network bonds cutting across it, when extended all the way to the boundary, connect between $A$ and $A^c$. This can be done and is explained in~\S\ref{sec:wedge}. }
\be
S_\mrm{EE}(A) = \# \log_p \left| \frac{x_A-y_A}{\epsilon_p} \right|_p .
\ee
The proportionality constant will be left undetermined for now.

\subsection{Perfect tensors and quantum error-correcting codes from finite fields}
\label{sec:codes}

Our goal in this subsection is to recall some features of quantum error-correcting codes associated to $2n$-index perfect tensors, as used by Pastawski, Yoshida, Harlow, and Preskill~\cite{HaPPY}. We will review the three-qutrit code and associated four-index perfect tensor that they construct, and then show how this is the case $q=3$ of a family of perfect tensors associated to powers $q=p^m$ of odd primes. While the corresponding quantum error-correcting codes are not new~\cite{GBR}, our goal is to highlight the properties of these particular codes that make them relevant to $p$-adic holography. In particular, as we recalled in~\S\ref{sec:Mumford}, each vertex of the Bruhat--Tits tree for a degree-$n$ unramified extension $k$ of $\Q_p$ is marked with a copy of the residue field~$\FF_{p^n}$. As such, finite fields appear as important ingredients both in the construction of holographic tensor networks and in the algebraic setting of the Bruhat--Tits tree. We feel that the codes discussed here are natural candidates to consider in connecting $p$-adic geometry to tensor network models, although of course this choice is not inevitable and any code with the right properties will define a holographic code. 

\subsubsection{The three-qutrit code}

In their paper, Pastawski \textsl{et~al} consider the following quantum error-correcting code, which encodes a one-qutrit logical Hilbert space in a three-qutrit physical Hilbert space: 
\begin{align*}
\ket{0} &\mapsto \ket{000} + \ket{111} + \ket{222} \\
\ket{1} &\mapsto \ket{012} + \ket{120} + \ket{201} \\
\ket{2} &\mapsto \ket{021} + \ket{102} + \ket{210} .
\end{align*}
The encoded data is protected against erasure of any single qutrit. If we represent the state by a tensor,
\[ \ket{a} \mapsto T_{abcd}\ket{bcd},\]
then $T_{abcd}$ is perfect in the sense of Pastawski \textsl{et~al}, and defines a perfect state:
\[ \ket{\psi} = T_{abcd} \ket{abcd}. \]
(Throughout, we use Einstein's summation convention.) To recall, a tensor with $2n$ indices, each representing a qudit Hilbert space of any chosen fixed size, is perfect when it satisfies any of the following equivalent conditions:
\begin{itemize}
\item Given any partition of the indices into two disjoint collections $A\sqcup B$, where $|A|\leq|B|$, the tensor defines an injection of Hilbert spaces $\Hilb_A \inj \Hilb_B$: a linear map that is a unitary isomorphism of its domain with its image (carrying the subspace norm).
\item The corresponding perfect state is maximally entangled between any tensor factors $\Hilb_{A,B}$ of equal size (each consisting of~$n$ qudits). That is, after tracing out $n$ of the $2n$ qudits, the remaining $n$-qudit density matrix is proportional to the identity operator. 
\end{itemize}
It is straightforward to check that the above tensor $T_{abcd}$ is an $n=2$ perfect tensor on qutrits. To rephrase the way it is constructed so as to make its generalization to larger codes more apparent, we notice that the particular states that appear in the encoding of a basis state $\ket{a}$ are lines of slope~$a$ in $\FF_3^2$: if the three qudits are labeled by an element $x$ of~$\FF_3$, then the states are of the form $\otimes_x \ket{f(x)}$, where $f(x) = ax + b$, and we sum over the three possible choices of $b\in\FF_3$. The result is a perfect tensor because a line is determined either by two of its points, or by one point and knowledge of the slope; conversely, given any two points, or any one point and one slope, exactly one corresponding line exists.

\subsubsection{Perfect polynomial codes}
We would like to generalize this to a family of perfect tensors in which the qudit Hilbert spaces are of size $q=p^m$, so that a basis can be labeled by the elements of~$\FF_q$. An obvious guess is to associate a function or collection of functions $f_a:\FF_q \rightarrow \FF_q$ to each logical basis state~$\ket{a}$, generalizing the collection of lines $f_a(x)=ax+b$ that were used when $q=3$. These functions should have the property that knowledge of some number of evaluations of $f_a$ will uniquely specify~$a$, whereas knowing any smaller number of evaluations will give no information about~$a$ whatsoever. The encoded states will then take the form $\sum_b \left(\otimes_x \ket{f_a(x)}\right)$, for some collection of $x$'s in~$\FF_q$. Here $b$ stands for a collection of numbers parameterizing the set of functions $f_a$.  

The simplest choice of such a class of functions are polynomials of fixed degree $d$:
\[ f_a(x) = ax^d+b_{d-1}x^{d-1}+\cdots+b_1x + b_0. \]
Over the real numbers, $d+1$ points determine such a polynomial. Over finite fields, one must be a little careful---by Fermat's little theorem, 
\[ x^q - x = 0, \ \forall x \in \FF_q. \]
As such, if $d\geq q$, we can't determine a polynomial uniquely by its evaluations---after all, there are at most $q$ possible evaluations over a finite field! However, polynomials of degree $d<q$ can be recovered uniquely; in fact, every function from $\FF_q$ to~$\FF_q$ is a polynomial function, uniquely represented by a polynomial of degree $d<q$ (there are exactly $q^2$ elements of each collection). 

However, if we choose $d$ too large, the resulting code will not have error-correcting properties: we will need almost all of the physical qudits to recover the logical one. We know that for codes obtained from $2n$-index perfect tensors, one logical qudit is encoded in $2n-1$ physical qudits, and is recoverable from any~$n$ of them.  This is sometimes called a $[[2n-1,1,n]]_q$ code.
For polynomial codes, we must have $2n-1\leq q$ (since there are at most $q$ possible evaluations of the code function), and furthermore $n = d+1$. Thus, the largest possible perfect tensor we can obtain from this class of codes has $q+1$ indices, corresponding to the $[[q,1,(q+1)/2]]_q$ code; the polynomials used in making this code are of degree $d=(q-1)/2$.\footnote{Recall that codes with $p = 2$ are not a part of this family. Rather, a $p=2$ code can be realized e.g. as $[[5,1,3]]_2$.} $q=3$ recovers the linear qutrit code that we discussed above.

To be concrete, when $q=5$, the code takes the following form: 
\[\ket{a} \mapsto \sum_{b_0,b_1 \in \FF_5} \ket{b_0, b_0+b_1+a,b_0+2b_1+4a,b_0+3b_1+4a,b_0+4b_1+a}.\]
The numbers that appear are just $x$ as the coefficient of~$b_1$ and $x^2$ as the coefficient of~$a$. This encoded state already contains 25 basis states, and the perfect state $\ket{\psi_5}$ constructed from this tensor is a combination of $5^3=125$ basis states. 

These $(q+1)$-index perfect tensors seem like logical candidates to use in constructing a family of quantum codes associated to Bruhat--Tits trees. In particular, they are naturally associated to the data of a finite field $\FF_q$, which appears at each vertex of the tree; moreover, they have $q+1$ qudit indices, which agrees with the valence of the tree. 

However, the exact way to combine these ingredients remains a little unclear. In particular, since paths in the tree correspond to geodesics in the $p$-adic hyperbolic space, it seems more natural to think of the legs of the tree as \textsl{cutting across} contractions of tensor indices, rather than representing them. We expand on this idea in the section that follows.

\subsection{Bruhat--Tits trees and tensor networks}
\label{treetensorsubsection}

We now investigate the connection between Bruhat--Tits trees and tensor networks. The gist of this section is that, while the tree corresponds to ``geometry,'' the tree alone cannot define a tensor-network topology in the most naive way (tensors at vertices with indices contracted along edges). This is because, in typical tensor-network models of holography, the Ryu-Takayanagi formula holds because each unit distance along a geodesic corresponds to a bond (i.e. contracted tensor index) which is ``cut'' by the path and contributes a fixed amount (the logarithm of the dimension of the qudit Hilbert space) to the entanglement entropy. Since the paths in the tree correspond to geodesics in the bulk, one cannot hope to connect the tree to HaPPY's holographic code without adding tensors in such a way that their indices are contracted \textsl{across} the edges of the tree.

The extra structure we need to account for the network can be as simple as grouping the vertices of the tree in some fashion, associating bulk indices to the groups, and demanding perfect tensor structure, as we now explain.

\begin{quote}
{\bf A basic set of rules for constructing entanglement:} Group the vertices in the tree in some way; to each grouping we associate one or more bulk vertices. If two groupings share two tree vertices, then there is a tensor network bond connecting the bulk vertices of the groupings. The resulting tensor network should be composed of perfect tensors. This constructs a tensor network mapping between the boundary and the bulk.
\end{quote}

It is not clear what the most general rules for associating the tensor vertices to the tree vertices should be. In particular, we are not demanding planarity (the Bruhat--Tits tree has no intrinsic planar structure), so the resulting network could be quite complicated, or even pathological. In order for the nice properties of a bulk-boundary tensor network to hold  
additional criteria should likely be imposed. We leave the general form of these criteria for future work; in the following, we focus on one specific set of rules that works.

\subsection{Bruhat--Tits spanning trees of regular HaPPY tilings}
\label{apptilingalgorithm}

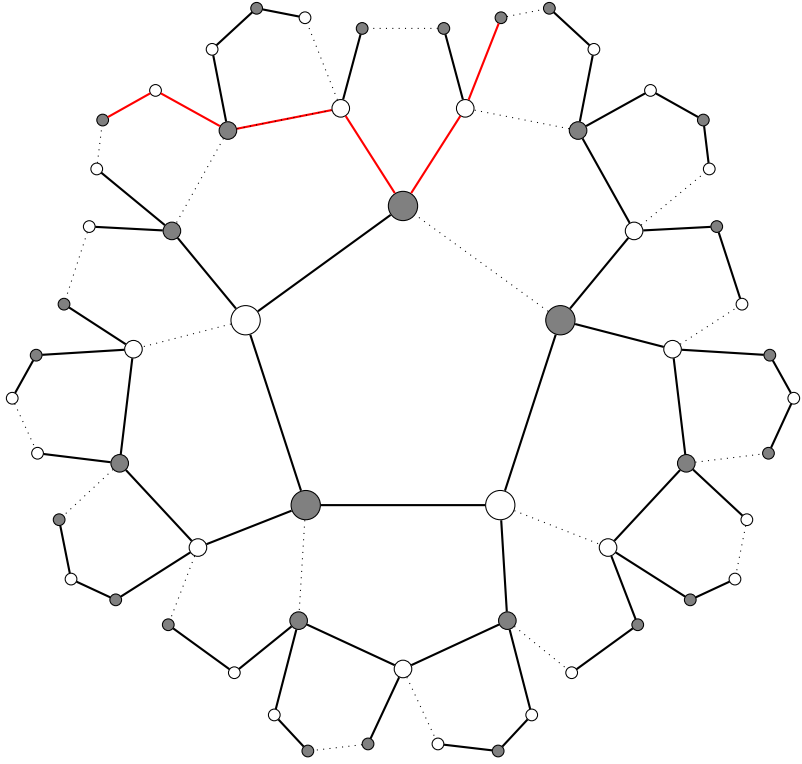
\begin{figure}[t]
\centering
\begin{tikzpicture}[scale=2.2]
\tikzstyle{vertex}=[draw,circle]
\tikzstyle{ver2}=[draw,scale=0.6,circle]
\tikzstyle{ver3}=[draw,scale=0.4,circle]
\draw (0*360/5 + 90:1) node[vertex,fill=gray] (v0) {};
\draw (1*360/5 + 90:1) node[vertex] (v1) {};
\draw (2*360/5 + 90:1) node[vertex,fill=gray] (v2) {};
\draw (3*360/5 + 90:1) node[vertex] (v3) {};
\draw (4*360/5 + 90:1) node[vertex,fill=gray] (v4) {};
\draw[thick] (v0) -- (v1) -- (v2) -- (v3) -- (v4);
\draw[dotted] (v4) -- (v0);

\draw (v0) + (0*360/5 + 0.5*65 + 90:0.7) node[ver2] (w01) {};
\draw (v0) + (0*360/5 + -0.5*65 + 90:0.7) node[ver2] (w02) {};
\draw (v1) + (1*360/5 + 0.5*65 + 90:0.7) node[ver2] (w11) {};
\draw (v1) + (1*360/5 + -0.5*65 + 90:0.7) node[ver2,fill=gray] (w12) {};
\draw (v2) + (2*360/5 + 0.5*65 + 90:0.7) node[ver2,fill=gray] (w21) {};
\draw (v2) + (2*360/5 + -0.5*65 + 90:0.7) node[ver2] (w22) {};
\draw (v3) + (3*360/5 + 0.5*65 + 90:0.7) node[ver2] (w31) {};
\draw (v3) + (3*360/5 + -0.5*65 + 90:0.7) node[ver2,fill=gray] (w32) {};
\draw (v4) + (4*360/5 + 0.5*65 + 90:0.7) node[ver2] (w41) {};
\draw (v4) + (4*360/5 + -0.5*65 + 90:0.7) node[ver2] (w42) {};

\draw (0.5*360/5 + 90:1.8) node[ver2,fill=gray] (x0) {};
\draw (1.5*360/5 + 90:1.8) node[ver2,fill=gray] (x1) {};
\draw (2.5*360/5 + 90:1.8) node[ver2] (x2) {};
\draw (3.5*360/5 + 90:1.8) node[ver2,fill=gray] (x3) {};
\draw (4.5*360/5 + 90:1.8) node[ver2,fill=gray] (x4) {};

\draw[thick,red] (v0) -- (w01) -- (x0); 
\draw[dotted] (w01) -- (x0) -- (w12); 
\draw[thick] (w12) -- (v1);

\draw[dotted] (v1) -- (w11); 
\draw[thick] (w11) -- (x1) -- (w22) -- (v2);

\draw[dotted] (v2) -- (w21); 
\draw[thick] (w21) -- (x2) -- (w32) -- (v3);

\draw[dotted] (v3) -- (w31); 
\draw[thick] (w31) -- (x3) -- (w42) -- (v4);

\draw[thick] (v4) -- (w41)-- (x4); 
\draw[dotted] (x4) -- (w02) ; 
\draw[thick,red] (w02) -- (v0);


\draw (0,0) node {};
\draw (0.5*360/5 + 90:1.3) node (c0) {};
\draw (1.5*360/5 + 90:1.3) node (c1) {};
\draw (2.5*360/5 + 90:1.3) node (c2) {};
\draw (3.5*360/5 + 90:1.3) node (c3) {};
\draw (4.5*360/5 + 90:1.3) node (c4) {};

\draw (w01) + (-15 + 0*360/5 + 90:0.5) node[ver3,fill=gray] (y01) {};
\draw (w02) + (+15 + 0*360/5 + 90:0.5) node[ver3,fill=gray] (y02) {};
\draw (w11) + (-15 + 1*360/5 + 90:0.5) node[ver3,fill=gray] (y11) {};
\draw (w12) + (+15 + 1*360/5 + 90:0.5) node[ver3] (y12) {};
\draw (w21) + (-15 + 2*360/5 + 90:0.5) node[ver3] (y21) {};
\draw (w22) + (+15 + 2*360/5 + 90:0.5) node[ver3,fill=gray] (y22) {};
\draw (w31) + (-15 + 3*360/5 + 90:0.5) node[ver3,fill=gray] (y31) {};
\draw (w32) + (+15 + 3*360/5 + 90:0.5) node[ver3] (y32) {};
\draw (w41) + (-15 + 4*360/5 + 90:0.5) node[ver3,fill=gray] (y41) {};
\draw (w42) + (+15 + 4*360/5 + 90:0.5) node[ver3] (y42) {};

\draw (x0) + (-25 + 0.5*360/5 + 90:0.5) node[ver3] (z01) {};
\draw (x0) + (+25 + 0.5*360/5 + 90:0.5) node[ver3] (z02) {};
\draw (x1) + (-25 + 1.5*360/5 + 90:0.5) node[ver3] (z11) {};
\draw (x1) + (+25 + 1.5*360/5 + 90:0.5) node[ver3,fill=gray] (z12) {};
\draw (x2) + (-25 + 2.5*360/5 + 90:0.5) node[ver3,fill=gray] (z21) {};
\draw (x2) + (+25 + 2.5*360/5 + 90:0.5) node[ver3] (z22) {};
\draw (x3) + (-25 + 3.5*360/5 + 90:0.5) node[ver3] (z31) {};
\draw (x3) + (+25 + 3.5*360/5 + 90:0.5) node[ver3,fill=gray] (z32) {};
\draw (x4) + (-25 + 4.5*360/5 + 90:0.5) node[ver3] (z41) {};
\draw (x4) + (+25 + 4.5*360/5 + 90:0.5) node[ver3] (z42) {};

\draw (c0) + (-30 + 0.5*360/5 + 90:1.15) node[ver3,fill=gray] (a01) {};
\draw (c0) + (+30 + 0.5*360/5 + 90:1.15) node[ver3,fill=gray] (a02) {};
\draw (c1) + (-30 + 1.5*360/5 + 90:1.15) node[ver3] (a11) {};
\draw (c1) + (+30 + 1.5*360/5 + 90:1.15) node[ver3] (a12) {};
\draw (c2) + (-30 + 2.5*360/5 + 90:1.15) node[ver3,fill=gray] (a21) {};
\draw (c2) + (+30 + 2.5*360/5 + 90:1.15) node[ver3,fill=gray] (a22) {};
\draw (c3) + (-30 + 3.5*360/5 + 90:1.15) node[ver3] (a31) {};
\draw (c3) + (+30 + 3.5*360/5 + 90:1.15) node[ver3] (a32) {};
\draw (c4) + (-30 + 4.5*360/5 + 90:1.15) node[ver3,fill=gray] (a41) {};
\draw (c4) + (+30 + 4.5*360/5 + 90:1.15) node[ver3,fill=gray] (a42) {};

\draw (c0) + (-45 + 0.5*360/5 + 90:1.1) node[ver3] (b01) {};
\draw (c0) + (+45 + 0.5*360/5 + 90:1.1) node[ver3] (b02) {};
\draw (c1) + (-45 + 1.5*360/5 + 90:1.1) node[ver3,fill=gray] (b11) {};
\draw (c1) + (+45 + 1.5*360/5 + 90:1.1) node[ver3,fill=gray] (b12) {};
\draw (c2) + (-45 + 2.5*360/5 + 90:1.1) node[ver3] (b21) {};
\draw (c2) + (+45 + 2.5*360/5 + 90:1.1) node[ver3] (b22) {};
\draw (c3) + (-45 + 3.5*360/5 + 90:1.1) node[ver3,fill=gray] (b31) {};
\draw (c3) + (+45 + 3.5*360/5 + 90:1.1) node[ver3,fill=gray] (b32) {};
\draw (c4) + (-45 + 4.5*360/5 + 90:1.1) node[ver3] (b41) {};
\draw (c4) + (+45 + 4.5*360/5 + 90:1.1) node[ver3,fill=gray] (b42) {};

\draw[thick] (x4) -- (z42) -- (a42);
\draw[dotted] (b42) -- (a42);
\draw[thick,red] (b42) -- (w02);

\draw[dotted] (y02) -- (y01);
\draw[thick] (w02) -- (y02) ;
\draw[thick] (y01) -- (w01);

\draw[dotted] (w01) -- (b01);
\draw[thick] (b01) -- (a01);
\draw[thick] (a01) -- (z01) -- (x0);

\draw[thick,red] (x0) -- (z02) -- (a02);
\draw[dotted] (b02) -- (a02);
\draw[thick] (b02) -- (w12);

\draw[thick] (w12) -- (y12); 

\draw[dotted] (y12) -- (y11);
\draw[thick] (y11) -- (w11);

\draw[thick] (w11) -- (b11);
\draw[thick] (b11) -- (a11);

\draw[dotted](a11) -- (z11);
\draw[thick] (z11) -- (x1);

\draw[dotted] (x1) -- (z12);
\draw[thick] (z12) -- (a12) -- (b12) -- (w22);

\draw[dotted] (w22) -- (y22); 
\draw[thick] (y22) -- (y21) -- (w21);

\draw[thick] (w21) -- (b21)-- (a21);
\draw[dotted] (z21) -- (a21);
\draw[thick] (z21) -- (x2);

\draw[dotted] (x2) -- (z22);
\draw[thick] (z22) -- (a22) -- (b22) -- (w32);

\draw[dotted] (w32) -- (y32); 
\draw[thick] (y32) -- (y31) -- (w31);

\draw[thick] (w31) -- (b31)--(a31);
\draw[dotted] (z31) -- (a31);
\draw[thick] (z31) -- (x3);

\draw[dotted] (x3) -- (z32);
\draw[thick] (z32) -- (a32) -- (b32) -- (w42);

\draw[dotted] (w42) -- (y42); 
\draw[thick] (y42) -- (y41) -- (w41);

\draw[dotted] (w41) -- (b41);
\draw[thick] (b41) -- (a41);
\draw[thick] (a41) -- (z41) -- (x4);

\end{tikzpicture}
\caption{Mapping between a $p=2$ Bruhat--Tits tree and the regular hyperbolic tiling~$\{5,4\}$. The first three regions constructed by the algorithm are shown. The red geodesic separates the causal wedge for the boundary region on which the geodesic ends from its complement in the tree. Since this is $p=2$, the number of edges in a tile is different than our standard choice in the arbitrary $q$ case. }
\label{treeandtiling}
\end{figure}

Although the most general set of rules for assigning tensors is unclear, \happies of uniform tiling can easily be constructed from the minimal proposal above with the addition of a few simple rules. These extra rules introduce planarity, so that the Bruhat--Tits tree becomes the spanning tree of the graph consisting of the edges of the \happy tiles. For $q>3$, we can construct a \happy associated to a $[[q,1,(q+1)/2]]_q$ code by grouping the vertices of the tree into sets of $q$, corresponding to tessellation tiles, and adding one bulk vertex to each tile. These tiles are organized into ``alleyways;'' each tile consists of vertices connected by a geodesic for tiles that are the starting points of alleyways, or of vertices living on two geodesics for tiles along the alleyway (see Fig.~\ref{treeandtiling}). The edges of each tile consist of either $q-1$ or~$q-2$ segments coming from the geodesics, and one or two fictitious segments (the dotted lines in Fig.~\ref{treeandtiling}) respectively, that we draw only to keep track of which tree vertices have been grouped. Furthermore, each vertex connects to exactly one dashed edge. Since the tree has valence $q+1$, the \happy tiling has $q+2$ tiles meeting at each vertex.

The description above works for  $q>3$. The case $q=2$ is special and can be obtained from the $[[5,1,3]]_2$ code; this is the case depicted in Fig.~\ref{treeandtiling}. In fact, any size polygon could be used; the only real constraint is that the tiling be hyperbolic of the form $\{n ,q+2 \}$, with~$q$ a prime power. The pathologies of low primes come from the difficulty in demanding the tiling be hyperbolic and requiring perfect tensors; for instance, the $p=2$ case would require a $3$ index perfect tensor, but all perfect tensors have an even number of indices by construction.

In this picture of tiles, the tensor network bonds can be thought of as cutting across the edges of the tiles. Indeed, because of planarity, each edge can be associated with the tensor network bond of its vertices, precisely reproducing the HaPPY construction for uniform tilings of the hyperbolic plane.\footnote{We should always remember, however, that in our construction, unlike in \cite{HaPPY}, bulk indices and tensor network connections are fundamentally associated to groups of tree vertices, and not to the geometric elements of a tile.}

An interesting feature of our construction is that it introduces a peculiar notion of distance on the boundary, in that points $x,y,\in\mathbb{Q}_p$ that are that are far apart (in terms of the norm $\abs{x-y}_p$) can belong to the same tile, or to neighboring tiles, so they can be ``close in entanglement''; this is a concrete manifestation of the dissociation between entanglement and geometry inherently present in our model.

\subsubsection{Explicit tree-to-tessellation mapping}

We now explicitly construct an identification between Bruhat--Tits trees and a \happy of uniform tiling. The end goal is to show that a planar graph of uniform vertex degree $v$ admits a spanning tree of uniform vertex degree $v-1$. Although both the degree of the tree and the size of the tile are constrained by the quantum error correcting code, for the sake of generality we will work with $n$-gonal tiles, $n\geq 5$, and trees of valence $k$, $k\geq3$. The algorithm constructing the mapping proceeds by starting with one $n$-gonal tile, then builds regions of tiles moving radially outward. Each region is built counterclockwise.\footnote{There are, of course, many variations of this algorithm that work; here we only exhibit one of them. For the purposes of constructing the mapping it does not matter which variation we use.} 

\begin{figure}
\centering
\begin{tikzpicture}[scale=2.5]
\tikzstyle{vertex}=[draw,circle]
\tikzstyle{ver2}=[draw,scale=0.6,circle]
\tikzstyle{ver3}=[draw,scale=0.4,circle]
\draw (0*360/5 + 90:1) node[vertex,fill=gray] (v0) {};
\draw (1*360/5 + 90:1) node[vertex] (v1) {};
\draw (2*360/5 + 90:1) node[vertex,fill=gray] (v2) {};
\draw (3*360/5 + 90:1) node[vertex] (v3) {};
\draw (4*360/5 + 90:1) node[vertex,fill=gray] (v4) {};
\draw[thick] (v0) -- (v1) -- (v2) -- (v3) -- (v4);
\draw[dotted] (v4) -- (v0);

\draw (v0) + (0*360/5 + 0.5*65 + 90:0.7) node[ver2] (w01) {};
\draw (v0) + (0*360/5 + -0.5*65 + 90:0.7) node[ver2] (w02) {};
\draw (v1) + (1*360/5 + 0.5*65 + 90:0.7) node[ver2] (w11) {};
\draw (v1) + (1*360/5 + -0.5*65 + 90:0.7) node[ver2,fill=gray] (w12) {};
\draw (v2) + (2*360/5 + -0.5*65 + 90:0.7) node[ver2] (w22) {};
\draw (v4) + (4*360/5 + 0.5*65 + 90:0.7) node[ver2] (w41) {};

\draw (0.5*360/5 + 90:1.8) node[ver2,fill=gray] (x0) {};
\draw (1.5*360/5 + 90:1.8) node[ver2,fill=gray] (x1) {};
\draw (4.5*360/5 + 90:1.8) node[ver2,fill=gray] (x4) {};

\draw[thick] (v0) -- (w01) -- (x0); 
\draw[dotted] (w01) -- (x0) -- (w12); 
\draw[thick] (w12) -- (v1);

\draw[dotted] (v1) -- (w11) node[pos=0.7, above] {$e_n$}; 
\draw[thick] (w11) -- (x1) -- (w22) node[pos=0.6,above] {$e_l$} ;
\draw[thick] (w22) -- (v2) node[pos=0.4, above] {$e_f$} ;

\draw[thick] (v4) -- (w41)-- (x4); 
\draw[dotted] (x4) -- (w02) ; 
\draw[thick] (w02) -- (v0);

\draw[line width=0.4mm] (90:2.15) [->] ++(-54:1.3) arc(54:198:1.3);

\draw[line width=0.4mm] (-165:1.25) [->] ++(0:0.3) arc(0:270:0.3);

\end{tikzpicture}
\caption{Mapping between a $p=2$ Bruhat--Tits tree and a pentagonal HaPPY tiling, after the third tile of the second region has been built. Edge $e_n$ is constrained to be dashed, so edges $e_f$ and $e_l$ are solid. The arrows represent the direction in which the algorithm constructs regions and tiles.}
\label{buildingtree}
\end{figure}
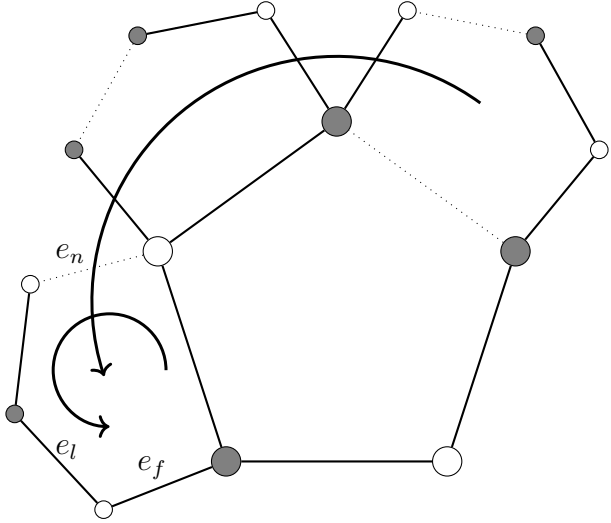

The purpose of this algorithm is to build a graph of dashed and solid edges such that every vertex has degree $k+1$ and exactly one dashed edge connecting to it. The \happy tiling is given by the solid and dashed edges, and the tree is given by the solid edges, as in Fig.~\ref{treeandtiling}. The steps of the algorithm are as follows (see also Fig.~\ref{buildingtree} for a pictorial representation):
\begin{enumerate}
\item Start with an $n$-gonal tile with one edge dashed and $n-1$ solid edges. This is region $r=1$. The left vertex of the dashed line is the current vertex.
\item To construct region $r+1$, for the current vertex, add an edge $e_f$ extending outward, then go counterclockwise around the tile being created, breaking off the edges shared with region $r$ as soon as a vertex with degree less than $k+1$ is encountered. Call the first new edge after breaking off $e_n$.  If either $e_f$ or $e_n$ are constrained (by the condition that in the graph we want to obtain each vertex has degree $k+1$ and precisely one dashed edge connecting to it) to be dashed, make them dashed, otherwise they are solid. If neither $e_f$ nor $e_n$ are dashed, make the ``farthest'' edge (call it~$e_l$) dashed; otherwise, leave it as a solid edge.\label{mostimportantpointtwo}
$e_l$ is chosen so that its distance to the existing graph is as large as possible, and so equal on both sides if possible; if the number of new edges is even, so that this prescription is ambiguous, the choice closer to~$e_f$ is taken.\label{step2}
\item
 Move counterclockwise to the next vertex on the boundary of the current region, skipping any vertices of degree $k+1$. This new vertex becomes the current vertex.
\item Repeat the step above until we have built an edge $e_f$ on all vertices of region $r$ that have degree less than $k+1$. This completes region $r+1$.
\item To start on the next region, set the current vertex to be the left vertex of the dashed line on the first $r+1$ tile that we built, then go to step \ref{step2} above.
\end{enumerate}

We now show why the algorithm works:
\begin{itemize}
\item By induction, there can be no neighboring vertices of degree greater than two on the boundary at any step, except when building a tile on the next-to-last edge of a tile from the previous region, in which case a vertex of degree $3$ neighbors a vertex of some degree. This is because if the boundary has no neighboring vertices of degree $k+1$, any tile we add shares with the boundary of the previously constructed tiles at most two edges, so (since $n\geq 5$) it will have at least three free edges, adding at least two vertices of degree $2$ between the vertices to which it connects.
\item From the previous point, when constructing any tile, the vertices to which $e_f$ and $e_n$ connect cannot both have degree $k$ before adding $e_f$ and $e_n$, so either $e_f$ or $e_n$ can be made solid.
\item If $e_f$ and $e_n$ are not dashed, then $e_l$ only connects to two solid edges, so it can be made dashed.
\item By the above, each new tile we add introduces exactly one dashed edge, so the graph of solid edges remains a tree at all steps.
\end{itemize}
This completes the proof.

\subsection{Bulk wedge reconstruction}
\label{sec:wedge}

In this subsection we discuss how bulk reconstruction, in the sense of \cite{HaPPY}, functions for our proposal. Although the construction in~\S\ref{apptilingalgorithm} replicates the tensor network tiling of \cite{HaPPY}, there are some differences of interpretation.

The algorithm outlined in Pastawski \textsl{et al.}'s paper prescribes that, starting with a given boundary region, one should add tiles one by one if the bulk qubits they carry can be reconstructed from the known data; i.e., if one knows (or can reconstruct) a majority of the edge qubits on the tile. When no further tiles can be added in this manner, the reconstruction is complete, and the boundary of the region is the ``greedy geodesic.''

In our case, for a given drawing of the tree, the first step to reconstructing a boundary open set $|x_A-y_A|_p$ is to identify a geodesic $\mathcal{G}$ that separates the causal wedge for the boundary region on which $\mathcal{G}$ ends from its complement in the tree. This is nontrivial, because due to the non-planarity of the tree, paths that end on the ball corresponding to the endpoints of the geodesic can be drawn outside the wedge.
A $\mathcal{G}$ with the desired property is a geodesic from which only one path leaves into the complement of the wedge; such an example is drawn in Fig.~\ref{treeandtiling} in red.
Given a choice of planar structure, there is a unique ``outermost'' separating geodesic associated to each open set.
 Once a separating geodesic has been identified, we can assign $x_A$ and $y_A$ to its endpoints, the inside of the ball to the tree inside the wedge, and the complement of the ball to the complement of the wedge in the tree. 

For the tilings $\{5,n\}$, it is straightforward to see that the greedy geodesic for a boundary open ball coincides with the tree geodesic $\mathcal{G}$ that forms the ``boundary'' of that open ball in the chosen planar structure. The alleyways in the diagram are sequences of tiles joined along dashed lines, such that fewer than half of the edges on each tile are exposed to either side; therefore, each alleyway forms a ``firebreak,'' which the greedy algorithm cannot jump across. If none of the dashed edges are known, none of the tiles in the alleyway can be reconstructed. Therefore, starting at a boundary open set, the greedy algorithm propagates up the alleyways whose ends lie inside the region, and fills out the region marked off by the tree geodesic. It cannot stop before the region is filled, since by construction each tile neighbors at least three tiles that are further away from the center than the tile itself is.

The only place where subtleties arise in this argument is at the two uppermost tiles in the boundary alleyways. Since these tiles form the tops of alleyways, they have one dashed edge and four contiguous solid edges; it is possible (as one can check by drawing a large enough picture) for the outermost geodesic to loop around three or more or of these solid edges before moving off down the boundary alleyway. If this happens, the greedy algorithm will jump over the separating geodesic at the top tile. One can remedy this problem by adopting a different choice of separating geodesic in these cases. For instance, instead of always turning left, one can turn right after going around two edges of the top tile if going left would create a problem. After this modification, one can resume turning left. 

When the tiles are larger than pentagons,
 a difficulty arises when one is building tiles for which two edges touch the previous part of the picture: one may be forced (by valence requirements) to build a dashed edge at~$e_n$ or~$e_f$, while another dashed edge exists in the previously built part, a distance of only one away. 
An instance where this occurs (although, of course, it causes no problems for pentagonal tiles) can be seen at the bottom right corner of Fig.~\ref{treeandtiling}.
This constructs an alley that could be jumped by the greedy algorithm. 

A simple fix for this problem would be to simply use a pentagon whenever this situation arises, resulting in a nonuniform tiling where the alleyways still function as firebreaks. While this will produce a valid holographic code, it is not immediate that the tiling is even regular in this case.

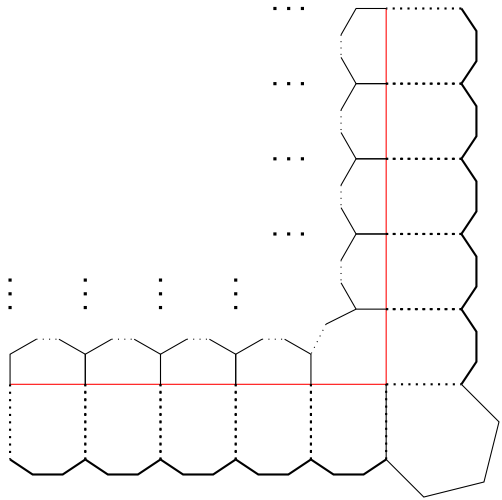
\begin{figure}
\centering
\begin{tikzpicture}
\draw[red] (0,0) -- (5,0) -- (5,5);
\foreach \x in {0,1,...,4} {
	\draw[dotted, thick] (\x,0) -- (\x,-1);
	\draw[thick] (\x,-1) -- (\x + 0.3, -1.2) -- (\x + 0.7, -1.2) -- (\x + 1, -1);
	\draw[dotted, thick] (\x + 1,-1) -- (\x + 1, 0);
	};
\foreach \x in {0,1,...,4} {
	\draw[dotted, thick] (5,\x) -- (6,\x);
	\draw[thick] (6,\x) -- (6.2,\x + 0.3) -- (6.2,\x + 0.7) -- (6,\x+1);
	\draw[dotted, thick] (6,\x+1) -- (5,\x + 1);
	};
\draw (5,0) -- (5,-1) -- (5.5,-1.5) -- (6.3,-1.3) -- (6.5,-0.5) -- (6,0);
\foreach \x in {1,2,...,4} {
	\draw[very thick, loosely dotted] (\x-1,1) -- (\x-1,1.5);
	\draw[very thick, loosely dotted] (3.5,\x+1) -- (4,\x+1);
	};
\foreach \x in {0,1,...,3} {
	\draw (\x,0) -- (\x,0.4) -- (\x + 0.35, 0.6);
	\draw (\x + 0.65,0.6) -- (\x + 1, 0.4) -- (\x + 1,0);
	\draw[dotted] (\x + 0.35,0.6) -- (\x + 0.65,0.6);
	};
\foreach \x in {1,2,3,4} {
	\draw (5,\x+1) -- (4.6,\x+1) -- (4.4,\x+0.65);
	\draw (4.4,\x+0.35) -- (4.6,\x) -- (5,\x);
	\draw[dotted] (4.4,\x+0.65) -- (4.4,\x+0.35);
	};
\draw (4.6,1) -- (4.2,0.8);
\draw[dotted] (4.2,0.8) -- (4,0.4);
\end{tikzpicture}
\caption{Mapping between the tree and tiling for reconstructing a chosen causal wedge, for the tiling $\{6,4\}$. The two central alleys partition the plane into two regions: the causal wedge of the red geodesic and its complement. Tilings can be constructed to either side of the shown alley by building tiles via step \ref{mostimportantpointtwo} of the algorithm of~\S\ref{treetensorsubsection}. The departure from the general algorithm at the turning point of the red geodesic is necessitated by the small value of~$n$.}
\label{wedgerecfig}
\end{figure}

Another option, if we are willing, is to use a different algorithm  that alters the tiling near a specifically chosen geodesic, so that the problem does not arise for that particular wedge.
We explain how to construct the tiling in this case. The idea is to construct two alleyways, with the sides with one edge per tile pointing towards the wedge. This separates the plane into two regions, that for the purposes of tile building don't talk to each other. Since the rules for building a tile from~\S\ref{apptilingalgorithm} are (almost) local, they have no information about the global structure of the row being built. It is thus possible to use them to cover the two regions, moving ``left'' and ``right'' to create rows, and ``up'' and ``down'' to stack the rows. We give the explicit steps of the algorithm (see Fig.~\ref{wedgerecfig}):
\begin{enumerate}
\item Build an alley of $k$-gons, by starting with a $k$-gon with one dashed edge (call this the root tile) and building the successive gons always on the dashed edges. For each $k$-gon except for the starting one, the number of solid edges on the two sides of the dashed edges should be $1$ and $k-3$ respectively, with all $1$'s occurring on one side, and all $(k-3)$'s on the other.
\item Build a second alley of $k$-gons, starting on the solid edge of the root tile that neighbors the root's dashed edge on the side of $1$'s. For each $k$-gon except the starting one, the number of solid edges on the two sides of the dashed edges should be $1$ and $k-3$ respectively, with all $1$'s facing the $1$'s of the first alley. The plane has now been split into two regions: wedge and complement.
\item To construct the tiling inside the wedge, start on some edge of the two central alleyways, and construct a tile using the rules from point \ref{mostimportantpointtwo} in the algorithm in~\S\ref{apptilingalgorithm}. Then move to the right, and construct new tiles rotating clockwise in the construction of each tile. To do the other side, start from the initial tile and move left, constructing tiles using the rules from point \ref{mostimportantpointtwo} in the algorithm while rotating counterclockwise. This fills a ``row''.

\item Once the ``row'' is complete, move to the row ``above'' it and repeat.

\item To construct the tiling of the complement of the wedge, run the two bullet points above, but having the clockwise and counter-clockwise rotations swapped, and moving ``below'' instead of ``above''. 
\end{enumerate}

This algorithm works because locally the construction is the same as the one in the algorithm of Sec~\ref{apptilingalgorithm}. The individual tile building procedure does not depend on whether it is going around a finite region (as in~\S\ref{apptilingalgorithm}), or along an infinite ``row''. And since inside the wedge more than half of each tile's neighbors are further away from the center than the tile is, the reconstruction covers the entire wedge.


While in the original HaPPY construction \cite{HaPPY} one tensor network suffices to reconstruct the causal wedge associated to any boundary interval, in our case 
the tree identifies a certain collection of open sets on the boundary when pentagonal tiles are used (and, in one possible generalization to larger tiles, may even treat one boundary region as special). Even for pentagons, there may be many ways to draw the spanning tree on the same tiling. Moreover, the boundary tiles are not treated on an equal footing: they form the ends of shorter and longer alleyways. The longer the alleyway in which a boundary tile appears, the larger the first open set that includes it. 
One can think of these extra choices as follows: If one were to draw all possible greedy geodesics on the tiling, all edges would be marked, and there would be no information. Marking the geodesics with a subgraph is only useful when there is a unique path between pairs of boundary points, so that one knows ``which way'' to turn in order to recover the geodesic. This means that the marked subgraph should have no closed loops: that is, it should be a tree, and the spanning tree is (by definition) a largest possible subtree.

\subsection{Discussion}

By choosing a planar assignment of tensors, we have found a way to think of the Bruhat-Tits tree as the spanning tree of a tensor network built from a tiling of the plane. Although this choice gives an example of a connection between these two constructions, it is also somewhat arbitrary. Moreover, the rules we have identified here still break symmetries of the tree, since a $\PGL(2,\Q_p)$ transformation need not preserve the grouping of vertices or the planar structure. One might hope to construct a tensor network associated to the tree with  more minimal auxiliary structure, so that the full symmetry group of the $p$-adic bulk spacetime is manifest for the network as well; however, for the reasons outlined above, it is difficult to understand how to do this while making contact with tensor network proposals existing in the current literature.
But it is at any rate reasonable to ask for a more exotic way of connecting the tree with tensors, that either weakens the requirements on the chosen planar embedding, or makes no reference to the choice of an embedding at all. 

The former idea (indeed, either of these) might obscure the geometric interpretation that is typical in familiar tensor networks, but it is expected that quantum gravity contains non-geometric states, so there is a sense in which at least small deviations from planarity should be physically acceptable. One simple example might be to connect two distant parts of the tree together through a non-planar tensor. If this can be done in a consistent way, one might interpret this non-planar defect as a bridge of entanglement between two points of the tree. It would be interesting to investigate whether a configuration with defects or more complicated non-planar structures can be understood in terms of the $\mrm{ER}=\mrm{EPR}$ proposal \cite{EREPR}.



To turn to the second of these suggestions, as we have already indicated, it would make sense from the point of view of $p$-adic AdS/CFT (where no planar structure is required at all, and the tree is viewed intrinsically as a geometric  space) to arrive at a construction of a tensor network that doesn't require any planar data whatsoever.
One such different possible approach to the construction of holographic codes on 
Bruhat--Tits trees along these lines can be developed using its algebro-geometric properties,
in particular the fact that the link of each vertex is a copy of the projective line over
a finite field. 
Indeed, this was part of our motivation for discussing an algebro-geometric construction of perfect tensors.
Using constructions of (classical) algebro-geometric codes 
associated to curves over finite fields, and an algorithm that associates quantum
codes to self-orthogonal classical codes, one can obtain holographic codes 
intrinsically associated to the geometry of Bruhat--Tits trees. We will develop
this additional viewpoints, and discuss its relation to usual tensor networks, in a separate paper.



\section{$p$-adic conformal field theories and holography for scalar fields}
\label{scalars}

In this section, we turn from tensor networks to genuine field theories defined on $p$-adic spacetimes: either in the bulk of the tree~$T_p$ (or possibly a quotient by a Schottky group) or on a $p$-adic algebraic curve at the boundary. We will find evidence for a rich holographic structure strongly reminiscent of ordinary AdS/CFT. The conformal theory on the fractal $p$-adic boundary is analogous to 1+1 dimensional field theory with a $p$-adic global conformal group; our main example is the $p$-adic free boson which permits a Lagrangian description. In the bulk, semi-classical massive scalar fields defined on the lattice model naturally couple to operators on the conformal boundary in a way that allows for precise holographic reconstruction. One can also interpret the radial direction in the tree as a renormalization scale. These observations unite discrete analogs of AdS geometry, conformal symmetry, and renormalization in a holographic way. 

\subsection{Generalities of $p$-adic CFT, free bosons, and mode expansions}
\label{sec41}

While non-archimedean conformal field theory has been considered in the literature from several different perspectives~\cite{Melzer, BrFr, ChekhovZabrodin}, it remains much less well-studied than ordinary two-dimensional CFT.  
Melzer~\cite{Melzer} defines these theories in general by the existence of an operator product algebra, where all operators in the theory are primaries with the familiar transformation law under the global conformal group $\SL(2,\mathbb{Q}_p)$. Descendants are absent in Melzer's formulation because there is no analogue of the derivative operators $\partial$ and~$\bar\partial$ acting on complex-valued functions over~$\Q_p$~\cite{Melzer}, and (correspondingly) no local conformal algebra.

In this formulation, the correlation function between two primary fields $\phi_m(x)$ and $\phi_n(y)$ inserted at points $x$ and $y$ and having scaling dimensions $\Delta_n$ is given (after normalization) by
\begin{equation}
\langle \phi_{m}(x) \phi_{n}(y)  \rangle = \frac{\delta_{m,n}}{|x-y|_p^{2 \Delta_n}}.
\end{equation}
(We will understand this formula holographically in what follows.) 
As in the archimedean case, as we take the points $x$ and $y$ to be close together ($p$-adically), we wish to expand the product as a sum of local field insertions: the operator product expansion. For two such primaries $\phi_m(x)$ and $\phi_n(y)$, there exists an $\epsilon > 0$ such that for $|x-y|_p < \epsilon$, the correlation function (perhaps with other primaries $\phi_{n_i}(x_i)$ inserted) is given by the expansion
\begin{equation}
\langle \phi_{m}(x) \phi_{n}(y) \phi_{n_1}(x_1) \dots \phi_{n_i}(x_i) \rangle = \sum_r \tilde{C}^r_{mn}(x,y) \langle \phi_{r}(y) \phi_{n_1}(x_1) \dots \phi_{n_i}(x_i) \rangle,
\end{equation}
where the sum runs over all primaries in the theory, and $\tilde{C}^r_{mn}(x,y)$ are real valued. This relation should hold whenever $|x-y|_p$ is smaller than the distances to the $x_i$'s. Invariance under $\SL(2,\mathbb{Q}_p)$ implies
\begin{equation}
\tilde{C}^r_{mn}(x,y) = C^r_{mn}|x-y|_p^{\Delta_r - \Delta_m - \Delta_m}
\end{equation}
with constant OPE coefficients $C^r_{mn}$.

Theories defined in this way enjoy a number of special properties not true of their archimedean counterparts. They are automatically unitary since they possess no descendant fields. Additionally, because $\mathbb{Q}_p$ is an \textit{ultrametric} field, all triangles are isosceles: for $x,y,z \in \mathbb{Q}_p$, from the $p$-adic norm we have 

\begin{equation}
\text{If  } |x-y|_p \neq |y-z|_p \text{  , then   } |x-z|_p = \max \left \{|x-y|_p, |y-z|_p \right \}.
\end{equation}
This fundamental property of the $p$-adic numbers implies that the three- and four-point functions are \textsl{exactly} determined by the conformal weights and OPE coefficients. In the case of the four-point function, after an $\SL(2,\mathbb{Q}_p)$ transformation which maps three points to $0, 1,$ and $\infty$, the only free parameter (the cross ratio of the original points) must be contained in a ball in the neighborhood of one of the other points. Since the OPE is exact in each neighborhood, one can compute the three possible cases and determine the full four-point function. 

In fact, all higher-point functions are constrained by global conformal symmetry alone; by contrast, the spectrum of OPE coefficients is less constrained than in familiar CFTs. A consistent model can be constructed using the structure constants of any unital commutative algebra, subject to one simple condition.
These features may be of interest in the study of conformal field theory and conformal blocks, but we do not pursue that direction here; the interested reader is referred to~\cite{Melzer}.


Let us now step back and consider the $p$-adic theory from the perspective of quantizing a classical field theory described by a Lagrangian. Many familiar objects from the study of quantum fields over normal (archimedean) spacetime have direct analogues in the $p$-adic setting. For example, one frequently makes use of the idea of a \textsl{mode expansion} of a field on flat spacetime in terms of a special class of basis functions, the plane waves:
\begin{equation}
\phi(x) = \int_\R dx\, e^{ikx} \tilde{\phi}(k).
\end{equation}
The functions $e^{ikx}$ are eigenfunctions of momentum, or equivalently of translations. Mathematically, we can think of these as \textsl{additive characters} of~$\R$: they are group homomorphisms 
$\chi: \R\rightarrow\C$, such that $\chi(x+y) = \chi(x)\chi(y)$. 

The additive characters of the fields~$\Q_p$ are also known: they take the form \cite{vladimirov-book}
\begin{equation}
\chi_k(x) = e^{2\pi i \{ k x \} }. \label{additivechars}
\end{equation}
Here $k,x\in\Q_p$, and the normalization factor $2\pi$ is included for convenience (in keeping with the typical math conventions for Fourier transforms). The symbol $\{\cdot\}: \Q_p \rightarrow \Q$ denotes the \textsl{fractional part} of the $p$-adic number.\footnote{As with other notations referring to the $p$-adics, we will sometimes use the subscript $\{ \cdot \}_p$ when it is necessary for emphasis or to make reference to a specific choice of prime.} It is defined by truncating the decimal expansion to negative powers of the prime, 
\begin{equation}
\left\{ \sum_{k=m}^\infty a_k p^k \right \} = \sum_{k=m}^{-1} a_k p^k,
\end{equation}
where the right-hand side is interpreted as an ordinary rational number, understood to be zero when the range of the sum is empty ($m$ is non-negative). Since a $p$-adic number and its fractional part differ (at least in a formal sense) by an integer, it makes sense that the complex exponential~\eqref{additivechars} should depend only on the fractional part of~$kx$. (However, care should be taken: in general, it is not true for rational~$x$ that $e^{2\pi i x}= e^{2\pi i \{ x \}_p}$. For instance, $0.1 = 1/10$ is a $3$-adic integer.)

A wide class of scalar fields on~$\Q_p$ can be expanded in a basis of the additive characters, just like a mode expansion in the archimedean setting: 
\begin{equation}
\phi(x) = \int d\mu(k)\,e^{2\pi i \{ k x \} } \, \tilde{\phi}(k). 
\end{equation}
Here $d\mu(k)$ is the Haar measure on~$\Q_p$. The theory of the $p$-adic Fourier transform is developed in more detail in Appendix~\ref{app:integrals}.


Our principal example (and also by far the most well-studied instance) of a $p$-adic conformal field theory is the free boson: a single (real or complex) scalar field on~$\PL(\Q_p)$ or another $p$-adic Riemann surface, with a massless quadratic action.
This theory was of interest in the context of $p$-adic string theory, in which the worldsheet is a $p$-adic space, but the target space (and hence all physically observable quantities) are ordinary. Many results were derived in that literature, including the well-known Freund--Olson--Witten tachyon scattering amplitudes~\cite{FreundOlson,FreundWitten,BFOW}. 

Our interpretation of the system in question will be somewhat different, as we will emphasize the holographic nature of the interplay between field theory defined on a Riemann surface (algebraic curve) and the study of its hyperbolic filling, a quotient of the Bruhat--Tits tree. (In the $p$-adic string literature, it was common to view the tree as playing the role of the ``interior'' of the worldsheet.)
Many of our results will parallel aspects of the $p$-adic string, but we will view this theory as a CFT on $\PL(\mathbb{Q}_p)$ without any reference to a target space. 

The $p$-adic free boson is considered here because it permits a Lagrangian description in terms of the nonlocal Vladimirov derivative, $\partial_{(p)}$, which acts on complex- or real-valued fields of a $p$-adic coordinate. This derivative is defined by
\begin{equation}
\partial_{(p)}^n f(x) = \int_{\mathbb{Q}_p} \frac{f(x') -f(x)}{|x'-x|_p^{n+1}} dx'.
\end{equation}
In the $p$-adic string literature, $\partial_{(p)}$ 
 is also known as a normal derivative, for reasons that will become clear in the following sections. Intuitively, the formula is similar to Cauchy's representation of the $n$-th derivative of a function by a contour integral. A more detailed explanation of its properties is given in Appendix~\ref{app:derivatives}. While the parameter~$n$ is often taken to be an integer, it may in principle assume any real value. 
 
 One can arrive at the following action either by ``integrating out'' the interior of the string worldsheet $T_p$ as done in \cite{Zabrodin}, or by hypothesis as the minimal ``quadratic'' action of a scalar over a $p$-adic coordinate~\cite{Dragovich}. The action for a single scalar is (setting the overall coupling to 1)~\cite{Spokoiny, Zhang, Grange}:

\begin{equation}
S_p[\phi] = -\int_{\mathbb{Q}_p} \phi(x) \partial_{(p)} \phi(x) dx.
\end{equation}
where 
$\partial_{(p)} \phi(x)$ is the first Vladimirov derivative of the field $\phi$. We take $\phi(x)$ to be a scalar representation of the conformal group (see~\cite{Sally} 
for discussion of representations of $\SL(2,\mathbb{Q}_p)$ in general). Under an element
\begin{equation*}
g=
\begin{pmatrix}
a & b \\
c & d
\end{pmatrix}
\end{equation*}
of the conformal group, where $a,b,c,d \in \mathbb{Q}_p$ and~$ad-bc = 1$, 
 quantities in the above expression transform as
\begin{align*}
x \rightarrow \frac{ax + b}{cx+d}, \hspace{3 mm} &x' \rightarrow \frac{ax' + b}{cx'+d}, \\
dx \rightarrow \frac{dx}{|cx+d|_p^2}, \hspace{3 mm} &dx' \rightarrow \frac{dx'}{|cx'+d|_p^2}, \\
|x'-x|_p^{-2} \rightarrow |(cx'+d)&(cx+d)|_p^2|x'-x|_p^{-2}.
\end{align*}
As in~\cite{Melzer}, a field $\phi_n(x)$ having conformal dimension $\Delta_n$ transforms as
\begin{equation}
\phi_n(x) \rightarrow \phi_n' \left(\frac{ax + b}{cx+d} \right) = |cx+d|_p^{2 \Delta_n}\phi_n(x)
\end{equation}
under the $p$-adic conformal group. For the free boson $\phi(x)$, we claim $\Delta = 0$. With this one can see the derivative $\partial_{(p)} \phi(x)$ carries a weight $|cx+d|_p^{2}$ and thus is a field of dimension $1$. It should now be clear that the action $S_p[\phi]$ is invariant under the global conformal group.

Given the action $S_p[\phi]$, we can define the partition function in the usual way by integrating over configurations with measure $\mathcal{D} \phi$. As in the case of the $p$-adic string, because $\phi$ is a complex (and not $p$-adic) valued field, this integration measure is exactly the one that appears in ordinary field theory:
\begin{equation}
Z_p = \int \mathcal{D} \phi \hspace{1mm} e^{-S_p[\phi]}.
\end{equation}

As many authors have noted~\cite{Spokoiny, Zhang, Parisi}, this action and the partition function actually describe a free theory. This means the saddle point approximation to the partition function is exact, and it can be computed by Gaussian integration exactly as in the case of a real free field. Of more interest in the present discussion is the two point function. To do this we introduce sources $J(x)$ to define the generating function:
\begin{equation}
Z_p[J] = \int \mathcal{D} \phi \hspace{1mm} \exp{\left (-S_p[\phi] + \int_{\mathbb{Q}_p} J(x')\phi(x') dx' \right )}.
\end{equation}
The sources for the 2-point function or propagator take the form of $p$-adic delta functions at the insertion points $x,y$ are $J(x) = \delta(x'-x) + \delta(x'-y)$. Just as in the real case, we vary with respect to $\phi(x)$ and find the classical solution which extremizes the above action. This is the Green's function for the Vladimirov  derivative $G(x-y)$, satisfying
\begin{equation}
\partial_{(p)} G(x-y) = - \delta(x-y).
\end{equation}
(Just as it is on the real line, the Dirac distribution on the $p$-adics is the integral kernel representing the linear functional $\ev_x: f \mapsto f(x)$.)

To solve for $G(x-y) = \langle 0 |\phi(x) \phi(y) |0 \rangle$, we apply the $p$-adic Fourier transform to both sides using techniques from Appendix~\ref{app:integrals}. In Fourier space the derivative brings down one power of the momentum and the delta function becomes an additive character:
\begin{equation}
\widetilde{G}(k) = - \frac{\chi(ky)}{|k|_p}.
\end{equation}
The 2-point function in position space can be obtained by inverse Fourier transform (with $u = x-y$):
\begin{align}
G(x-y) &= - \int_{\mathbb{Q}_p} \frac{\chi(k(y-x))}{|k|_p}dk\\
&= - \int_{\mathbb{Q}_p} \frac{\chi(ku)}{|k|_p}dk.
\end{align}
This integral is divergent as $k \rightarrow 0$. We compute two similar integrals in Appendix~\ref{app:derivatives}, where the apparent divergence is canceled by the numerator. Unlike in those examples, this integral really does diverge logarithmically, just as the two-point function of a dimension-zero operator in 2d conformal field theory has a logarithmic divergence. Proceeding as in that case, we introduce a regulator to extract the finite part by computing
\begin{equation}
\lim_{\alpha \rightarrow 0} \int_{\mathbb{Q}_p} \chi(ku)|k|^{\alpha -1}_pdk.\\
\end{equation}
This appears in the second integral computed in the appendix; in terms of the $p$-adic gamma function $\Gamma_p(x)$ it is:
\begin{equation}
\lim_{\alpha \rightarrow 0} \int_{\mathbb{Q}_p} \chi(ku)|k|^{\alpha -1}_p\, dk
= \lim_{\alpha \rightarrow 0} \Gamma_p(\alpha)|u|_p^{-\alpha}.
\end{equation}
As $\alpha \rightarrow 0$ the gamma function has a simple pole and the norm has a log piece:
\begin{align}
 \lim_{\alpha \rightarrow 0} & \Gamma_p(\alpha) \approx \frac{p-1}{p \ln p}\frac{1}{\alpha}, \\
\lim_{\alpha \rightarrow 0} &|u|_p^{-\alpha} \approx 1 - \alpha \ln|u|_p.
\end{align}
Finally we restore $u=x-y$ and find the 2-point function up to normalization is:
\begin{equation}
\langle 0 |\phi(x) \phi(y) |0 \rangle \sim \ln\left |\frac{x-y}{a}\right|_p, \hspace{3mm} a\rightarrow 0.
\end{equation}
This is exactly analogous to the correlator for the ordinary free boson in two dimensions.




\subsection{The Laplacian and harmonic functions on~$T_p$}
\label{sec42}

In addition to boundary scalar fields, we will be interested in scalar fields in the ``bulk,'' i.e., defined on the Bruhat--Tits tree. Such a field is a real- or complex-valued function on the set of vertices. We will also consider fields that are functions on the set of edges; as we will discuss later, such functions will be analogous to higher-form fields or metric degrees of freedom in the bulk. For now we mention them for completeness and to fix some standard notation. For more information about fields in the tree, the reader can consult~\cite{Zabrodin} and references given therein. 

We think of the tree as the 1-skeleton of a simplicial complex, and make use of standard notations and ideas from algebraic topology. The two types of fields mentioned above are just 0- and 1-cochains; we will refer to the space of such objects as $C^*(T_p)$, where $*=0$ or~$1$. 

If an orientation is chosen on the edges of the tree, the boundary operator acts on its edges by $\partial e = t_e - s_e$, where $s$ and~$t$ are the source and target maps. The corresponding coboundary operator acts on fields according to the rule
\begin{equation}
d: C^0(T_p) \rightarrow C^1(T_p), \quad (d\phi)(e) = \phi(t_e) - \phi(s_e).
\end{equation}
The formal adjoint of this operator is
\begin{equation}
d^\dagger: C^1(T_p) \rightarrow C^0(T_p), \quad (d^\dagger \psi)(v) = \sum_e \pm \psi(e),
\end{equation}
where the sum is over the $p+1$ edges adjacent to vertex~$v$, with positive sign when $v$ is the source and negative sign when it is the target of~$e$. Whether or not $d^\dagger$ is actually an adjoint to~$d$ depends on the class of functions being considered; the $L^2$ inner product must be well-defined, and boundary conditions at infinity must be chosen to avoid the appearance of a boundary term.

Upon taking the anticommutator $\{ d, d^\dagger \}$, we obtain an operator of degree zero, which is the proper analogue of the Laplacian. We will most often use its action on the 0-cochains, which can be represented by the formula 
\begin{equation}
\Laplace\phi(v) = \sum_{d(v,v')=1} \phi(v') - (p+1)\phi(v).
\end{equation}
This is sometimes written using the notation $\Laplace_p = t_p - (p+1)$, where~$t_p$ is the Hecke operator on the tree. 
The analogous formula for 1-cochains is
\begin{equation}
\Laplace \psi(e) = \sum_{e'} \pm\psi(e') - 2 \psi(e), \label{edge-laplacian}
\end{equation}
where the sum goes over the $2p$ edges adjacent to~$e$ at either side. Unlike for the vertices, there is a dependence on the choice of orientation here: an edge in the sum~\eqref{edge-laplacian} enters with positive sign when it points in the ``same direction'' as~$e$, i.e., points out from $t_e$ or into $s_e$. Edges enter with negative sign when the opposite is true. In the standard picture of the tree with $\infty$ at the top and all finite points of~$\Q_p$ at the bottom (e.g.\ Fig.~\ref{ryufigure}), oriented with all edges pointing downward, we therefore have $(p-1)$ negative and $(p+1)$ positive terms in~\eqref{edge-laplacian}; however, there is no obstruction to a choice of orientation for which any desired collection of signs  appears.
Notice that, for general~$p$, the Laplacian acting on edges (unlike on vertices) will not have a zero mode; this makes sense, since the tree is a contractible space. The exception is $p=2$, for which the standard choice of vertical orientations defines a Laplacian which annihilates constant functions of the edges. (Of course, the $p=2$ tree is still contractible.) 

We should remark on one important point: the entire analysis of this paper treats the case where the edges of the tree have uniform lengths, and argues that this is analogous to a maximally symmetric vacuum solution in ordinary gravity. It is natural to wonder what the correct analogues of the metric degrees of freedom actually are. One might speculate that allowing the edge lengths to be dynamical (breaking the $\PGL(2,\Q_p)$ symmetry) should correspond to allowing the metric to vary; after all, this would vary the lengths of paths in the tree, which are the only data that seem logically connected to the metric. By analogy with the Archimedean case, it would then make sense to assume that the edge Laplacian \eqref{edge-laplacian} will play a role in the linearized bulk equations of motion for edge lengths around a background solution. However, we will relegate further investigation of this idea to future work. 

\subsubsection{Action functional and equation of motion for scalar fields}
\label{sec:modes}

Equipped with these ingredients, it is now straightforward to write down action functionals and equations of motion for free scalar fields. The massless quadratic action is
\begin{equation}
S[\phi] = \sum_e \left|d\phi(e)\right|^2.
\end{equation}
In what follows, we will study properties of solutions to the ``wave equation'' $\Laplace \phi = 0$, and its massive generalization $(\Laplace - m^2) \phi = 0$, on the tree. These have been considered in~\cite{Zabrodin}. 

There is a family of basic solutions to the Laplace equation, labeled by a choice of a boundary point~$x$ and an arbitrary complex number~$\kappa$. The idea is as follows: Given an arbitrary vertex~$v$ in the bulk of the tree, a unique geodesic (indeed, a unique path) connects it to~$x$. As such, exactly one of its $p+1$ neighbors will be closer (by one step) to~$x$, and the other $p$ will be farther by one step. Therefore, the function
\begin{equation}
\varepsilon_{\kappa,x}(v) = p^{-\kappa \, d(x,v)} \label{planewave}
\end{equation}
will be an eigenfunction of the Hecke operator, with eigenvalue $(p^\kappa + p^{1-\kappa})$.

The catch in this is that the distance~$d(x,v)$ is infinite everywhere in the bulk. We need to regularize it by choosing a centerpoint $C$ in the tree, and declaring that $d(x,C)=0$. (This just scales the eigenfunction~\eqref{planewave} by an infinite constant factor). Then $d(x,v) \goesto -\infty$ as $v\goesto x$, but we have a well-defined solution to the Laplace equation everywhere in the tree. These solutions are analogous to plane waves; the solution varies as the exponent of the (regularized) distance to a boundary point, which in the normal archimedean case is just the quantity $\mathbf{k}\cdot\mathbf{r}$. 

The corresponding eigenvalue of the Laplacian is 
\begin{equation}
\Laplace \varepsilon_{\kappa, x} = m_\kappa^2 \varepsilon_{\kappa,x} = \left[ (p^\kappa + p^{1-\kappa}) - (p+1) \right] 
\varepsilon_{\kappa,x}. \label{masses}
\end{equation}
It is therefore immediate that the harmonic functions on the tree (solutions to the massless wave equation) are those with $\kappa = 0$ or~$1$; $\kappa=0$ is the zero mode consisting of constant functions, whereas $\kappa=1$ is the nontrivial zero mode. The eigenvalues~\eqref{masses} are invariant under the replacement $\kappa \goesto 1 - \kappa$, due to the inversion symmetry of the boundary theory. 

If we are considering a real scalar field, we must be able to write a basis of real solutions. Of course, when~$\kappa$ is real, we will always be able to do this. More generally, if $\kappa = \kappa_0 + i \gamma$, our solutions look like
\begin{equation}
\varepsilon\sim p^{-\kappa_0\, d}e^{-i\gamma\ln(p) \, d}, \quad
p^{(\kappa_0-1)d} e^{i\gamma\ln(p)\,d}.
\end{equation}
Thus, to construct a basis of real solutions, the following possibilities can occur:
\begin{itemize}
\item $\kappa = 1/2$. In this case, there is no restriction on~$\gamma$, and the solutions look like cosines and sines of~$\gamma \ln(p) \, d(x,v)$, modulated by~$p^{d(x,v)/2}$.
\item $\kappa > 1/2$.\footnote{Due to the $\kappa\mapsto1-\kappa$ symmetry, such a choice is always possible.} In this case, the amplitude parts of the two solutions are linearly independent, and so $\exp(i\gamma\ln(p)\, d)$ must be real. Since~$d$ is an integer, the choices are $\gamma \ln(p) = 0$ or~$\pi\pmod{2\pi}$.
\end{itemize}
While it would be interesting to consider solutions with nonzero~$\gamma$, we will consider only the one-parameter family of solutions with real~$\kappa$ in the sequel. The parameter $m^2_\kappa$ then attains its minimum value  for $\kappa = 1/2$. Considering only solutions of this plane-wave form, we therefore have a bound
\begin{equation}
m^2_\kappa \geq - (\sqrt{p}-1)^2,
\end{equation}
in perfect analogy with the BF bound on the mass of fields in ordinary AdS.
Note that we could also rewrite~\eqref{masses} in the form
\begin{equation}
m^2_\kappa = -(p+1) + 2\sqrt{p} \cosh\left[ \left(\kappa - \frac{1}{2} \right)\ln p \right].
\end{equation}

\subsection{Bulk reconstruction and holography}
\label{sec43}

It is clear from the definition that, when the real part of~$\kappa$ is positive, the plane wave solution~\eqref{planewave} tends to zero everywhere on the boundary, except at the point~$x$ (where it tends to infinity). So we can think of it as representing the solution to the Laplace equation (taking $\kappa=1$) in the bulk, with specified Dirichlet-type boundary conditions that look like a delta function centered at~$x$. By linearity, we can therefore reconstruct the solution to more general Dirichlet problems by superposition: if the boundary value is to be a certain function $\phi_0(x)$ on~$\partial T_p = \PL(\Q_p)$, then the required bulk harmonic function is
\begin{equation}
\phi(v) = \frac{p}{p+1}\int d\mu_0(x)\, \phi_0(x) \, \varepsilon_{1,x}(v). \label{kernel}
\end{equation}
Here $d\mu_0(x)$ is the Patterson-Sullivan measure on~$\PL(\Q_p)$. The normalization factor can be fixed by taking the boundary value to be the characteristic function of any $p$-adic open ball in the boundary.

We can perform the analogous calculation for massive fields as well, but the sense in which $\phi(v)$ will approach $\phi_0(x)$ as $x\rightarrow v$ will be more subtle (since the equation of motion will have no constant mode). Using notation from~\cite{Zabrodin}, let $\delta(a\goesto b, c\goesto d)$ be the overlap (with sign) of the two indicated oriented paths in the tree, and let
\begin{equation}
\abr{v,x} = \delta(C\goesto v, C\goesto x) + \delta(v\goesto x, C\goesto v). \label{abr}
\end{equation}
This expression makes sense for any bulk vertex~$v$; $x$ may be either a boundary or a bulk point. Note that~$\abr{z,x}$ is just the negative of the ``regularized distance'' occurring in our previous discussion.

We would like to compute the bulk solution to the massive equation of motion obtained by integrating our primitive solution~\eqref{planewave} over its boundary argument, weighted by a boundary function. As a simple choice of boundary function, pick the characteristic function of the $p$-adic open ball on the boundary consisting of points below a chosen vertex~$w$ in the tree:\footnote{The notion of ``below'' of course depends on the data of a centerpoint $C\in T_p$ having been fixed. Then any open ball that is sufficiently small in the Patterson--Sullivan measure (again defined with respect to~$C$) corresponds to a unique bulk vertex~$w$, and $C$ is above $w$ by definition. But the requirement that the sets are small presents no problem. One can see this either by noting that any open ball can be carried to any other by a M\"obius transformation, or by recalling that the expressions are linear, so that a big open set can be thought of as built out of smaller ones.}
\begin{equation}
\phi_w(v) = \int_{\partial B_w} d\mu_0(x)\, p^{\kappa\abr{v,x}}.
\end{equation} 
The integral is straightforward to calculate. There are two cases:
\begin{enumerate}[leftmargin=0.7in]
\item[$v\not\in B_w$] Here, the integrand is constant, and is just equal to $p^{\kappa\abr{v,w}}$. The measure of the set over which the integral is performed is $\mu_0(\partial B_w)=p^{-d(C,w)}$, so that the final result is
\begin{equation}
\phi_w(v) = p^{\kappa\abr{v,w}-d(C,w)}.
\end{equation}
Note that, if~$v$ moves towards the boundary along a branch of the tree, $\abr{v,w}$ differs from $-d(C,v)$ by a constant, so that the solution scales as $p^{-\kappa\, d(C,v)}$.

\item[$v\in B_w$] There are now two cases to consider: $x\in B_v$ or $x\not\in B_v$. In the first scenario, the integrand is again constant; its value is~$p^{ \kappa\, d(C,v)}$, and the measure is $\mu_0(B_v)=p^{-d(C,v)}$.

In the second scenario, the geodesic $x\goesto C$ will meet the geodesic $v\goesto C$ at a distance $h$ above $v$; by assumption, $1\leq h \leq d(v,w)$. For each value of~$h$, the integrand takes the constant value $p^{\kappa \left( d(C,v) - 2 h \right)}$, and the measure of the corresponding set is 
\begin{equation}
\mu(h) = \frac{p-1}{p} p^{-d(C,v)+h}.
\end{equation}
The factor $(p-1)/p$ enters because $p-1$ of the $p$ vertices one step below the meeting vertex correspond to meeting height $h$ (one of them is closer to~$v$).
Putting the pieces together, the result is
\begin{align}
\phi_w(v) &= p^{(\kappa-1)d(C,v)} \left( 1  + 
\frac{p-1}{p} \sum_{h=1}^{d(w,v)} \left( p^{1-2\kappa} \right) ^h \right) 
\nonumber \\
&= \left( \frac{p^{-2\kappa} - 1}{p^{1-2\kappa} - 1} \right)
 p^{ ( \kappa - 1 )  d(C,v)}
+  \frac{p-1}{p} 
\left( \frac{p^{ (2\kappa - 1) d(C,w) }}{p^{2\kappa-1}-1} \right)
p^{-\kappa \, d(C,v)} . \label{massive-Green}
\end{align}
\end{enumerate}

The reader can check that we recover the correct answer in the massless case, $\kappa\goesto 1$. Furthermore, our result is a superposition of the asymptotic behavior of the two eigenfunctions corresponding to the mass determined by our original choice of~$\kappa$. To resolve the ambiguity, we will choose $\kappa > 1/2$.

At this point, we have accumulated enough understanding of scalar fields on the tree to point out how the simplest version of holography will work: namely, classical scalar fields in a non-dynamical AdS background, neglecting backreaction and metric degrees of freedom. In the archimedean case, this version of holography was neatly formulated by Witten~\cite{Witten} in terms of a few simple key facts. Firstly, the coupling between bulk scalar fields and boundary operators must relate the asymptotics (and hence the mass) of the bulk fields to the conformal dimension of the corresponding boundary operators; massless bulk scalars should couple to marginal operators in the boundary CFT. Secondly, the crucial fact that allows the correspondence to work is the existence of a unique solution to the generalized Dirichlet problem for the bulk equations of motion with specified boundary conditions.

Luckily, as we have now shown, all of the important features of the problem persist in the $p$-adic setting, and Witten's analysis can be carried over kit and caboodle to the tree.
In particular, we make his holographic ansatz:
\begin{equation}
\left\langle \exp \int_{\PL(\Q_p)} d\mu_0\, \phi_0 \mathscr{O} \right\rangle_\text{CFT} = e^{-I_\text{bulk}[\phi]}, \label{holo-ansatz}
\end{equation}
where the bulk field~$\phi$ is the unique classical solution extending the boundary condition~$\phi_0$, and~$\mathscr{O}$ is a boundary operator to which the bulk field couples. In the massless case, where one literally has $\phi_0(x) = \lim_{v\goesto x} \phi(v)$, $\mathscr{O}$ is an exactly marginal operator in the CFT.

Given our result~\eqref{massive-Green}, it is simple to write down the correctly normalized bulk-reconstruction formula for massive fields: 
\begin{align}
\phi(v) &= \frac{p^{1-2\kappa}-1}{p^{-2\kappa}-1} \int d\mu_0(x) \phi_0(x) p^{\kappa \abr{v,x}} , \label{massive-Dirichlet} \\ 
\phi(v) &\sim p^{(\kappa -1) d(C,v)} \phi_0(x) \text{ as } v\goesto x.\nonumber
\end{align}
When the point~$v$ approaches the boundary, the exponent in the kernel becomes
\begin{equation}
\abr{v,x} = -d(C,v) + 2 \ord_p(x-y),
\end{equation}
where $y$ is any boundary point below~$v$. (To see why this is true, the reader may again find it useful to look ahead to Fig.~\ref{ryufigure}.) \eqref{massive-Dirichlet} then becomes
\begin{equation}
\phi(v)= \left( \frac{p^{1-2\kappa}-1}{p^{-2\kappa}-1}\right)p^{-\kappa\,d(C,v)}
 \int d\mu(x) \frac{\phi_0(x)}{ \abs{x-y}_p^{2\kappa}} .
\end{equation}
We can now understand why the Vladimirov derivative is a ``normal'' derivative on the boundary; it measures the rate of change in the holographic direction of the reconstructed bulk function. In particular, 
we have that
\begin{align}
 \lim_{v\goesto y} (\phi(v) - \phi(y) ) p^{\kappa d(C,v)}
&= \left( \frac{p^{1-2\kappa}-1}{p^{-2\kappa}-1}\right)
 \int d\mu(x) \frac{\phi_0(x) - \phi_0(y)}{ \abs{x-y}_p^{2\kappa}} \nonumber \\
 &= \left( \frac{p^{1-2\kappa}-1}{p^{-2\kappa}-1}\right) \partial_{(p)}^{2\kappa - 1} \phi_0(y).
\end{align}
An argument precisely akin to Zabrodin's demonstration~\cite{Zabrodin} that the bulk action may be written (upon integrating out the interior) as a boundary integral of the nonlocal Vladimirov action shows that we can write $I_\text{bulk}[\phi]$ in exactly this form. This demonstrates, exactly as in Witten's archimedean analysis, that a massive field~$\phi$ corresponds to a boundary operator of conformal dimension~$\kappa$, where~$\kappa>1/2$ is the larger of the two values that correspond to the correct bulk mass. Moreover, the boundary two-point function is proportional to~$\abs{x-y}_p^{-2\kappa}$, as expected.


\subsection{Scale dependence in bulk reconstruction of boundary modes}
\label{reconstruction}

Let us consider how the mode expansion of a boundary scalar field interacts with the reconstruction of the corresponding bulk harmonic function. We will be interested in developing the interpretation of the extra, holographic direction as a renormalization scale in our $p$-adic context. The idea that moving upward in the tree corresponds to destroying information or coarse-graining is already suggested by the identification of the cone above $\Z_p$ (or more generally any branch of the tree) with the inverse limit
\begin{equation}
\Z_p = \varprojlim \Z/p^n\Z,
\end{equation}
where the set of vertices at depth~$n$ corresponds to the elements of~$\Z/p^n\Z$, and the maps of the inverse system are the obvious quotient maps corresponding to the unique way to move upwards in the tree. A nice intuitive picture to keep in mind is that $p$-adic integers can be thought of as represented on an odometer with infinitely many $\FF_p$-valued digits extending to the left. $\Z/p^n\Z$ is then the quotient ring obtained by forgetting all but~$n$ digits, so that there is integer overflow; the maps of the inverse system just forget successively more odometer rings. Since digits farther left are smaller in the $p$-adic sense, we can think of this as doing arithmetic with finite (but increasing) precision. The parallel to the operation of coarse-graining is apparent; however, we will be able to make it more precise in what follows. 

Let's consider a boundary field that is just given by an additive character (plane wave), $\phi_0(x) \sim \exp(2\pi i \{ k x\}_p)$. 
Just as in the complex case, a plane wave in a given coordinate system won't define a solution of fixed wavelength everywhere on~$\PL(\C)$; the coordinate transformation (stereographic projection) will mean that the wavelength tends to zero as one moves away from the origin, and the function will become singular at infinity. Therefore, we should instead consider a boundary function of \textsl{wavepacket} type, that looks like a plane wave, but supported only in a neighborhood of the origin. 

A nice choice to make in the $p$-adic setting is to take the boundary function to be 
\begin{equation}
\phi_0(x) = e^{2\pi i \{ k x \}} \cdot \Theta(x, \Z_p),
\end{equation}
where~$\Theta(x,S)$ is the characteristic function of the set $S\subset \Q_p$.
The transformation~\eqref{measures} is actually trivial inside~$\Z_p$, so no distortion of the wavepacket occurs at all (unlike for a similar setup in~$\C$). Of course, we ought to  take $|k|_p>1$, so that $\{kx\}$ is not constant over the whole of~$\Z_p$. 

Given this choice of boundary function, the corresponding solution to the bulk equations of motion can be reconstructed using the integral kernel~\eqref{kernel}:
\begin{equation}
\phi(v) = \frac{p}{p+1}\int_{\Z_p} d\mu(x)\, e^{2\pi i \{ k x \} } p^{-d_C(x,v)}.
\end{equation}
Recall that $d_C(x,v)$ is the distance from~$v$ to~$x$, regularized to be zero at the centerpoint~$v=C$ of the tree. We will calculate this integral when~$v$ is inside the branch of the tree above~$\Z_p$. 

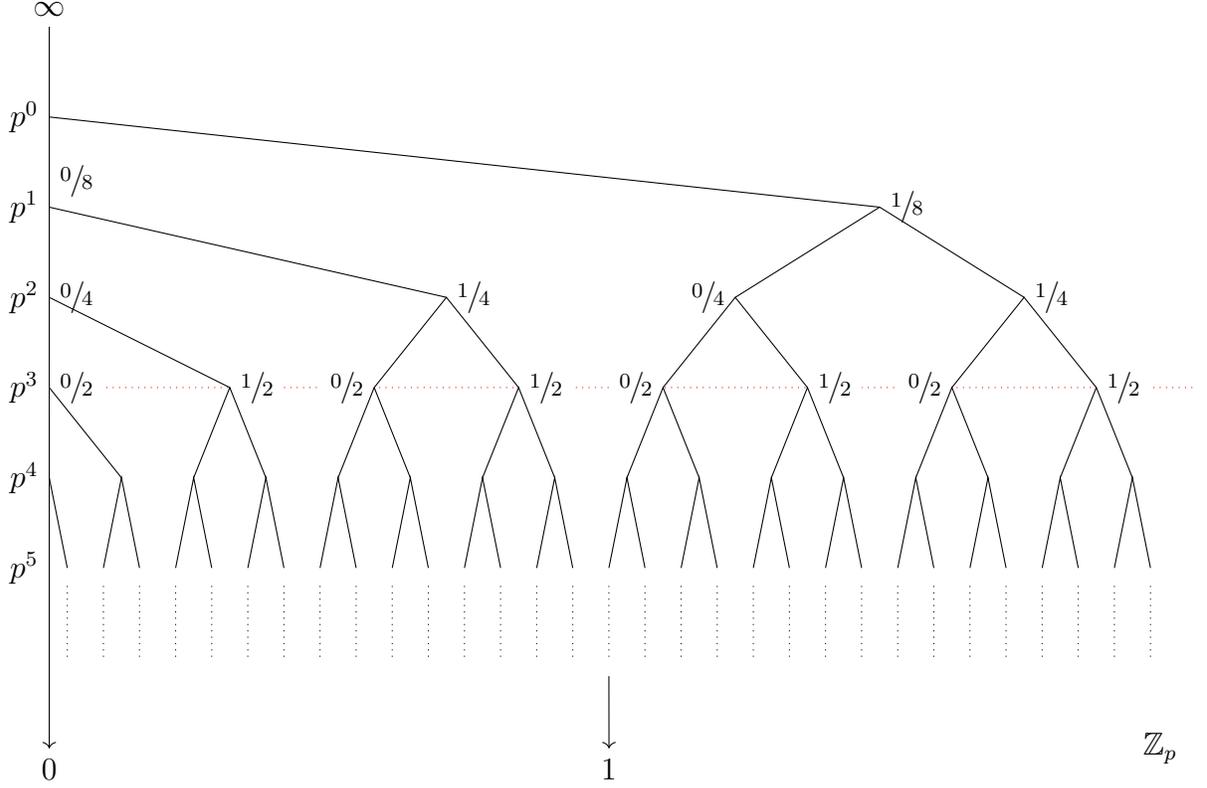
\begin{figure}[t]
\centering
\begin{tikzpicture}[scale=1.2]
\draw (0,1) node[anchor=south] (top) {$\infty$};
\draw (0,-7) node[anchor=north] (bottom) {$0$};
\path[->] (top) edge (bottom);
\draw (0,0) node[anchor=east] {$p^0$};
\draw (0,-1) node[anchor=east] {$p^{1}$};
\draw (0,-2) node[anchor=east] {$p^{2}$};
\draw (0,-3) node[anchor=east] {$p^{3}$};
\draw (0,-4) node[anchor=east] {$p^{4}$};
\draw (0,-5) node[anchor=east] {$p^{5}$};

\draw (0,-4) -- (0.2,-5);
\foreach \x in {1,2,...,15} {
	\draw (\x*0.8 - 0.2,-5) -- (\x*0.8,-4) -- (\x*0.8 + 0.2,-5);
	};

\foreach \x in {0,1,2,...,30} {
	\draw[dotted] (\x*0.4 + 0.2,-5.2) -- (\x*0.4 + 0.2,-6);
	};

\draw (15*0.4 + 0.2,-7) node[anchor=north] (1) {$1$};
\path[->] (15*0.4 + 0.2,-6.2)  edge (1);

\draw (12,-7) node[anchor=west] {$\Z_p$};

\draw (0,-2) -- (2.0,-3);
\foreach \x in {1,2,3} {
	\draw (\x*3.2+0.4,-3) -- (\x*3.2 + 1.2,-2) -- (\x*3.2 + 2.0,-3);
	};

\draw[red, dotted]
(0,-3) node[anchor=west,black,fill=white] {$\sfrac{0}{2}$} --
(2,-3) node[anchor=west,black,fill=white] {$\sfrac{1}{2}$} --
(3.6,-3) node[anchor=east,black,fill=white] {$\sfrac{0}{2}$} --
(5.2,-3) node[anchor=west,black,fill=white] {$\sfrac{1}{2}$} --
(6.8,-3) node[anchor=east,black,fill=white] {$\sfrac{0}{2}$} --
(8.4,-3) node[anchor=west,black,fill=white] {$\sfrac{1}{2}$} --
(10.0,-3) node[anchor=east,black,fill=white] {$\sfrac{0}{2}$} --
(11.6,-3) node[anchor=west,black,fill=white] {$\sfrac{1}{2}$} -- (12.7,-3);

\draw (0,-3) -- (0.8,-4);
\foreach \x in {1,2,3,...,7} {
	\draw (\x*1.6,-4) -- (\x*1.6 + 0.4,-3) -- (\x*1.6+ 0.8,-4);
	};

\draw (0,-1) -- (4.4,-2);
\draw  (7.6,-2) -- (9.2,-1) -- (10.8,-2);

\draw (0,-2) node[anchor=west] {$\sfrac{0}{4}$};
\draw (4.4,-2) node[anchor=west] {$\sfrac{1}{4}$};
\draw (7.6,-2) node[anchor=east] {$\sfrac{0}{4}$};
\draw (10.8,-2) node[anchor=west] {$\sfrac{1}{4}$};

\draw (0,0) -- (9.2,-1);

\draw (0,-1) node[anchor=south west] {$\sfrac{0}{8}$};
\draw (9.2,-1) node[anchor=west] {$\sfrac{1}{8}$};

\end{tikzpicture}
\caption{A drawing of~$\Z_p$ ($p=2$ for simplicity). Here $k=p^{-4}$. The marked fractions at vertices indicate contributions to $\{kx\}$, which are summed along the geodesic ending at~$x$.}
\label{calc}
\end{figure}

\begin{proposition}
Let $v$ be a vertex in the branch above~$\Z_p$, at a depth~$\ell$ (i.e., since $v\in\Z_p$, distance from the centerpoint) such that $0\leq \ell < -\ord_p(k)-2$. Then the reconstructed bulk function $\phi(v)$ is zero. 
\begin{proof}
The claim relies on the simple fact that the sum of all $p$-th roots of unity is zero. Since $v$ is above the red line in Fig.~\eqref{calc} (at depth equal to $-\ord_p(k)-1$), both terms in the integrand are locally constant below the line, and the integral may be evaluated as a sum along the vertices at the height of the red line. Furthermore, the factor $p^{-d_C(v,x)}$ is constant for each of the $p$ vertices on the line that descend from the same ancestor. Since the measure of each branch is equal, the integral is proportional to the sum of all $p$-th roots of unity, and hence to zero. Notice that this also demonstrates that the reconstructed bulk function is zero everywhere \textsl{outside} $\Z_p$: it is zero at the central vertex, and zero on the boundary of the open ball complementary to~$\Z_p$. 
\end{proof}
\end{proposition}

Even without calculating the explicit form of the bulk function for vertices below the screening height, this simple argument already allows us to make our physical point: in $p$-adic holography, the qualitative features of ordinary AdS/CFT persist in a setting where the bulk geometry is discrete, and in some cases are even sharpened. For instance, we have shown explicitly that modes for which $|k|_p$ is large (i.e. the short-wavelength behavior of the boundary conditions) must drop out of the reconstructed bulk field, making exactly zero contribution to it above a height in the tree precisely determined by~$|k|_p$. The usual intuition that moving into the bulk along the holographic direction corresponds to integrating out UV modes is thus neatly confirmed. 

The explicit form of the reconstructed bulk function at vertices below the screening height is easy to calculate, but less central to our discussion; we leave the computation as an exercise for the reader.

\subsection{The possibility of higher-spin fields}
\label{sec:spin}

We now wish to propose an analogue of higher-spin fields that could be defined in the $p$-adic case. While we will motivate our proposal here, we do not investigate any properties of $p$-adic CFT with fields other than scalars. We will return to this question in future work. 

We proceed by analogy with two-dimensional CFT, in which the conformal dimension and spin together describe a character of the multiplicative group $\C^\times$:
\begin{equation}
\phi(re^{i\theta}\cdot z) = r^\Delta e^{is\theta} \phi(z). 
\end{equation}
The group $\C^\times \simeq \R_{>0}^\times \times \U(1)$; the conformal dimension determines a character of the first factor, and the spin a character of the second, which can be thought of as scale transformations and rotations of the coordinate respectively. The existence of the logarithm function means that we can think of the multiplicative group $\R_+$ as isomorphic to the additive group~$\R$.

The structure of the group of units of any local field is understood (see~\cite{Neukirch} for details). In particular, for the field~$\Q_p$, the result is that 
\begin{equation}
\Q_p^\times\simeq p^\Z \times \FF_p^\times \times U^{(1)}, \label{multiplicative}
\end{equation}
where $U^{(1)}$ is the group of ``principal units'' of the form $1+p\cdot a$, with $a\in\Z_p$. 
This decomposition just reflects the structure of the $p$-adic decimal expansion: since the $p$-adic norm 
is multiplicative, any number $x \neq 0$ can be written in the form
\begin{equation}
x = p^{\ord_p(x)} \left( x_0 + x_1 p + \cdots \right),
\end{equation}
where $x_0\neq 0$ (so that $x_0 \in \FF_p^\times \simeq C_{p-1}$) and the other $x_i$ may be any digits chosen from~$\FF_p$. Dividing through by~$x_0$, one gets 
\begin{equation}
x = p^{\ord_p(x)} \cdot x_0 \left( 1 + \sum_{i\geq 1} \tilde{x}_i p^i \right),
\end{equation}
where $\tilde{x}_i = x_i/x_0$, and the factor in parentheses is a principal unit. 
A character of~$\Q_p^\times$ is therefore a triple of characters, one for each factor in~\eqref{multiplicative}. The first factor, as in the normal case, corresponds to the scaling dimension of the field; the last two factors are therefore analogous to the spin. Obviously, the second factor corresponds to a $\Z/(p-1)\Z$ phase. 
It is also known~\cite{Neukirch} that the set of characters of~$U^{(1)}$ is countable and discrete.

In fact, we can naively understand a broader class of the characters of~$\Z_p^\times = \FF_p^\times \times U^{(1)}$. Recall the description of $\Z_p$ as the inverse limit of its finite truncations:
\begin{equation}
\Z_p = \varprojlim \Z/p^n\Z.
\end{equation}
Since this is an inverse limit of rings, there are projection maps between the respective multiplicative groups:
\begin{equation}
\Z_p^\times \goesto (\Z/p^n\Z)^\times \simeq C_{p^{n-1}(p-1)}.
\end{equation}
Therefore, any multiplicative character of a cyclic group $C_{p^{n-1}(p-1)}$ (i.e., any finite root of unity of order $p^{n-1}(p-1)$, for arbitrary~$n$) will give a character of~$(\Z/p^n\Z)^\times$, which will in turn pull back to define a multiplicative character of the spin part of~$\Q_p^\times$. Spin in the $p$-adic case is therefore both similar to and interestingly different from ordinary two-dimensional CFT.

\section{Entanglement entropy}
\label{EE}



The entanglement entropy in quantum field theories is a notoriously difficult and subtle quantity to compute, and much effort has been expended in developing a toolbox of techniques that provide exact results. One of the first systems in which the computation became tractable was two-dimensional conformal field theory, and in particular the theory of free bosons. Since we are primarily considering the free boson in our discussion, one might hope that the same techniques can be applied in the $p$-adic case. While we believe that this is the case, and plan to give a full calculation of the entanglement entropy in future work, there are subtleties that arise in each technique and prevent it from being used naively. We will demonstrate these techniques, illustrate the subtle issues that arise, and justify to some extent our conjecture for the entanglement entropy in what follows. The discussion in this section should all be understood as speculative; while we include it due to tantalizing analogies with ordinary AdS/CFT and in the hopes of encouraging further investigation, all currently remains unsupported by rigorous proof.

As in the real case, we expect the entanglement entropy to have UV divergences. These are normally thought of as localized to the ``boundary'' of the region under consideration. Care must be used in defining what we mean by interval and boundary; the $p$-adic numbers have no ordering, and every element of an open set is equally (or equally not) a boundary element. Whenever possible, we must think in terms of open sets. Over the reals, the open sets are intervals with measure or length given by the norm of the separation distance of the endpoints; as the reader will recall, $p$-adic open sets are perhaps best visualized using the Bruhat--Tits tree. Once a center $C$ of the tree is picked, we can pick any other vertex $v$ and consider the cone of points below $v$ extending out towards the boundary, which is an open neighborhood in $\PL(\mathbb{Q}_p)$. A perhaps surprising fact which follows from the definition of the $p$-adic norm $|x-y|_p$  ($x,y \in \PL(\mathbb{Q}_p)$) is that it is related to the height of the cone required to connect $x$ to $y$ (see Fig.~\ref{ryufigure}).

We can take the boundary region $V$ to be the open set defined by the cone below a chosen bulk vertex $v$. Following standard arguments, say of~\cite{Srednicki, CasiniHuerta}, the total Hilbert space on $\mathbb{Q}_p$ splits into Hilbert spaces on $V$ and its complement, $\mathcal{H} = \mathcal{H}_V \otimes \mathcal{H}_{-V}$. The entanglement entropy is defined by $S(V) = -\Tr (\rho_V \log \rho_V)$ and by construction satisfies $S(V) = S(-V)$. As there are an infinite number of points $x_i \in V \in \PL(\mathbb{Q}_p)$, there is an unbounded number of local degrees of freedom $\phi(x_i)$ (as is typical of quantum field theories). In the continuum case, this implies logarithmic divergences arising from entanglement between modes in $V$ and those in $-V$. We expect the same to be true in the $p$-adic case.

In the works of Cardy and Calabrese~\cite{CaCa1, CaCa2}, the entanglement entropy for intervals in $1+1$-dimensional conformal field theories are explicitly calculated. The $p$-adic field theories considered here are exactly analogous to the two-dimensional free boson; in both, the scalar $\phi(x)$ has conformal dimension zero and (as we have shown above) a logarithmically divergent propagator. We wish to understand how much of their calculation can be duplicated in the $p$-adic case. These authors generally follow a series of steps beginning with the replica trick, which is the observation that $n$ powers of the reduced density matrix $\rho_V$ can be computed by evaluating the partition function on a Riemann surface obtained by gluing $n$ copies of the theory together along the interval $V$. The entanglement entropy follows from analytic continuation of these results in~$n$, followed by the limit $n \rightarrow 1$, according to the formula
\begin{equation}
\Tr(\rho_V^n) = \frac{Z_n(V)}{Z_1^n}, \hspace{3mm} S_V = - \lim_{n \rightarrow 1}\frac{\partial}{\partial n}\frac{Z_n(V)}{Z_1^n},
\end{equation}
where $Z_n(V)$ is the $n$-sheeted partition function and $Z_1$ is the partition function of $1$ sheet with no gluing, which is required for normalization.

In $1+1$ dimensions, the $n$-sheeted partition function can be viewed as a Riemann surface, and the holomorphic properties of this surface make the calculation tractable. In particular, if the interval has the boundary points $x$ and $y$, the complicated world sheet topology can be mapped to the target space by defining multi-valued \textit{twist fields} $\Phi_n(x), \Phi_n(y)$ on the plane whose boundary conditions implement the $n$ sheeted surface.  One can find that $\Tr(\rho_V^n)$ behaves exactly like the $n^{th}$ power of a two point function of the twist fields, once their conformal dimension has been determined using Ward identities:
\begin{equation}
\Tr(\rho_V^n) \sim \langle 0 |\Phi_n(x) \Phi_n(y) |0 \rangle^n \sim \left| {x-y} \right|^{-\frac{c}{6}(n-\frac{1}{n})},
\end{equation}
where $c$ is the central charge and $\epsilon$ is a normalization constant from $Z_1$. When $n =1$ exactly, the twist fields have scaling dimension 0 and the above correlator no longer makes sense. Instead, taking the limit as $n \rightarrow 1$, the linear term is $-n \frac{c}{3} \ln\left ( \frac{x-y}{\epsilon}\right )$. Taking the derivative gives the famous universal formula for the entanglement entropy~\cite{Holzhey}.


The difficulty in performing the same calculation over the $p$-adics consists in fixing the dimensions of the twist operators. It seems plausible that these operators can be defined similarly to the normal case; after all, all that they do is implement certain boundary conditions at branch points on the fields in a theory of $n$ free bosons. However, the usual arguments that fix their dimension rely on the existence of a uniformizing transformation $z\mapsto z^n$ that describes the relevant $n$-sheeted branched cover of~$\PL$ by~$\PL$; the Schwarzian of this holomorphic (but not M\"obius) transformation then appears as the conformal dimension. The argument using the OPE with the stress tensor is identical in content. Both cases rely on the existence of holomorphic (but not fractional-linear) transformations, and a measure---the Schwarzian or conformal anomaly---of their ``failure'' to be M\"obius. 

In the $p$-adic case, this is related to the question of local conformal transformations; it has been suggested~\cite{Melzer} that no such symmetries exist. Moreover, since~$\Q_p$ is not algebraically closed, a transformation like $z\mapsto z^n$ need not even be onto. Nevertheless, we can still define the twist operators, and we suppose that they transform as primaries with some conformal dimensions $\Delta_n$. Their two-point function then gives the density matrix. This function is:
\begin{equation}
 \langle 0 |\Phi^{(p)}_n(x) \Phi^{(p)}_n(y) |0 \rangle^n \sim \left| {x-y} \right|_p^{-2n \Delta_n},
\end{equation}
where $\Delta_n$ are the model-dependent (and unknown) conformal dimensions. Inserting this ansatz into $ - \lim_{n \rightarrow 1}\frac{\partial}{\partial n}\Tr(\rho_V^n)$ and taking the limit $n \rightarrow 1$, $\Delta_n \rightarrow 0$ gives:
\begin{equation}
\left (2 n \frac{\partial \Delta_n}{\partial n}\big |_{n=0}  \right )\ln \left |\frac{x-y}{\epsilon} \right |_p .
\end{equation}
While this is not a proof, it provides some evidence for the expected logarithmic scaling of the entropy. We expect that the dimensions $\Delta_n \goesto 0$ as $n\goesto 1$, since of course the twist operator on one sheet is just the identity. If we could fix the conformal dimension without using the conformal anomaly, this calculation would fix not only the logarithmic form of the answer, but also the coefficient that plays the role of the central charge. It may be possible to do this by examining the path integral with twist-operator insertions directly.

A possible way around this difficulty might be to consider a harder problem first: to think about $N>1$ intervals rather than one. The genus of the Riemann surface that appears in the replica trick is then $g=(n-1)(N-1)$; thus, for one interval, we are considering a branched cover of~$\PL$ over itself, and the conformal anomaly is a necessary ingredient. However, one might hope that for two intervals, we can simply compute the partition function on a series of higher-genus Riemann surfaces (which is understood in the $p$-adic case), and take the limit as the genus approaches zero. Discussions of the entanglement entropy in terms of Schottky uniformization---which therefore appear tailored to our needs---have appeared in the literature~\cite{Faulkner}. 

Two difficulties appear in this case: the first is matching the moduli of the Riemann surface in question to the lengths of the intervals; the second is more subtle, and reflects the fact that, over the $p$-adics, not every branched cover of the $p$-adic projective line is a Mumford curve~\cite{Bradley}. We believe that one of the strategies outlined here will succeed in producing a rigorous computation of the entanglement entropy, but we must relegate that computation to future work.


\subsection{Ryu-Takayanagi formula}
\label{sec:RT}

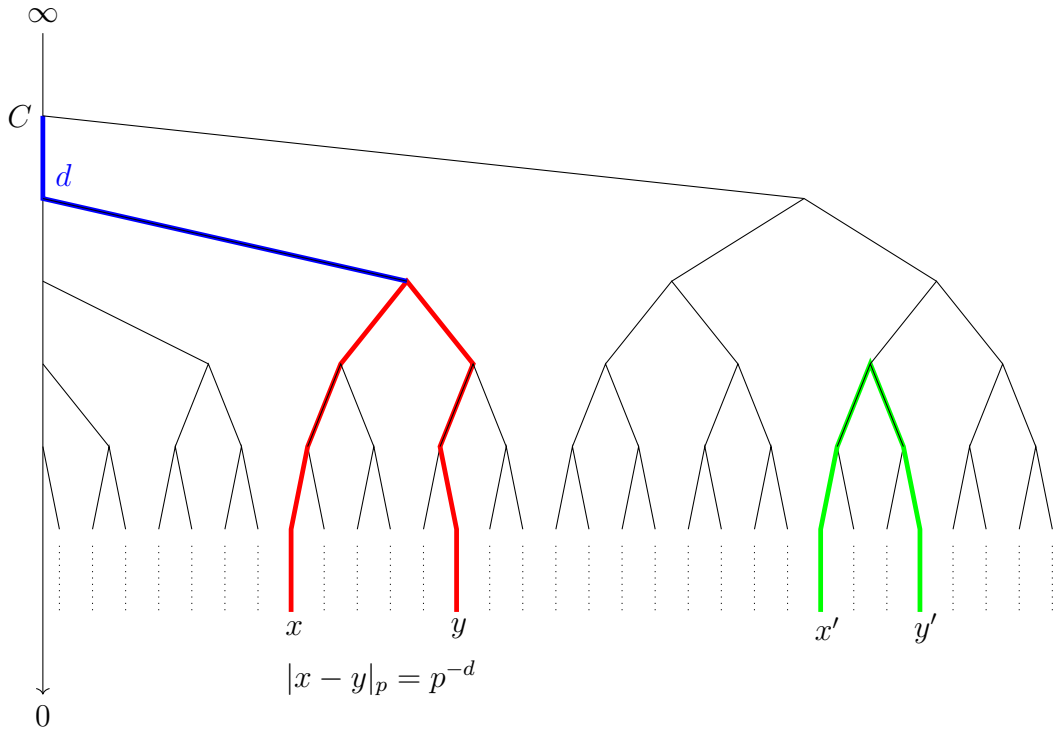
\begin{figure}
\centering
\begin{tikzpicture}[scale=1.1]
\draw (0,1) node[anchor=south] (top) {$\infty$};
\draw (0,-7) node[anchor=north] (bottom) {$0$};
\path[->] (top) edge (bottom);
\draw (0,0) node[anchor=east] {$C$};
\draw (0,-1) node[anchor=east] {};
\draw (0,-2) node[anchor=east] {};
\draw (0,-3) node[anchor=east] {};
\draw (0,-4) node[anchor=east] {};
\draw (0,-5) node[anchor=east] {};

\draw (0,-4) -- (0.2,-5);
\foreach \x in {1,2,...,15} {
	\draw (\x*0.8 - 0.2,-5) -- (\x*0.8,-4) -- (\x*0.8 + 0.2,-5);
	};

\foreach \x in {0,1,2,...,30} {
	\draw[dotted] (\x*0.4 + 0.2,-5.2) -- (\x*0.4 + 0.2,-6);
	};

\draw (2.8,-6.2) node[anchor=west] {$x$}; 
\draw (4.8,-6.2) node[anchor=west] {$y$}; 
\draw (2.8,-6.8) node[anchor=west] {$|x-y|_p = p^{-d}$}; 

\draw (9.2,-6.2) node[anchor=west] {$x'$}; 
\draw (10.4,-6.2) node[anchor=west] {$y'$}; 
\draw[green, line width=1.8pt]
(9.4,-6.0) node[anchor=west] {} --
(9.4,-5) node[anchor=west] {} --
(9.6,-4) node[anchor=east] {} --
(10,-3) node[anchor=west] {} --
(10.4,-4) node[anchor=east] {} --
(10.6, -5)node[anchor=east]{} -- (10.6, -6);

\draw (0,-2) -- (2.0,-3);
\foreach \x in {1,2,3} {
	\draw (\x*3.2+0.4,-3) -- (\x*3.2 + 1.2,-2) -- (\x*3.2 + 2.0,-3);
	};

\draw[red, line width=1.8pt]
(3,-6.0) node[anchor=west] {} --
(3,-5) node[anchor=west] {} --
(3.2,-4) node[anchor=east] {} --
(3.6,-3) node[anchor=west] {} --
(4.4,-2) node[anchor=east] {} --
(5.2,-3) node[anchor=west] {} --
(4.8,-4) node[anchor=east] {} --
(5,-5) node[anchor=west,black,fill=white] {} -- (5,-6);

\draw[blue, line width=1.8pt]
(4.4,-2) node[anchor=east] {} --
(0,-1) node[anchor=south west] {$d$} --
(0,0);

\draw (0,-3) -- (0.8,-4);
\foreach \x in {1,2,3,...,7} {
	\draw (\x*1.6,-4) -- (\x*1.6 + 0.4,-3) -- (\x*1.6+ 0.8,-4);
	};

\draw (0,-1) -- (4.4,-2);
\draw  (7.6,-2) -- (9.2,-1) -- (10.8,-2);

\draw (0,-2) node[anchor=west] {};
\draw (4.4,-2) node[anchor=west] {};
\draw (7.6,-2) node[anchor=east] {};
\draw (10.8,-2) node[anchor=west] {};

\draw (0,0) -- (9.2,-1);

\draw (0,-1) node[anchor=south west] {};
\draw (9.2,-1) node[anchor=west] {}; 
\end{tikzpicture}

\caption{Boundary anchored geodesics in $T_p$ have a natural interpretation in terms of the $p$-adic norm. Once the arbitrary position of the center $C$ is fixed, the norm of open sets in $\mathbb{Q}_p$ is given by $p^{-d}$, where $d$ is the integer number of steps from $C$ required before the path to the endpoints splits. In this example, $|x-y|_p$ is described by the red geodesic and the value is $p^{-2}$. The set corresponding to the green geodesic has a smaller norm by a factor of $p$ because the vertex is 1 step further down the tree. 
As in the case of real AdS, the length of the geodesic is formally infinite. One can truncate the tree at a fixed distance from the center, then take the limit as this cutoff is removed. It should be apparent that the (formally infinite) red geodesic is longer than the green one by two steps. Up to constant factors, the length of any boundary-anchored geodesic is a universal infinite term minus $d$. This explains the logarithmically divergent scaling of geodesic length with $p$-adic norm.}
\label{ryufigure}
\end{figure}

Let us take as given the conjecture from the previous section that the entanglement entropy of a region in the boundary CFT should be computed as the logarithm of its $p$-adic size. We take our interval to be the smallest $p$-adic open ball which contains points $x$ and $y$. This interval has size $|x-y|_p$. To understand the Ryu-Takayanagi formula, it remains to compute the length of the unique geodesic connecting $x$ to $y$. The tree geometry for this setup is depicted in Fig.~\ref{ryufigure}. Since there are an infinite number of steps required to reach the boundary, the geodesic length is formally infinite, just as in the real case. We regulate this by cutting off the tree at some finite tree distance $a$ from the center $C$, which can be thought of as~$\ord_p(\epsilon)$ for some small $p$-adic number $\epsilon$. We will then take this minimum number $\epsilon \rightarrow 0$ ($p$-adically). This limit will push the cutoff in the tree to infinite distance from $C$.

An $\SL(2, \mathbb{Q}_p)$ transformation can always be used to move the points $x$ and $y$ to the $\mathbb{Z}_p$ part of the tree first to simplify the argument. Then introducing the distance cutoff $a$ effectively truncates the decimal expansions of $x$ and $y$ to the first $a$ decimal places. In the case where $|x-y|_p =1$, the geodesic connecting the two points passes through $C$ and has length $2a$. If $|x-y|_p < 1$, the geodesic is shorter by a factor of $2d$, where $|x-y|_p = p^{-d}$. Roughly speaking, as can be seen in Fig.~\ref{ryufigure}, smaller boundary regions are subtended by shorter geodesics in the tree. 

We see that the cutoff-dependent distance is
\begin{equation}
d(x,y)_a = 2a + \frac{2}{\ln p}\ln|x-y|_p .
\end{equation}

We would like to take $a \rightarrow \infty$. Up to the factor of $\ln p$, we can define $a$ to be the logarithm of a $p$-adic cutoff $\epsilon$ such that $a \rightarrow \infty$ as $\epsilon \rightarrow 0$. Using this definition, we find the length of a boundary-anchored geodesic to be
\begin{equation}
d(x,y) = \lim_{\epsilon \rightarrow 0}  \hspace{.5 mm} \frac{2}{\ln p}\ln \left |\frac{x-y}{\epsilon} \right |_p .
\end{equation}
Up to the overall factor in front (which presumably depended on our choice of the length of each leg of the tree), we see the geodesic length is logarithmically divergent in interval size.

\label{ryutakayanagi}

\subsection{An adelic formula for entanglement?}
We have argued that the general form of entanglement entropy scaling for the boundary theory is dual to a geodesic length in the bulk. At the present time we lack a $p$-adic notion of central charge $c$ or theory dependent quantity which counts boundary degrees of freedom. Nevertheless, we claim the general form is
\begin{equation}
S_p(x-y) = c_p \ln \left |\frac{x-y}{\epsilon_p} \right |_p .
\end{equation}

We now wish to speculate about the possibility of an \textit{adelic} formula for the entanglement entropies.

In the study of $p$-adic numbers, there exists a surprising formula which relates the various $p$-adic valuations of a rational number to its real norm. This is a different form of the fundamental theorem of arithmetic, and is sometimes known as an adelic formula:
\begin{equation}
\prod_{p}^{\infty} |x|_p = 1.
\end{equation}
Here $x \in \mathbb{Q}$ and the product is taken over all primes. The ``prime at $\infty$'' corresponds to the usual archimedean norm $|x|_{\infty} = |x|$. This equality follows by considering the unique prime factorization of $x$ into a product of prime powers. When $x$ contains a factor $p^n$, then $|x|_p = p^{-n}$. This means the infinite product over primes is well defined because only finitely many terms are not equal to 1. In fact, the product over the finite primes gives exactly the inverse of the real norm $|x|$. Therefore the product over all finite places and the infinite place is unity. 

We now wish to again recall the familiar formula for the universal entanglement entropy of an interval in a 1+1 dimensional conformal field theory; written suggestively in the ``prime at infinity'' notation:
\begin{equation}
S_{\infty}(x-y) = \frac{c}{3} \ln \left |\frac{x-y}{\epsilon_{\infty}} \right |_{\infty} .
\end{equation}
One might hope through a better understanding of $p$-adic conformal field theory or the holographic dual, the value of the proportionality constant or the $p$-adic central charge might be determined. In the (perhaps unlikely) event that the central charges of the $p$-adic theory agree with the real case, then using the adelic formula for the interval, we propose:
\begin{align}
\text{If } c_p =c/3 & \text{ for all } p\text{ and $x-y \in \mathbb{Q}$, then} \nonumber \\
&\sum_{p}^{\infty} S_p(x-y) \propto \ln \left (\prod_{p}^{\infty} \left |\frac{x-y}{\epsilon_p} \right |_p  \right) = 0 .
\end{align}
One might be suspicious about this formula; each of the entropies $S_p$ are formally divergent. Additionally, since these quantities are entropies they are expected to be positive. Therefore care must be taken in interpreting the above. 

One possible resolution is the erroneous application of the adelic formula to the cutoffs $\epsilon_p$. In computing $S_p$ holographically, we assumed $|\epsilon|_p \rightarrow 0$. However, if $|\epsilon_p|_p \rightarrow 0$ in one norm, it is not generally true that $|\epsilon_p|_{p'} \rightarrow 0$ for another choice $p'$. Therefore, we require a numerically different cutoff parameter $\epsilon_p$ for each system over $p$. As all these parameters are taken to $0$ in their respective norms, the corresponding entanglement entropies diverge. 

Understanding that the cutoffs $\epsilon_p$ do not cancel on the left or right hand sides, we are left with the divergent pieces of the entropy being equal on both sides. However, if we vary the length of $|x-y|_{\infty}$ in the real physical system, we see that the entropy difference associated with this interval is distributed across the $S_p$'s such that the sum is zero. Put another way, varying the real interval length will cause some values of $|x-y|_p$ for different $p$ to increase and others to decrease. This causes some $S_p$ to increase and others to decrease such that the total change of entropy over all finite and infinite places is 0. 

We will leave it to future work to try to derive or understand this relation further.

\section{$p$-adic bulk geometry: Schottky uni\-formi\-za\-tion and non-archi\-me\-dean black holes}


\subsection{Holography for Euclidean higher-genus black holes}
\label{matilde-tree}

The first explicit form of AdS/CFT correspondence for the asymptotically $\text{AdS}_3$ higher genus black holes,
in the Euclidean signature, was obtained in \cite{ManMar}, where the computation of the Arakelov Green function
of \cite{Man91} is shown to be a form of the holographic correspondence for these black holes, where the
two-point correlation function for a field theory on the conformal boundary $X_\Gamma$ is written in terms
of gravity in the bulk $\cH_\Gamma$, as a combination of lengths of geodesics.

At the heart of Manin's holographic formula lies a simple identity relating conformal geometry
on $\bP^1(\bC)$ and hyperbolic geometry on $\bH^3$, namely the fact that the cross ratio of four points
on the boundary $\bP^1(\bC)$ can be written as the length of an arc of geodesic in the bulk $\bH^3$.
More precisely, consider the two point correlation function $g(A,B)$ on $\bP^1(\bC)$.
This is defined by considering, for a divisor $A=\sum_x m_x\, x$, the Green function of the Laplacian
$$ \partial \bar\partial g_A = \pi i ( \deg(A) d\mu -\delta_A ), $$
with $\delta_A$ the delta current associated to the divisor, $\delta_A(\varphi)=\sum_x m_x \varphi(x)$,
and $d\mu$ a positive real-analytic 2-form. The Green function $g_A$ has the property
that $g_A - m_x \log |z|$ is real analytic for $z$ a local coordinate near $x$, and is normalized
by $\int g_A d\mu=0$.  For two divisors $A,B$, with $A$ as above and $B=\sum_y n_y\, y$ 
the two point function is given by $g(A,B)=\sum_y n_y g_A(y)$. For degree zero divisors it
is independent of the form $d\mu$ and is a conformal invariant. If $w_A$ is a meromorphic
function on $\bP^1(\bC)$ with $Div(w_A)=A$, and $C_B$ is a $1$-chain with boundary 
$B$, the two point function satisfies
$$ g(A,B) = Re \int_{C_B} \frac{dw_A}{w_A}. $$
For $(a,b,c,d)$ a quadruple of points in $\bP^1(\bC)$, the cross ratio 
$\langle a,b,c,d \rangle$ satisfies 
$$ \langle a,b,c,d \rangle = \frac{w_{(a)-(b)}(c)}{w_{(a)-(b)}(d)} , $$
where $(a)-(b)$ is the degree zero divisor on $\bP^1(\bC)$ determined by the points $a,b$,
and the two point function is
$$ g((a)-(b),(c)-(d)) = \log \frac{| w_{(a)-(b)}(c) |}{| w_{(a)-(b)}(d) |}. $$
Given two points $a,b$ in $\bP^1(\bC)$, let $\ell_{\{ a,b \}}$ denote the unique geodesics 
in $\bH^3$ with endpoints $a$ and $b$. Also given a geodesic $\ell$ in $\bH^3$ and a point 
$c\in \bP^1(\bC)$ we write $c * \ell$ for the point of intersection between $\ell$ and the unique
geodesic with an endpoint at $c$ and intersecting $\ell$ orthogonally. We also write $\lambda(x,y)$
for the oriented distance of the geodesic arc in $\bH^3$ connecting two given points $x,y$ 
on an oriented geodesic. Then the basic
holographic formula identifies the two point function with the geodesic length
$$ g((a)-(b),(c)-(d)) = - \lambda(a *\ell_{\{c,d\}}, b* \ell_{\{c,d\}}). $$
One can also express the argument of the cross ratio in terms of angles between bulk geodesics (see \cite{Man91}, \cite{ManMar}).
This basic formula relating the two point correlation
function on the boundary to the geodesic lengths in the bulk is adapted to
the higher genus cases by a suitable procedure of averaging over the action
of the group that provides an explicit construction of a basis of meromorphic
differentials on the Riemann surface $X_\Gamma$ in terms of cross ratios on $\bP^1(\bC)$. 
A basis of holomorphic differentials on $X_\Gamma$, with 
$$ \int_{A_k} \omega_{\gamma_j} = 2\pi i \delta_{jk}, \ \ \ \ \  \int_{B_k} \omega_{\gamma_j} = \tau_{jk} $$
the period matrix, is given by
$$ \omega_{\gamma_i}  = \sum_{h \in S(\gamma_i)} d_z \log \langle z^+_h, z^-_h, z, z_0 \rangle, $$
for $z,z_0\in \Omega_\Gamma$, with $S(\gamma)$ the conjugacy
class of $\gamma$ in $\Gamma$. The series converges absolutely when $\dim_H(\Lambda_\Gamma)<1$.
Meromorphic differentials associated to a divisor $A=(a)-(b)$ are similarly obtained as averages
over the group action
$$ \nu_{(a)-(b)}= \sum_{\gamma \in \Gamma} d_z \log\langle a,b,\gamma z, \gamma z_0\rangle $$
and the Green function is computed as a combination $\nu_{(a)-(b)} - \sum_j X_j(a,b) \omega_{\gamma_j}$
with the coefficients $X_j(a,b)$ so that the $B_k$-periods vanish. Since in the resulting formula
each crossed ratio term is expressible in terms of the length of an arc of geodesic in the bulk,
the entire Green function is expressible in terms of gravity in the bulk space. 
We refer the reader to~\S\S2.3, 2.4, and~2.5 of \cite{ManMar} and to \cite{Mar} 
for a more detailed discussion and the resulting explicit formula of the holographic
correspondence for arbitrary genus.

\subsection{Holography on $p$-adic higher genus black holes}
\label{sec:Drinfeld}

\begin{figure}[t]
\centering
\begin{tikzpicture}[scale=1.5]
\tikzstyle{vertex}=[draw,circle];
\tikzstyle{ver2}=[draw,scale=0.6,circle];
\tikzstyle{ver3}=[];

\draw (0*60:1) node[vertex] (a0) {};
\draw (1*60:1) node[vertex] (a1) {};
\draw (2*60:1) node[vertex] (a2) {};
\draw (3*60:1) node[vertex] (a3) {};
\draw (4*60:1) node[vertex] (a4) {};
\draw (5*60:1) node[vertex] (a5) {};

\draw (a0) + (0*60 + 30:1) node[ver2] (b0) {};
\draw (a0) + (0*60 - 30:1) node[ver2] (c0) {};
\draw (a1) + (1*60 + 30:1) node[ver2] (b1) {};
\draw (a1) + (1*60 - 30:1) node[ver2] (c1) {};
\draw (a2) + (2*60 + 30:1) node[ver2] (b2) {};
\draw (a2) + (2*60 - 30:1) node[ver2] (c2) {};
\draw (a3) + (3*60 + 30:1) node[ver2] (b3) {};
\draw (a3) + (3*60 - 30:1) node[ver2] (c3) {};
\draw (a4) + (4*60 + 30:1) node[ver2] (b4) {};
\draw (a4) + (4*60 - 30:1) node[ver2] (c4) {};
\draw (a5) + (5*60 + 30:1) node[ver2] (b5) {};
\draw (a5) + (5*60 - 30:1) node[ver2] (c5) {};

\foreach \x in {-1,0,1} {
	\foreach \n in {0, 1, 2, 3, 4, 5} {
		\draw (b\n) + (\n*60 + 20 + \x*20:1) node[ver3] (d\n\x) {};
		\draw (c\n) + (\n*60 - 20 + \x*20:1) node[ver3] (e\n\x) {};
		\draw (b\n) + (\n*60 + 20 + \x*20:1.5) node[ver3] (D\n\x) {};
		\draw (c\n) + (\n*60 - 20 + \x*20:1.5) node[ver3] (E\n\x) {};
		\draw[thick] (d\n\x) -- (b\n);
		\draw[thick] (e\n\x) -- (c\n);
		\draw[very thick, loosely dotted] (d\n\x) -- (D\n\x) ;
		\draw[very thick, loosely dotted] (e\n\x) -- (E\n\x) ;
	};
};

\draw[thick] (b0) -- (a0) -- (c0);
\draw[thick] (b1) -- (a1) -- (c1);
\draw[thick] (b2) -- (a2) -- (c2);
\draw[thick] (b3) -- (a3) -- (c3);
\draw[thick] (b4) -- (a4) -- (c4);
\draw[thick] (b5) -- (a5) -- (c5);

\draw[thick] (a0) -- (a1) -- (a2) -- (a3) -- (a4) -- (a5) -- (a0);

\draw[very thick, loosely dotted] (0,0) circle (3.5);

\end{tikzpicture}
\caption{The $p$-adic BTZ black hole. (As pictured, $p=3$). \label{BTZpadicFig}}
\end{figure}
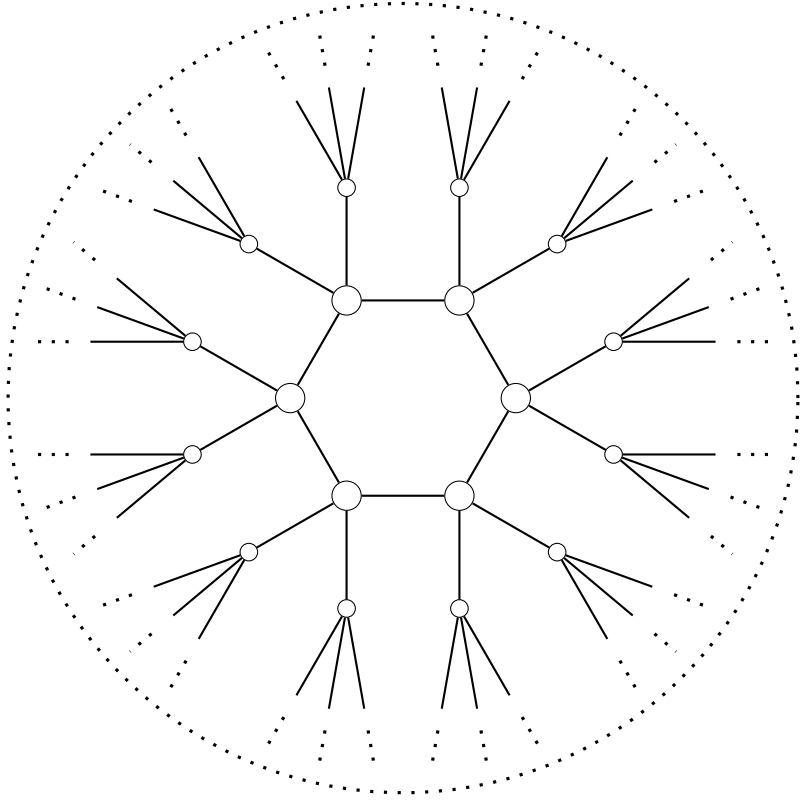

In the special case of a genus-one curve, the relevant Schottky group is isomorphic to $q^\bZ$, for some $q\in \bK^*$, 
and the limit set consists of two points, which we can identify with $0$ and $\infty$ in $\bP^1(\bK)$.
The generator of the group acts on the geodesic in $T_\bK$ with endpoints $0$ and $\infty$
as a translation by some length $n =\log |q|=v_\fm(q)$, the valuation. 
The finite graph $G_\bK$ is then a polygon with
$n$ edges, and the graph $T_\bK / \Gamma$ consists of this polygon with infinite trees
attached to the vertices. The boundary at infinity of $T_\bK / \Gamma$ is a Mumford curve
$X_\Gamma(\bK)$ of genus one with its $p$-adic Tate uniformization. The graph 
$T_\bK / \Gamma$ is the $p$-adic BTZ black hole, with the central polygon $G_\bK$
as the event horizon (see Fig.~\ref{BTZpadicFig}).

The higher genus cases are $p$-adic versions of the higher genus black holes
discussed above, with the finite graph $G_\bK$ as event horizon.

Given a set of generators $\{ \gamma_1, \ldots, \gamma_g \}$ of a $p$-adic
Schottky group, let $n_{\gamma_i}$ be the translation lengths that describe the action of each
generator $\gamma_i$ on its axis $\ell_{\gamma_i}$. More precisely, if an element
$\gamma$ is conjugate in $\PGL(2,\bK)$ to an element of the form
$$ \begin{pmatrix} q & 0 \\ 0 & 1 \end{pmatrix}, $$
then the translation length is $n_\gamma =v_\fm(q)= \ord_\bK(q)$, 
the order (valuation) of $q$. The translation lengths $\{ n_{\gamma_i} \}$
are the Schottky invariants of the $p$-adic Schottky group $\Gamma$. It is shown in \cite{CMR}
that the Schottky invariants can be computed as a spectral flow.

The Drinfeld--Manin holographic formula of \cite{DriMan} for $p$-adic black holes of arbitrary genus is
completely analogous to its archimedean counterpart of \cite{Man91}. 
There is a good notion of $\bK$-divisor on $\bP^1(\bK)$, as a function $\bP^1(\bar\bK)\to \bZ$, with 
$z\mapsto m_z$, with the properties that $m_{z_1}=m_{z_2}$ if $z_1$ and $z_2$ are conjugate over $\bK$;
that all points $z$ with $m_z\neq 0$ lie in the set of points of $\bP^1$ over a finite extension of $\bK$; and
that the set of points with $m_z\neq 0$ has no accumulation point. As before we write such
a divisor as $A=\sum_z m_z\, z$. Given a $\Gamma$-invariant divisor $A$ of degree zero, 
there exists a meromorphic function on $\Omega_\Gamma(\bK)$ with divisor $A$. It is given
by a Weierstrass product 
$$ W_{A,z_0}= \prod_{\gamma \in \Gamma} \frac{w_A(\gamma z)}{w_A(\gamma z_0)}, $$
where $w_A(z)$ is a $\bK$-rational function on $\bP^1(\bK)$ with divisor $A$. The convergence
of this product is discussed in Proposition 1 of \cite{DriMan}: the non-archimedean
nature of the field $\bK$ implies that the product converges for all $z\in \Omega_\Gamma \setminus \cup_\gamma \gamma( \supp(A))$. The function $W_{A,z_0}$ is a $p$-adic automorphic
function (see \cite{Man76}) with $W_{A,z_0}(\gamma z)=\mu_A(\gamma) W_{A,z_0}(z)$,
with $\mu_A(\gamma)\in \bK^*$, multiplicative in $A$ and $\gamma$. One obtains a
basis of $\Gamma$-invariant holomorphic differentials on $X_\Gamma(\bK)$ by taking
$$ \omega_{\gamma_i}= d \log W_{(\gamma_i-1)z_0,z_1}, $$
where
$$ W_{(\gamma-1)z_0,z_1}(z) = \prod_{h\in C(\gamma)} \frac{w_{hz^+_\gamma - h z^-_\gamma}(z)}{w_{hz^+_\gamma - h z^-_\gamma} (z_0)}, $$
for $C(g)$ a set of representatives of $\Gamma/\gamma^\bZ$. 

It is shown in \cite{DriMan} that the order of the cross ratio on $\bP^1(\bK)$ is given by
$$ \ord_\bK \frac{w_A(z_1)}{w_A(z_2)} = \# \{ \ell_{z_1,z_2}, \ell_{a_1,a_2} \}, $$
for $A=a_1 - a_2$ and $\ell_{x,y}$ the geodesic in the Bruhat--Tits tree with endpoints $x,y\in \bP^1(\bK)$, with
$\# \{  \ell_{z_1,z_2}, \ell_{a_1,a_2} \}$ the number of edges in common to the two geodesics in $T_\bK$.
This is the basic $p$-adic holographic formula relating boundary two point function to gravity in the bulk.

\begin{figure}[t]
\includegraphics[width=11cm]{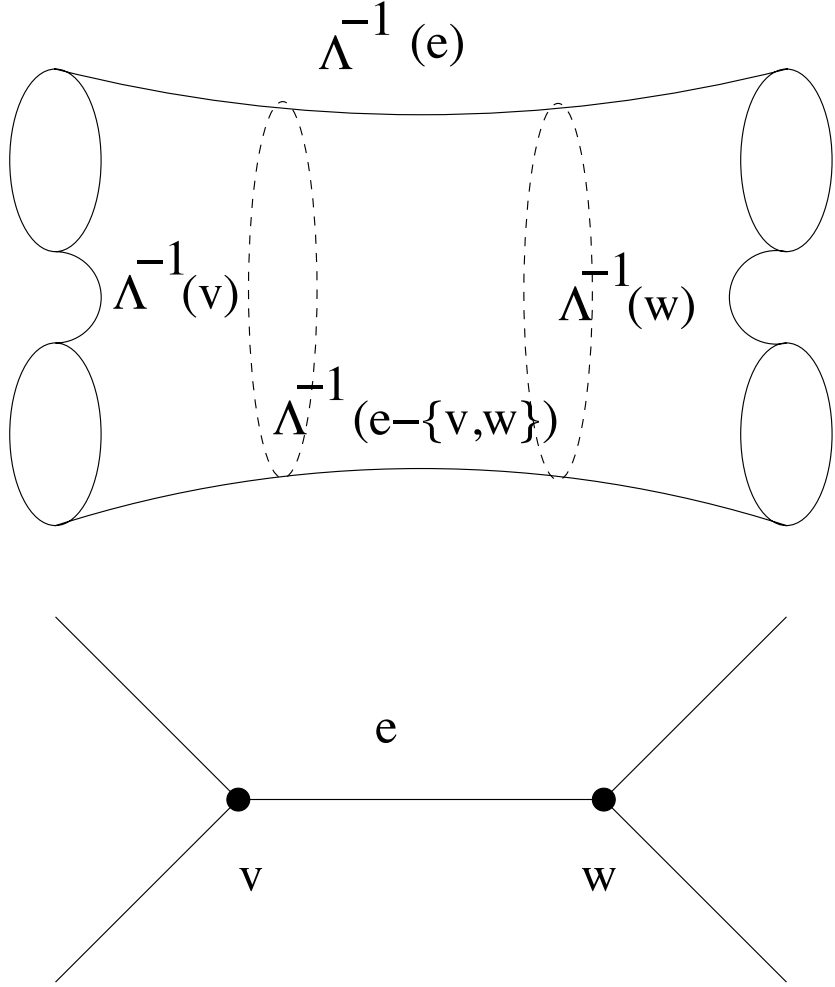} 
\centering
\caption{Drinfeld's $p$-adic upper half plane and the Bruhat--Tits tree. \label{DriFig}}
\end{figure}

A difference with respect to the Archimedean case is that, over $\bC$, both the absolute value and
the argument  of the cross ratio have an interpretation in terms of geodesics, with the absolute
value expressed in terms of lengths of geodesic arcs and the argument in terms of angles between
geodesics, as recalled above. In the $p$-adic case, however, it is only the valuation of the two point
correlation function that has an interpretation in terms of geodesic lengths in the Bruhat--Tits tree.
The reason behind this discrepancy between the archimedean and non-archimedean cases
lies in the fact that the Bruhat--Tits tree $T_\bK$ is the correct analog of the hyperbolic handlebody $\bH^3$
only for what concerns the part of the holographic correspondence that involves the absolute
value (respectively, the $p$-adic valuation) of the boundary two point function. There is a more
refined $p$-adic space, which maps surjectively to the Bruhat--Tits tree, which captures the
complete structure of the $p$-adic automorphic forms for the action of a $p$-adic Schottky group
$\Gamma$: Drinfeld's $p$-adic upper half plane, see Chapter I of \cite{BoutCar}.
Given $\bK$ as above, let $\bC_p$ denote the completion of the algebraic closure of $\bK$.
Drinfeld's $p$-adic upper half plane is defined as $\bH_\bK = \bP^1(\bC_p) \setminus \bP^1(\bK)$.
One can view this as an analog of the upper and lower half planes in the complex case, with
$\bH^+\cup \bH^-= \bP^1(\bC)\setminus \bP^1(\bR)$. There is a surjection $\lambda: \bH_\bK \to T_\bK$,
defined in terms of the valuation, from Drinfeld's $p$-adic upper half plane $\bH_\bK$ to the 
Bruhat--Tits tree $T_\bK$. For vertices $v,w\in V(T_\bK)$ connected by an edge $e\in E(T_\bK)$,
the preimages $\lambda^{-1}(v)$ and $\lambda^{-1}(w)$ are open subsets of $\lambda^{-1}(e)$,
as illustrated in Fig.~\ref{DriFig}. The map $\lambda: \bH_\bK \to T_\bK$ is equivariant
with respect to the natural actions of $\PGL(2,\bK)$ on $\bH_\bK$ and on $T_\bK$. In particular,
given a $p$-adic Schottky group $\Gamma \subset \PGL(2,\bK)$, we can consider the quotients
$\tilde \cH_\Gamma=\bH_\bK/\Gamma$ and $\cH_\Gamma =T_\bK/\Gamma$ and the induced
projection $\lambda: \tilde \cH_\Gamma \to \cH_\Gamma$. Both quotients have conformal boundary
at infinity given by the Mumford curve $X_\Gamma =\Omega_\Gamma(\bK)/\Gamma$, with
$\Omega_\Gamma(\bK)=\bP^1(\bK)\smallsetminus \Lambda_\Gamma$, the domain of
discontinuity of the action of $\Gamma$ on $\bP^1(\bK)=\partial \bH_\bK=\partial T_\bK$.
One can view the relation between $\bH_\bK$ and $T_\bK$ illustrated in Fig.~\ref{DriFig},
and the corresponding relation between $\tilde\cH_\Gamma$ and $\cH_\Gamma$, by
thinking of $\tilde\cH_\Gamma$ as a ``thickening'' of the graph $\cH_\Gamma$, just as in
the Euclidean case one can view the union of the fundamental domains of the action of
$\Gamma$ on $\bH^3$, as illustrated in Fig.~\ref{FundDomH3Fig}, as a thickening
of the Cayley graph (tree) of the Schottky group $\Gamma$, embedded in $\bH^3$.

Thus, when considering the non-archimedean holographic correspondence and 
$p$-adic black holes of arbitrary genus, one can choose to work with either $\cH_\Gamma$
or with $\tilde\cH_\Gamma$ as the bulk space, the first based on Bruhat--Tits trees and
the second (more refined) based on Drinfeld's $p$-adic upper half spaces. 
In this paper we will be focusing on those aspects of the
non-archimedean AdS/CFT correspondence that are captured by the Bruhat--Tits tree, while
we will consider a more refined form of non-archimedean holography, based on Drinfeld's
$p$-adic upper half planes, in forthcoming work. 

\subsection{Scalars on higher-genus backgrounds: sample calculation}

\begin{figure}
\centering
\begin{tikzpicture}[scale=1.7]
\tikzstyle{vertex}=[draw,circle];
\tikzstyle{ver2}=[draw,scale=0.6,circle];
\tikzstyle{ver3}=[];

\foreach \n in {0,1,...,8} {
	\draw (\n,0) node[vertex] (a\n) {\n};
	\draw (\n,0) + (20 - 90:0.7) node[ver2] (b\n) {};
	\draw (\n,0) + (-20 -90:0.7) node[ver2] (c\n) {};
	\draw[thick] (b\n) -- (a\n) -- (c\n);
	\foreach \x in {-1,0,1} {
		\draw (b\n) + (0 - 90 + \x*15:0.7) node[ver3] (b\n\x) {};
		\draw (c\n) + (-0 -90 + \x*15:0.7) node[ver3] (c\n\x) {};
		\draw[thick] (b\n\x) -- (b\n);
		\draw[thick] (c\n\x) -- (c\n);
	};
};
\draw (c20) node[anchor=north] {$v$};
\draw (c20) node[draw,scale=0.3,fill=black,circle,anchor=south] {};
\draw[thick] (-0.5,0) -- (a0); \draw[thick] (a8) -- (8.5,0);
\draw[very thick, loosely dotted] (-0.5,0) -- (-1,0) node[anchor=east] {$0$};
\draw[very thick, loosely dotted] (8.5,0) -- (9,0) node[anchor=west] {$\infty$};
\draw[thick] (a0) -- (a1) -- (a2) -- (a3) -- (a4) -- (a5) -- (a6) -- (a7) -- (a8);

\draw[dashed,->] (0,0.3) -- (2,0.3) node[anchor=south] {$\ell = 4$} -- (4,0.3);
\end{tikzpicture}
\caption{The action of a rank-one Schottky group (translation by~$\ell$ along a fixed geodesic) on the Bruhat--Tits tree. As pictured, $n=h=2$.}
\label{g1tree}
\end{figure}
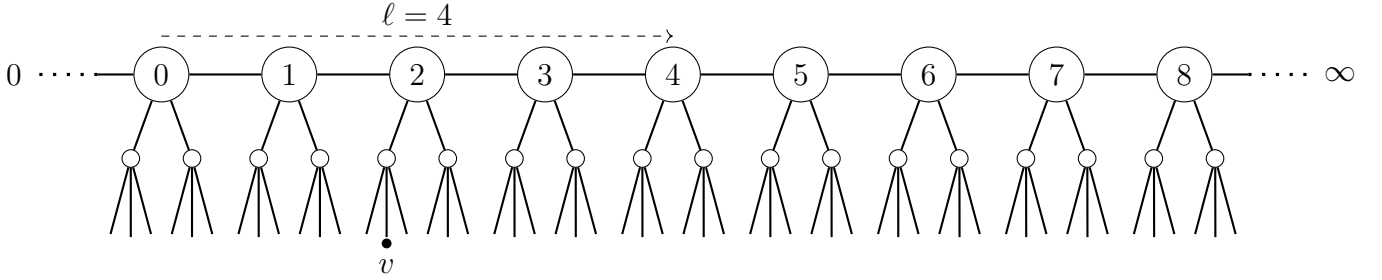

In light of this discussion of higher-genus holography in the $p$-adic case, it is easy to understand how to generalize the arguments and calculations we discussed  for scalar fields in~\S\ref{scalars} to the BTZ black hole, or to higher-genus hyperbolic handlebodies, the $p$-adic analogues of Krasnov's Euclidean black holes. One can simply think of the higher-genus geometry as arising from the quotient of the tree~$T_p$ (and its boundary $\PL(\Q_p)$) by the action of a rank-$g$ Schottky group. Any quantity that can then be made equivariant under the action of the Schottky group will then descend naturally to the higher-genus setting. 

As a simple example, it is easy to construct the genus-1 analogue of our basic Green's function~\eqref{planewave}, using the method of images.  
We perform this calculation in the following paragraphs.
The result makes it easy to perform the reconstruction of bulk solutions to the equations of motion in a BTZ background, with specified boundary conditions at infinity along the genus-1 conformal boundary.

Without loss of generality, we can label the distance along the geodesic which is translated by the chosen Schottky generator by integers, and imagine that the source is attached at a boundary point~$x$ connected to the vertex~$0$. The bulk vertex~$v$ at which we want to evaluate the Green's function will be attached to vertex~$n$ ($0\leq n < \ell$), at a depth~$h$ from the central geodesic. The quantity to be calculated is simply
\begin{equation}
\varepsilon^{(g=1)}_{\kappa,x}(v) = \sum_{g\in\Z} p^{\kappa \abr{v,gx}},
\end{equation}
where the sum ranges over the images of~$x$ under the Schottky group. We take the integrand to be normalized to 1 at the vertex where the branch containing~$x$ meets the central geodesic.
The cases $n=0$ and~$n\neq 0$ are different, and we will treat them separately. 
\begin{enumerate}[leftmargin=0.7in]
\item[$n=0$:]
In this case, the sum becomes
\begin{align}
\varepsilon_{\kappa,x}(v) &= p^{\kappa \abr{v,x}} + 2 \sum_{m>0} \left( p^{-\kappa \ell} \right) ^m \nonumber \\
&= p^{\kappa \abr{v,x}} + \frac{2 p^{-\kappa h}}{p^{\kappa\ell} - 1}.
\end{align}
\item[$n\neq 0$:]
In this case, the sum becomes
\begin{align}
\varepsilon_{\kappa,x}(v) &= \sum_{m\leq 0}  p^{\kappa ( - n - h - \abs{m}  \ell ) } 
 +  \sum_{m>0}  p^{\kappa (  n - h - m  \ell ) }  \nonumber \\
&= p^{-\kappa h} \left( \frac{p^{\kappa(\ell-n)}+p^{\kappa n}}{p^{\kappa\ell}-1} \right).
\end{align}
\end{enumerate}
In  both cases, the result has the expected boundary behavior: it falls off asymptotically as $p^{-\kappa h}$ when~$v$ approaches any boundary point other than~$x$ itself.

\section{Conclusion}
In this work we have proposed an algebraically motivated way to discretize the AdS/CFT correspondence. The procedure of replacing real or complex spacetimes by $\mathbb{Q}_p$ introduces a nontrivial discrete bulk and boundary structure while still preserving many desirable features of the correspondence. The boundary conformal field theory lives on an algebraic curve in both the ordinary and non-archimedean examples; the $\PL(\mathbb{Q}_p)$ theory naturally enjoys the $p$-adic analogue of the familiar global conformal symmetry, $\PGL(2,\mathbb{Q}_p)$. This same group comprises the isometries of the lattice bulk spacetime $ T_p =  \PGL(2, \mathbb{Q}_p) / \PGL(2, \mathbb{Z}_p)$, a maximally symmetric coset space analogous to Euclidean AdS.

In analogy with the BTZ black hole and higher genus examples in $\text{AdS}_3$, higher genus bulk spaces in the $p$-adic case are obtained by Schottky uniformization. One takes quotients of the geometry by $p$-adic Schottky groups $\Gamma \subset \PGL(2,\Q_p)$, producing Mumford curves at the boundary. These curves holographically correspond to bulk geometries consisting of discrete black holes, which appear automatically and do not need to be put in by hand.

Having found a discretization which does not break any symmetries of the problem, we then proposed one way of obtaining a holographic tensor network from a Bruhat--Tits tree. We roughly identify the tree as a space of discrete geodesics in the network. Following Pastawski \textsl{et~al.}'s holographic error-correcting code, the entanglement entropy of a deleted region is reproduced by counting geodesic lengths in the bulk. This perspective puts a stronger notion of bulk geometry into tensor networks, and suggests that the $p$-adic systems considered here may be closer to tensor network models than their archimedean counterparts. This construction might have further applications in entangled bulk states, nongeometric bulk states, and other more exotic features of quantum gravity not present in many existing tensor network models.

After discrete bulk Hilbert spaces in tensor networks, we then turned our attention to continuous Hilbert spaces of scalar fields in the tree. In the semiclassical analysis, massless and massive scalar solutions to the lattice model couple naturally to CFT operators at the boundary, just as in the archimedean case. We identified boundary/bulk propagators in the discrete analog of empty AdS, as well as in the $p$-adic BTZ black hole; the method of images can be used to generalize these results to arbitrary higher-genus bulk backgrounds. We are led to believe that the semiclassical physics of the bulk ``gravity'' theory is dual to an exotic conformal field theory living on the fractal $p$-adic boundary. At the present time, little is known about these $p$-adic conformal field theories outside of $p$-adic string theory; we hope the connection to holography may draw attention to this area. Viewed as a renormalization scale, we have shown that moving up the tree corresponds to exact course graining of boundary mode expansions. The intimate relation between conformal symmetry, AdS~geometry, and renormalization still holds in this entirely discrete setting.

Motivated by the tensor network models, we suggest that the entanglement entropy of regions of the field theory is computed by the unique geodesic lengths in the bulk space. While as of yet we have no formal proof in the free-boson field theory, a number of arguments have been presented which support this conjecture. Under very specific circumstances, it might even be possible to learn certain properties of the archimedean entanglement entropies from their corresponding $p$-adic counterparts with the help of adelic formulas. 

While we have established some essential features of $p$-adic holography, ranging from algebraic curves to tensor networks and from bulk/boundary propagators and renormalization scales to entanglement, much about these exotic systems remains to be understood. We propose a number of ideas to be explored in future work.

One major ingredient missing from our story is a proper description (and quantization) of the gravitational degrees of freedom. The bulk geometries (with or without black holes in the interior) can loosely be described as $p$-adic discretizations of asymptotically AdS spacetimes. One way to add dynamical metric degrees of freedom without spoiling the asymptotic behavior might be to make the edge lengths of the Bruhat--Tits tree dynamical. The $p$-adic version of empty AdS might correspond to a solution with uniform edge lengths like the system considered here; thermal or black hole states seem to require topology change in the interior. 

If we believe that the full quantum gravity Hilbert space of the interior involves fluctuating edge lengths and graph topology, one might ask if tensor network models could be adapted to this picture. More complicated tensor networks might be used to study objects such as black holes, EPR pairs, and nongeometric states. The role of planarity of the tensor network may play an important role in this story.

From the point of view of the $p$-adic conformal field theory, one might ask for more interesting examples than the free boson. We have already offered some speculations about higher spin fields based on representation theory of the $p$-adic conformal group; it would be nice to formulate these models explicitly and search for interesting gravity duals. Additionally, the models we have studied so far do not appear to have extended conformal symmetry or a central charge. These important ingredients of 1+1 dimensional CFT's might appear with the more careful inclusion of finite extensions of $\mathbb{Q}_p$. These finite extensions might also be linked to the passage to Lorentzian signature.

To briefly mention another  important idea to which we will return in future work:  one should note that the path integral of a bulk field is itself very much like a tensor network, albeit one where the local Hilbert spaces are infinite-dimensional. A boundary state can be obtained by doing the bulk path integral with fixed boundary conditions; this defines a wave-functional on the space of boundary field configurations. The state is then computed by concatenating many copies of a single universal linear map, defined by integrating out one field and adding the terms in the action that couple it to $p$ of its neighbors. There is a sense, therefore, in which the bulk theories we have discussed already are tensor networks---recipes for constructing states that are built out of infinite hierarchical networks of concatenated, homogeneous, locally similar linear maps. 
Many questions leap to mind about this---for example, what are the error-correcting properties of the basic linear map defined by our path integral? Can one imagine truncating the path integral in a sensible way to define a tensor network of finite local dimension? We look forward to investigating these questions in detail.

We finally address future work for the entanglement entropy in a $p$-adic holographic theory. As already mentioned, the single and multiple interval entanglement entropies will likely require a detailed replica computation. This may be possible through a more detailed study of branched covers of the $p$-adic plane as Mumford curves. With entanglement entropies in hand, one might ask for new and old proofs of entropy inequalities; these are expected to be simplified by the ultrametric nature of the $p$-adics. Finally, it remains to be seen how much can be learned about real AdS/CFT from studying these systems adelically over every prime.

\subsection*{Note added}

As this work was being completed, we became aware of~\cite{Gubser}, which treats similar ideas from a somewhat different viewpoint, and in which some of our results in~\S\ref{sec41}--\ref{sec43} were independently obtained.

\subsection*{Acknowledgements}

The authors wish to thank N.~Bao, H.~Kim, M.~Kolo\u{g}lu, T.~McKinney, N.~Hunter-Jones, B.~Michel, M.~M.~N\u{a}st\u{a}sescu, H.~Ooguri, and A.~Turzillo for helpful conversations as this work was being prepared. I.A.S.\ is also grateful to the Partnership Mathematics and Physics of Universit\"at Heidelberg, the University of Bristol, and the Gone Fishing meeting at the University of Colorado, Boulder for hospitality. We are especially grateful to the anonymous referee, for careful reading and thorough and useful comments.

The work of M.H., I.A.S., and B.S.\ is supported by the United States Department of Energy under the grant DE-SC0011632, as well as by the Walter Burke Institute for Theoretical Physics at Caltech. M.M. is partially supported by NSF grants
DMS-1201512 and PHY-1205440, and by the Perimeter Institute for Theoretical Physics.

\begin{appendices}

\section{$p$-adic integration}
\label{app:integrals}

Here we review some aspects of $p$-adic integration, including basic properties and examples, the Fourier transform, and the $p$-adic gamma function $\Gamma_p$. A more comprehensive review is found in~\cite{BrFr}. For formal proofs, as well as extensive integration tables, the reader may consult \cite{Vladimirov}.

As already discussed, the unique additive Haar measure $dx$ on $\mathbb{Q}_p$ is normalized so that
\begin{equation}
\int_{\mathbb{Z}_p} dx = 1.
\end{equation}

To find the volume of the set $B^{r}$, which consists of $x \in \{\mathbb{Q}_p ,  |x|_p \leq p^r \}$, we may  scale the measure and reduce this to the integral above on $\mathbb{Z}_p$ as:
\begin{equation}
\int_{B^r} dx = p^r \int_{\mathbb{Z}_p} dx = p^r.
\end{equation}
As $r \rightarrow \infty$, the volume diverges as in the real case. Compactifying the point at infinity amounts to switching from the Haar measure to the Patterson-Sullivan measure $d \mu_0(x)$; these measures agree on $\mathbb{Z}_p$ and differ in the complement by $d \mu_0(x) = dx/|x|_p^2$.

With this measure the volume is computed with a change of variables:
\begin{align}
\int_{\mathbb{Q}_p} d \mu_0(x) &= \int_{\mathbb{Z}_p} dx + \int_{\mathbb{Q}_p-\mathbb{Z}_p} |x|_p^{-2} dx \\
&= 1 + \frac{1}{p} \int_{\mathbb{Z}_p} du, \hspace{3mm} u= \frac{1}{px}, \hspace{2mm} du = \frac{p\, dx}{|x|_p^2}\\
&= \frac{p+1}{p}.
\end{align}
A large class of elementary integrals may be evaluated using these methods; see the above references for complete details.

We now turn our attention to the $p$-adic Fourier transform of a function $f(x): \mathbb{Q}_p \rightarrow \mathbb{C}$. As discussed in section~\ref{sec:modes}, this involves integrating the function against the additive character  $\chi(x) = e^{2 \pi i \{ kx \} }$ over all $\mathbb{Q}_p$. This generates a new complex valued function in terms of the $p$-adic momentum $k \in \mathbb{Q}_p$:
\begin{align}
\widetilde{f}(k) &= \int_{\mathbb{Q}_p} \chi(kx) f(x) dx, \\
f(x) &= \int_{\mathbb{Q}_p} \chi(-kx) \widetilde{f}(k) dk.
\end{align}

The analogy with the real Fourier transform should be clear. In practice evaluating this kind of integral often requires one to divide $\mathbb{Q}_p$ into spheres consisting of points with $|x|_p = p^n$ and performing the integral on each sphere. This can be seen in the example:
\begin{equation}
\int_{B^r} \chi(kx) dx = 
\begin{cases}
    p^r, &\hspace{.5mm} |k|_p \leq p^{-r}\\
    0,  & \text{otherwise.}
\end{cases}
\end{equation}
As in the real case, one may find tables with numerous $p$-adic Fourier transforms of elementary functions in the literature.

The final integral expression is that of the Gelfand-Graev-Tate $\Gamma$ function:

\begin{equation}
\Gamma_p (\alpha) =  \int_{\mathbb{Q}_p} \chi(x) |x|_p^{\alpha -1} dx = \frac{1-p^{\alpha-1}}{1-p^{-\alpha}}.
\label{Gammap}
\end{equation}
This function has some similar properties to the ordinary gamma function. It is fairly ubiquitous in certain $p$-adic integral calculations, and we refer the reader to literature on $p$-adic string theory for details.

\section{$p$-adic differentiation}
\label{app:derivatives}


As already discussed, complex fields living on the boundary $\PL(\mathbb{Q}_p)$ are maps
\begin{equation}
f(x): \PL(\mathbb{Q}_p) \rightarrow \mathbb{C}.
\end{equation} 
In the archimedean case of 2d conformal field theory, we have $f(z, \bar{z}): \PL(\mathbb{C}) \rightarrow \mathbb{C}$ and it makes sense to define holomorphic and antiholomorphic derivatives $\frac{\partial f}{\partial z}$ and $\frac{\partial f}{\partial \bar{z}}$, using the normal definition of derivative. In the $p$-adic case, the analogous differentiation expressions no longer make sense, as we would be dividing a complex number by a $p$-adic number and  no such operation is  \textit{a priori} defined.

The only notion of derivative we may use is the \textit{Vladimirov derivative}~\cite{Vladimirov, Dragovich}, which is a nonlocal pseudo-differential operator. Roughly speaking, this operation is the $p$-adic analog of Cauchy's Differentiation Formula, in which the derivative of a function at a point is expressed as a weighted integral of the function over a curve. It is also known as a \textit{normal derivative} \cite{Zabrodin} in the context of the $p$-adic string, where it is interpreted as the derivative of the embedding coordinates $X^{\mu}$ normal to the boundary of the worldsheet. Because this operator is defined on $\mathbb{Q}_p$ without any reference to an embedding or worldsheet, we opt to refer to it as a Vladimirov derivative. The $n^{th}$ Vladimirov derivative is defined by the expression
\begin{equation}
\partial_{(p)}^n f(x) = \int_{\mathbb{Q}_p} dx'\, \frac{f(x') -f(x)}{|x'-x|_p^{n+1}} .
\label{vlad}
\end{equation}
In this expression, $n$ is frequently an integer, but it may in principle assume any real value. This is a regularized way of writing the convolution of~$f$ with the kernel $|x|_p^{-(1+n)}$; the convolution integral often does not converge, whereas~\eqref{vlad} (which is the same thing up to a shift by an infinite constant) is always well-defined and finite.
Some authors may choose a different normalization constant; the most common is
\begin{equation}
D^n f (x) = \frac{1}{\Gamma_p(-n)} \int dy\, \frac{f(y)-f(x)}{|y-x|_p^{1+n}},
\label{vlad-norm}
\end{equation}
where~$\Gamma_p$ is the $p$-adic gamma function. This is done so that Vladimirov derivatives obey the expected composition law on the nose: 
\begin{equation}
D^a D^b = D^b D^a = D^{a+b}.
\end{equation}
 
At first sight the expression above may not resemble any familiar notions of differentiation. We may see this as a good notion for derivative in two ways; in the case of the $p$-adic string this expression is the boundary limit of the normal derivative on $T_p$, as shown in \cite{Zabrodin}. We may also compute the Vladimirov derivative of some functions and compare with the real case. This is done in the following section.

\subsection{Examples}
We wish to first compute the derivative of the additive character, $\chi(kx)$. This function is the $p$-adic analog of a plane wave with momentum $k$, so we expect it to be an eigenfunction of the derivative with eigenvalue related to $k$. 
We can change variables to $y = k(x'-x)$ (for which $dy = |k|_p dx'$) and simplify the integral appearing in the Vladimirov derivative:
\begin{align}
\partial_{(p)}^n \chi(kx) &= \int_{\mathbb{Q}_p} \frac{ \chi(kx') - \chi(kx)}{|x'-x|_p^{n+1}}  dx' \\
&= |k|_p^n \int_{\mathbb{Q}_p} \frac{ \chi(y + kx) - \chi(kx)}{|y|_p^{n+1}}  dy \\
&= |k|_p^n \chi(kx) \int_{\mathbb{Q}_p} \frac{ \chi(y) - 1}{|y|_p^{n+1}}  dy ,
\end{align}
where we used the additive property of the character to extract the $x$ dependence. The integrand appears to diverge as $y \rightarrow 0$; however, the numerator is actually zero in an open neighborhood of the origin, so that the integral is finite.
(This integral is discussed in detail in~\cite{Bikulov} and~\cite{Ghoshal}.) The result is
\begin{equation}
 \int_{\mathbb{Q}_p} \frac{ \chi(y) - 1}{|y|_p^{n+1}}  dy = \frac{1-p^{-n-1}}{1-p^n} = \Gamma_p (-n),
\end{equation}
where we have used the definition of the $p$-adic gamma function in Eq. (\ref{Gammap}). So the end result is 
\begin{equation}
\partial_{(p)}^n \chi(kx) = \Gamma_p (-n) |k|_p^n \chi(kx).
\end{equation}
The additive character $\chi(kx)$ is therefore an eigenfunction of the Vladimirov derivative, with eigenvalue given (up to the factor of the gamma function) by the $p$-adic norm of its ``momentum'' $k$. For the derivatives with normalization~\eqref{vlad-norm}, we would have precisely
\begin{equation}
D^n \chi(kx) = |k|_p^n\, \chi(kx).
\end{equation}

Another example we may wish to compute is the $n^{th}$ derivative of $|x|_p^s$ for some $s \in \mathbb{C}$. This may be most easily be computed by Fourier transform and serves as an example of an alternative representation of the Vladimirov derivative:
\begin{equation}
\partial_{(p)}^n |x|_p^s = \int \chi(-kx)|k|_p^n \widetilde{|x|^s}_p dk,
\end{equation}
where $\widetilde{|x|^s}_p$ is the $p$-adic Fourier transform of $|x|_p^s$, given in~\cite{Zhang, Vladimirov}:
\begin{equation}
\widetilde{|x|^s}_p =  \int \chi(kx) |x|^s_p dx = \Gamma_p(s+1)|k|_p^{-s-1}
\end{equation}
everywhere it is defined. Applying this formula twice to the derivative we wish to compute, we arrive at
\begin{equation}
\partial_{(p)}^n |x|_p^s = \Gamma_p(s+1) \Gamma_p(n-s) |x|_p^{s-n},
\end{equation}
which should resemble the ordinary $n^{th}$ derivative of a polynomial function.




\end{appendices}
\end{spacing}



\end{document}